\documentclass[notitlepage,twocolumn,letterpaper,natbib,aps,prd,amsmath,amsfonts,nofootinbib,preprintnumbers,superscriptaddress,secnumarabic]{revtex4-1}
\pdfoutput=1
\usepackage{amssymb,amsmath,latexsym,mathrsfs}
\usepackage{url}
\usepackage{enumitem}
\usepackage{graphicx}
\usepackage{subfig}
\usepackage{booktabs}
\usepackage[usenames,dvipsnames]{color}
\usepackage[breaklinks,colorlinks,urlcolor=blue,citecolor=blue,linkcolor=magenta]{hyperref}
\usepackage{multirow}
\usepackage{float}
\usepackage{cases}
\usepackage{blindtext}
\usepackage{pifont}
\usepackage{hhline}
\usepackage{mathtools}
\usepackage[normalem]{ulem}
\usepackage{amsmath}
\usepackage{orcidlink}
\usepackage{ragged2e}
\usepackage{cancel}

\usepackage{xcolor}
\definecolor{linkcolor}{rgb}{0.0, 0.47, 0.75}
\definecolor{citecolor}{rgb}{1.0, 0.5, 0.0}
\hypersetup{
  linkcolor  = linkcolor,
  citecolor  = linkcolor,
  urlcolor   = linkcolor,
  colorlinks = true
}

\begin{document}

\title{How large could CP violation in neutral $B$ meson mixing be? \\ Implications for baryogenesis and upcoming searches}

\preprint{CERN-TH-2024-172}

\author{Carlos Mir\'o\,\,\orcidlink{0000-0003-0336-9025}}
\email{carlos.miro@uv.es}
\affiliation{Departament de F\'isica Te\`orica and Instituto de F\'isica Corpuscular (IFIC),\\ Universitat de Val\`encia -- CSIC, E-46100 Valencia, Spain.}

\author{Miguel Escudero\,\,\orcidlink{0000-0002-4487-8742}}
\email{miguel.escudero@cern.ch}
\affiliation{Theoretical Physics Department, CERN, 1211 Geneva 23, Switzerland}

\author{Miguel Nebot\,\,\orcidlink{0000-0001-9292-7855}\,}
\email{miguel.nebot@uv.es}
\affiliation{Departament de F\'isica Te\`orica and Instituto de F\'isica Corpuscular (IFIC),\\ Universitat de Val\`encia -- CSIC, E-46100 Valencia, Spain.}


\begin{abstract}
CP violation in neutral $B$ meson oscillations is an experimental observable that could be directly related to the baryon asymmetry of the Universe through the $B$-Mesogenesis mechanism. As this phenomenon is highly suppressed in the Standard Model, it could also be a sensitive probe for many new physics scenarios that modify neutral meson mixing. Motivated by these facts, and the timely $B$ physics program at the LHC and Belle II, we analyze how large CP violation in the mixing of neutral $B_d$ and $B_s$ meson systems could be. We answer this question, in light of current experimental data, within three different scenarios, namely: (i) generic heavy new physics only affecting the mass mixing $M_{12}^q$, (ii) vector-like quark extensions that introduce deviations of 3$\times$3 CKM unitarity, and (iii) light new physics modifying the decay mixing $\Gamma_{12}^q$. We find that enhancements of the semileptonic asymmetries, that measure the amount of CP violation in mixing, at the level of $10^{-3}$ for the $B_d$ system and $10^{-4}$ for the $B_s$ system can be achieved within scenarios (i) and (ii), while they are much more suppressed in realistic UV completions triggering scenario (iii). With respect to cosmology, the difficulty of finding large CP asymmetries in our analysis puts the $B$-Mesogenesis mechanism in tension. Finally, we conclude that upcoming experimental searches for CP violation in $B$ meson mixing at LHCb and Belle II are unlikely to detect a new physics signal for the most generic models.
\end{abstract}

\maketitle
\tableofcontents

\section{Introduction}\label{SEC:Introduction}

CP violation in the Standard Model is a consequence of the existence of three families, as predicted by Kobayashi and Maskawa \cite{Kobayashi:1973fv}, and as beautifully confirmed by an array of measurements in the quark sector, see~\cite{Workman:2022ynf}. However, CP violation has not yet been observed in all quark flavor transitions. In this study, we will focus on the yet to be measured CP violation in neutral $B$ meson mixings.

Neutral $B$ mesons are eigenstates of the strong and electromagnetic interactions with quark flavor content $B_q = \bar{b}q$ and $\bar{B}_q = b\bar{q}$, with $q = d,\,s$. Like kaons and $D$ mesons, they are not mass eigenstates and they form an ultra-degenerate oscillating system $B_q \leftrightarrow \bar{B}_q$, with $\Delta M_{B_d}/M_{B_d} \simeq 10^{-13}$ and $\Delta M_{B_s}/M_{B_s} \simeq 10^{-12}$. We know that kaon mass eigenstates ($K_S$ and $K_L$) are not CP eigenstates and that CP is violated in kaon-antikaon oscillations at the level of $10^{-3}$. On the other hand, the Standard Model (SM) predicts that CP is violated in $B$ meson oscillations at the level of $10^{-5}$~\cite{Artuso:2015swg,Albrecht:2024oyn}:
\begin{subequations}\label{eq:SMpredictionCP}
    \begin{align}
    A_{\rm SL}^{d,\mathrm{SM}} \simeq -4\times 10^{-4}\quad [{\rm SM \,\,prediction}]\, ,\\ 
    A_{\rm SL}^{s,\mathrm{SM}} \simeq +2\times 10^{-5} \quad [{\rm SM\,\, prediction}]\, .
\end{align}
\end{subequations}
Here, as it is common practice, $A_{\mathrm{SL}}^q$ encodes the amount of CP violation in neutral $B$ meson mixing, with SL standing for ``semileptonic'' since these are the typical modes where it is searched for.

CP violation in neutral $B$ meson oscillations is important for two reasons: 1) it is highly suppressed in the Standard Model, meaning that, if it is measured, it would constitute a clear indication of physics beyond the Standard Model (BSM); 2) as this type of CP violation arises in mixing, its effects are imprinted into \emph{any} $B$ meson decay. The latter is key because it could also be imprinted into novel additional BSM decays of $B$ mesons. In this context, the $B$-Mesogenesis mechanism~\cite{Elor:2018twp,Nelson:2019fln,Alonso-Alvarez:2021qfd} (see also~\cite{Ghalsasi:2015mxa,McKeen:2015cuz,Aitken:2017wie,Alonso-Alvarez:2019fym,Elor:2020tkc,Elahi:2021jia,Elor:2024cea}) combined precisely this feature (CP violation in $B$ meson oscillations) with a new decay mode of $B$ mesons into a baryon and a dark sector antibaryon, aiming to generate both the observed asymmetry between matter and antimatter as well as the dark matter abundance in the early Universe. Notably, within this mechanism, \emph{the baryon asymmetry of the Universe is directly proportional to the CP violation in neutral $B$ meson mixing}. In particular, in order to generate the observed baryon asymmetry of the Universe it is required that, at least~\cite{Alonso-Alvarez:2021qfd}:\footnote{It has been recently pointed out that radiative decays and inverse decays of resonant vector mesons $B_q^{*} \leftrightarrow B_q + \gamma$ will act as to suppress the generation of the baryon asymmetry within $B$-Mesogenesis, and probably one would need a semileptonic asymmetry larger than Eq.~\eqref{eq:minAsymmetry}~\cite{Alonso-Alvarez:2024inprep}. \label{footnote1}}
\begin{align}\label{eq:minAsymmetry}
A_{\rm SL}^{q} > +10^{-4}\quad[\text{required for Baryogenesis}].
\end{align}
Hence, the magnitude of CP violation in neutral $B$ meson mixing can constitute a sensitive probe of New Physics (NP), and also be at the origin of the observed asymmetry between matter and antimatter in the Universe. Unfortunately, it turns out that the CP asymmetry in the SM is not enough by at least an order of magnitude (compare Eqs.~\eqref{eq:SMpredictionCP} with Eq.~\eqref{eq:minAsymmetry}) to generate the observed baryon asymmetry, thus requiring the presence of new CP violating effects affecting $B$ meson oscillations.

There have been many studies of models of new physics that modify $B$ meson mixing and its CP violation --for global analyses see~\cite{Bauer:2010dga,Ligeti:2010ia,Lenz:2011zz,Bobeth:2011st,Lenz:2012az,Bevan:2013kaa,Charles:2013aka,Artuso:2015swg,Charles:2020dfl,DeBruyn:2022zhw} and for specific models see~\cite{Dorsner:2016wpm,Buttazzo:2017ixm,DiLuzio:2019jyq,DiLuzio:2019jyq,Allanach:2023uxz,Athron:2023hmz,Dobrescu:2010rh,Trott:2010iz,Bai:2010kf,Botella:2014qya,Iguro:2018qzf,Crivellin:2019dun,Athron:2021auq,Bonilla:2022qgm,Li:2024thq,Zheng:2022ssr,Datta:2022zng,Davighi:2023xqn}. However, the extent to which these new physics models can affect the CP violation in $B$ meson mixing has not been recently addressed. The latest detailed analyses date from over a decade, when the D0 collaboration reported an anomalously large CP asymmetry involving both $B_d$ and $B_s$ decays, see~\cite{D0:2010sht,D0:2011hom,D0:2013ohp}. After new measurements by LHCb, Belle and BaBar, it has become clear that global averages are compatible with the absence of CP violation in neutral $B$ meson mixing~\cite{HFLAV:2022esi}:
\begin{subequations}\label{eq:CP_averages}
\begin{align}
    A_{\rm SL}^{d,\mathrm{Exp}} &= (-21\pm 17)\times 10^{-4}\quad [{\rm world\,average\,`24}]\, ,\\ 
    A_{\rm SL}^{s,\mathrm{Exp}} &= (\,\,\,-6\pm 28)\times 10^{-4}\quad [{\rm world\,average\,`24}]\, .
\end{align}
\end{subequations}
In this context, we believe there are four reasons to explore this subject again, namely: 1) with the LHC there have been many new measurements of CP violating observables in the $B$ meson sector, which have implications for these types of models; 2) these CP asymmetries will be measured with higher precision at LHCb as well as in Belle II in the upcoming years; 3) the models of new physics which can affect the CP asymmetries face now many novel bounds from the LHC as a result of the plethora of stringent searches for new phenomena that in many cases require $\Lambda_{\rm NP} > 1\,{\rm TeV}$; and 4) there exists now a cosmological link between CP violation in neutral $B$ meson mixing and the baryon asymmetry of the Universe. 

Therefore, our main goal in this work is to understand how large CP violation in neutral $B$ meson mixing can be. We believe this is important in order to highlight what the relevance of future measurements of this quantity at collider experiments is, its implications for BSM models (that is, what sort of new physics could be at play if a non-standard value is measured), and finally to understand the conditions needed BSM to generate a large enough CP asymmetry to have successful baryogenesis through neutral $B$-Mesogenesis (see Eq.~\eqref{eq:minAsymmetry}).

In order to achieve these goals, we consider three different scenarios. 1) Models where the main modification to $B$ meson mixing arise from contributions to \emph{mass mixing}, $M_{12}^q$. These are the most standard modifications, as $M_{12}^q$ is dominated by virtual states, and heavy particles inaccessible to the LHC could contribute to it. 2) New physics models with additional vector-like quarks inducing deviations from $3\times 3$ CKM unitarity. The motivation for this follows from the original purpose of introducing 3 quarks in the Standard Model to allow for CP violation, and we want to understand how this type of scenario can contribute to or change these CP asymmetries. 3) BSM models that contribute to the \emph{decay width mixing}, $\Gamma_{12}^q$. These scenarios necessarily need to involve rather light states, and as we will check, these are highly constrained by LHC searches. 

Overall, our main conclusions are that upcoming measurements of CP violation in $B$ mixing will not be able to test the most generic new physics models, and that the $B$-Mesogenesis mechanism is to some extent theoretically depreciated because it is not easy to obtain large enough semileptonic CP asymmetries BSM. A visual description of these results is displayed in Fig.~\ref{fig:ASL_global_BMesogenesis}.

Our study is structured as follows. In Sec.~\ref{SEC:BMesongeneral}, we present an overview of neutral $B$ meson systems as well as the current status of our knowledge of their mixing. In Sec.~\ref{SEC:NPM12q}, we perform a global analysis of new physics models which affect mass mixing. In Sec.~\ref{SEC:VLQ}, we present the impact on the CP asymmetries from models including vector-like quarks that allow for deviations of $3\times 3$ unitarity in the CKM matrix. Sec.~\ref{SEC:G12q} is dedicated to BSM scenarios that can modify decay mixing. In Sec.~\ref{sec:implications} we discuss our results and compare them with the minimal requirements needed for successful $B$-Mesogenesis as well as upcoming sensitivities from LHCb and Belle II. Finally, in Sec.~\ref{sec:conclusions} we draw our conclusions. We also refer the practitioners to the appendices. There we discuss in depth the effective 2$\times$2 Hamiltonian for $B_q$--$\bar{B}_q$ meson systems in App.~\ref{App:DetailsSM}, we provide a detailed discussion on how the various contributions to $A_{\rm SL}^q$ arise in the global analyses in App.~\ref{app:G12M12analysis}, we outline the interactions and mixing pattern of scenarios with vector-like quark singlets in App.~\ref{App:ModelsVLQ}, their contributions to $M_{12}^q$ in App.~\ref{app:M12VLQ}, and loop functions related to $B$-Mesogenesis in App.~\ref{app:formulaeBMesogenesis}.

\section{Neutral $B$ meson mixing}\label{SEC:BMesongeneral}
\subsection{General framework and experimental status}
In this section, we introduce the fundamental concepts related to the phenomenon of neutral $B$ meson oscillations, and take the opportunity to set the notation that will be used throughout the manuscript. In addition, we compare the SM prediction with the latest experimental results on the mixing observables that are of interest for the subsequent analyses. Practitioners may want to skip this section as it is introductory, and we also refer to App.~\ref{App:DetailsSM} where the effective Hamiltonian is calculated in the Standard Model with great level of detail.

The mesons $B_q = \bar{b}q$ and  $\bar{B}_q = b\bar{q}$ ($q = d, s$) are eigenstates of the strong and electromagnetic interactions with opposite flavor content. Once the weak interaction is considered, they both mix and decay to other states. 

In the Weisskopf-Wigner approximation \cite{Weisskopf:1930ps,Weisskopf:1930au}, the time evolution\footnote{The time $t$ is measured in the rest frame of the $B_q$--$\bar{B}_q$ system.} of a superposition of neutral $B$ mesons, 
\begin{equation}
    |\psi(t)\rangle = a(t) |B_q\rangle + b(t) |\bar{B}_q\rangle\,,
\end{equation}
is controlled by the Hamiltonian
\begin{equation}\label{eq:Hamiltonian2x2}
    \!\!\!\! \mathcal{H}^q = M^q - i \frac{\Gamma^q}{2} = \begin{pmatrix}
M_{11}^q - i\Gamma_{11}^q/2 & M_{12}^q - i\Gamma_{12}^q/2 \\
M_{21}^q - i\Gamma_{21}^q/2 & M_{22}^q - i\Gamma_{22}^q/2
\end{pmatrix} \,,
\end{equation}
that is,
\begin{equation}
\mathcal H^q |\psi(t)\rangle = i\frac{d}{dt}|\psi(t)\rangle,\qquad |\psi(t)\rangle = \exp(-i\mathcal H^q t)|\psi(0)\rangle\,,
\end{equation}
with $M^q = M^{q\dagger}$ and $\Gamma^q = \Gamma^{q\dagger}$. The Hamiltonian is not diagonal in the $\{|B_q\rangle, |\bar{B}_q\rangle\}$ basis, meaning that neutral $B$ mesons will \emph{oscillate}, i.e., have a time-dependent varying flavor content in terms of $B_q$ and $\bar B_q$. Furthermore, CPT invariance, which is assumed hereafter, leads to $M_{11}^q = M_{22}^q$ and $\Gamma_{11}^q = \Gamma_{22}^q$. 

The off-diagonal elements are precisely the ones responsible for the $B_q \leftrightarrow \bar{B}_q$ transitions in two possible ways: (i) through intermediate virtual (off-shell) states encoded in the \textit{mass mixing} $M_{12}^q$, or (ii) through intermediate real (on-shell) states related to the \textit{decay width mixing} $\Gamma_{12}^q$. For further details on the physical meaning of these quantities and their relation to the underlying fundamental interactions, see App.~\ref{App:DetailsSM}.

On that respect, the mass and decay width differences between the heavy ($H$) and light ($L$) eigenstates of the Hamiltonian can be expressed,\footnote{The eigenvalues of $\mathcal H^q$ are $\mu_H^q=M_H^q-i(\Gamma_H^q/2)$ and $\mu_L^q=M_L^q-i(\Gamma_L^q/2)$, with $M_H^q, M_L^q, \Gamma_H^q, \Gamma_L^q \in \mathbb{R}$, and $M_H^q>M_L^q$ by definition. Notice the flipped $H\leftrightarrows L$ convention in $\Delta\Gamma_q$ with respect to $\Delta M_q$, commonly used in the literature and in agreement with the Particle Data Group (PDG) notation \cite{Workman:2022ynf}.} up to $\mathcal{O}(|\Gamma_{12}^q / M_{12}^q|^2)$, as
\begin{align}
    &\Delta M_q \equiv M_{H}^q - M_{L}^q \simeq 2|M_{12}^q|,\\
    &\Delta \Gamma_q \equiv \Gamma_{L}^q - \Gamma_{H}^q \simeq 2|\Gamma_{12}^q|\cos{\phi_{12}^q}\,,
\end{align}
where
\begin{equation}
    \phi_{12}^q \equiv \mathrm{arg}\left(-\frac{M_{12}^q}{\Gamma_{12}^q}\right)\,.
\end{equation}
This relative phase is directly related to the phenomenon of CP violation in $B$ meson mixing, which will translate into a deviation of $\Gamma_{12}^q / M_{12}^q$ from being a real quantity. 

More precisely, $\phi_{12}^q$ emerges in the computation of CP asymmetries involving flavor-specific decays characterized by a final state $f$ such that $B_q \nrightarrow \bar{f}$, $\bar{B}_q \nrightarrow f$ (i.e., $B_q$ cannot decay into $\bar f$ and $\bar B_q$ cannot decay into $f$) and $\langle f|\mathcal{T}|B_q \rangle = \langle \bar{f}|\mathcal{T}|\bar{B}_q \rangle$, being $\mathcal{T}$ the transition matrix describing the underlying fundamental interactions controlling these processes. In particular, one can consider the semileptonic decay modes $B_q \rightarrow X \ell^{+}\nu_\ell$ and $\bar{B}_q \rightarrow X \ell^{-}\bar{\nu}_\ell$, where the charge of the final state lepton projects the flavor content of the $B$ meson at the time of decay (the $\Delta F=\Delta Q$ rule of charged current weak interactions).\footnote{The $\bar{b}$ quark of the $B_q$ meson decays into a positive charged lepton by emitting a $W^{+}$ boson.} In those cases, the semileptonic asymmetries are defined as
\begin{equation}
    A_{\mathrm{SL}}^q \equiv \frac{\Gamma(\bar{B}_q(t) \rightarrow f)-\Gamma(B_q(t) \rightarrow \bar{f})}{\Gamma(\bar{B}_q(t) \rightarrow f)+\Gamma(B_q(t) \rightarrow \bar{f})}\,.
\end{equation}
Notice that, in the previous equation, we have explicitly included the time dependence of the meson states to make clear that, although they are produced as a $B_q$ ($\bar{B}_q$) meson at the initial time, they may oscillate into a $\bar{B}_q$ ($B_q$) and then decay to $\bar{f}$ ($f$). Therefore, the semileptonic asymmetries are actually measuring the difference between the probability of a $\bar{B}_q$ mixing into $B_q$ and a $B_q$ mixing into $\bar{B}_q$, that is, CP violation in mixing. Furthermore, it is interesting to remark that a positive value of $A_{\mathrm{SL}}^q$ indicates that it is more likely that a $\bar{B}_q$ oscillates into a $B_q$ rather than the opposite, which is precisely the scenario where the $B$-Mesogenesis mechanism can generate the observed baryon asymmetry of the Universe.

Taking into account the time evolution of $B$ mesons, it is straightforward to check that, up to $\mathcal{O}(|\Gamma_{12}^q / M_{12}^q|^2)$,
\begin{equation}\label{eq:ASL_def}
    A_{\mathrm{SL}}^q = \mathrm{Im}\left(\frac{\Gamma_{12}^q}{M_{12}^q}\right) = \left|\frac{\Gamma_{12}^q}{M_{12}^q}\right|\sin{\phi_{12}^q}\,,
\end{equation}
so that a signal of CP violation in mixing, i.e., $A_{\mathrm{SL}}^q \neq 0$, is due to a non-vanishing imaginary part of the ratio $\Gamma_{12}^q / M_{12}^q$. For completeness, we point out that, up to $\mathcal{O}(|\Gamma_{12}^q / M_{12}^q|^2)$, one also has
\begin{equation}
     \frac{\Delta \Gamma_q}{\Delta M_q} = -\mathrm{Re}\left(\frac{\Gamma_{12}^q}{M_{12}^q}\right) = \left|\frac{\Gamma_{12}^q}{M_{12}^q}\right|\cos{\phi_{12}^q}.
\end{equation}

All in all, we have related the three relevant observables for the study of $B_q$--$\bar{B}_q$ meson mixing with the parameters $|M_{12}^q|$, $|\Gamma_{12}^q|$ and $\phi_{12}^q$. The most recent experimental measurements of these quantities are \cite{HFLAV:2022esi}:
\begin{subequations}\label{eq:Mass-diff-Exp}
\begin{align}        
    \Delta M_d^\mathrm{Exp} &= 0.5065(19)\,\mathrm{ps}^{-1}\,,\\
    \Delta M_s^\mathrm{Exp} &= 17.765(6)\,\mathrm{ps}^{-1}\,,
\end{align}
\end{subequations}
\vspace{-0.5cm}
\begin{subequations}\label{eq:Decay-width-diff-Exp}
\begin{align}        
    \frac{\Delta \Gamma_d^\mathrm{Exp}}{\Gamma_d^\mathrm{Exp}} &= 0.001(10)\,,\\
    \Delta \Gamma_s^\mathrm{Exp} &= 0.084(5)\,\mathrm{ps}^{-1}\,,
\end{align}
\end{subequations}
\vspace{-0.5cm}
\begin{subequations}\label{eq:ASLq-Exp}
\begin{align}        
    A_{\mathrm{SL}}^{d,\mathrm{Exp}} &= (-21\pm 17)\times 10^{-4}\,,\\
    A_{\mathrm{SL}}^{s,\mathrm{Exp}} &= (\,\,\,-6 \pm 28)\times 10^{-4}\,,
\end{align}
\end{subequations}
with a correlation coefficient for the semileptonic asymmetries of $\rho(A_{\mathrm{SL}}^{d,\mathrm{Exp}},A_{\mathrm{SL}}^{s,\mathrm{Exp}}) = -0.054$. Notice that the semileptonic asymmetries are still compatible with zero. The projected $1\sigma$ sensitivities $\delta A_{\mathrm{SL}}^q$ for the semileptonic asymmetries from LHCb and Belle II according to~\cite{LHCb:2018roe,Alonso-Alvarez:2021qfd,Belle-II:2018jsg,Aihara:2024zds} are the following:
\begin{subequations}\label{eq:ASL_sensitivities_future}
\begin{align}
    \delta A_{\rm SL}^s &= 10\times 10^{-4} \,\,\, [\text{LHCb}\, (23\,\text{fb}^{-1}) -2025]\,,\\
        \delta A_{\rm SL}^s &= 3\times 10^{-4} \,\,\,\,\,\, [\text{LHCb}\, (300\,\text{fb}^{-1})-2040]\,,\\
            \delta A_{\rm SL}^d &= 8\times 10^{-4} \,\,\,\,\,\, [\text{LHCb}\, (23\,\text{fb}^{-1})-2025]\,,\\
        \delta A_{\rm SL}^d &= 2\times 10^{-4} \,\,\,\,\,\, [\text{LHCb}\, (300\,\text{fb}^{-1})-2040]\,,\\
        \delta A_{\rm SL}^d &= 5\times 10^{-4} \,\,\,\,\,\, [\text{Belle II}\, (50\,\text{ab}^{-1})-2035]\,,
\end{align}
\end{subequations}
where here the year has been taken according to the expected date such luminosity will be collected. Looking forward ahead, it is expected that FCC-ee~\cite{FCC:2018evy} will be able to measure these asymmetries with high precision. While there are no dedicated final sensitivity studies, it has been suggested that FCC-ee could reach a sensitivity $\sim 10^{-5}$ for $A_{\rm SL}^s$, and hence to be capable of testing the SM prediction (see~\cite{Monteil:2021ith}). 

\subsection{SM prediction for the mixing observables}
\begin{figure}[h!tb]
\begin{center}
\subfloat[\label{sfig:Box1-SM}]{\includegraphics[width=0.45\textwidth]{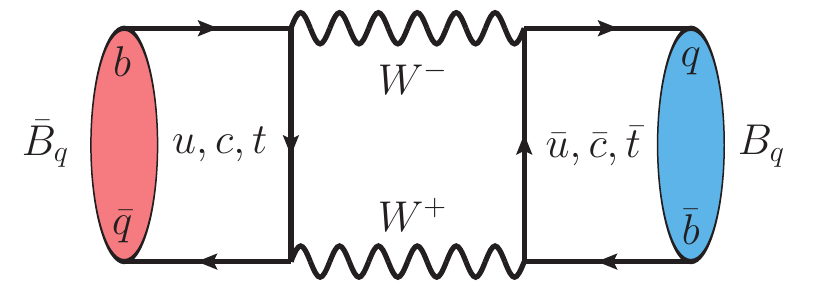}}\\
\subfloat[\label{sfig:Box2-SM}]{\includegraphics[width=0.45\textwidth]{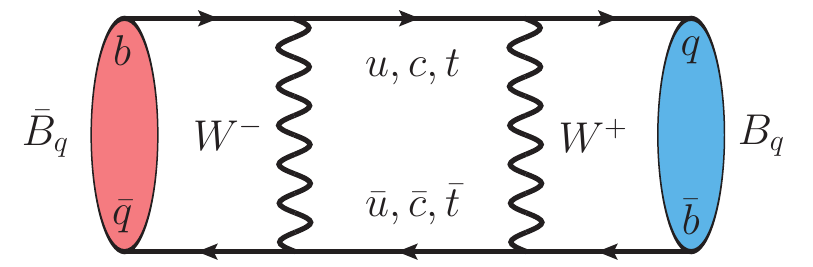}}
\caption{SM box diagrams mediating $B_q$--$\bar{B}_q$ mixing.}\label{fig:Box-SM}
\end{center}
\end{figure}

In the SM, the transitions $\bar{B}_q \leftrightarrow B_q$ occur at a fundamental level via box diagrams with $W$ boson exchange, as depicted in Fig.~\ref{fig:Box-SM} (see App.~\ref{App:DetailsSM} for details). 
On the one hand, $M_{12}^{q,\mathrm{SM}}$ arises from the dispersive part of these diagrams and is dominated by the virtual top quark contribution~\cite{Branco:1999fs,Buras:2020xsm}:
\begin{equation}
    M_{12}^{q,\mathrm{SM}} = \frac{G_F^2 M_W^2}{12\pi^2} (\lambda_{bq}^t)^2 S_0 (x_t) M_{B_q} f_{B_q}^2 B_{B_q} \hat{\eta}_B,
\end{equation}
with $G_F$ the Fermi constant and $M_W$ the mass of the $W$ boson. In general, we will use the short-hand notation $\lambda^\alpha_{bq} \equiv V_{\alpha b}V_{\alpha q}^{*}$ for this combination of CKM elements and $x_\alpha \equiv m_\alpha^2/M_W^2$. The Inami-Lim function \cite{Inami:1980fz} is defined as
\begin{equation}
    S_0 (x) = \frac{x}{(1-x)^2}\left[1 - \frac{11x}{4} + \frac{x^2}{4} - \frac{3x^2 \ln{x}}{2(1-x)} \right],
\end{equation}
and is evaluated at $x_t = \bar{m}_t^2(\bar{m}_t)/M_W^2$, where $\bar{m}_t(\bar{m}_t)$ is the top quark mass in the $\overline{\mathrm{MS}}$ scheme \cite{Bardeen:1978yd}, so that $S_0(x_t) \simeq 2.29$. $M_{B_q}$ is the common mass of the $B_q$ and $\bar{B}_q$ mesons under the strong and electromagnetic interactions. The factor $\hat{\eta}_B$ \cite{Buras:1990fn} includes two-loop perturbative QCD corrections and is chosen to be renormalization scale and scheme independent. It accounts for the running from the $\bar{m}_t$ scale to the $\bar{m}_b$ scale. Non-perturbative QCD and hadronization effects are encoded in the product $f_{B_q}^2 B_{B_q}(\bar{m}_b)$, where $f_{B_q}$ is the $B_q$ meson decay constant and $B_{B_q}$ the bag parameter evaluated at the $\bar{m}_b$ scale.\footnote{An alternative choice for the product of the two-loop perturbative QCD correction factor times the bag parameter can be found in the literature, namely, $\hat{\eta}_B B_{B_q} = \eta_B \hat{B}_{B_q}$. On the left hand side, all renormalization scale and scheme dependence enters on the bag parameter $B_{B_q}$, while, on the right hand side, it is translated into $\eta_B$. The second possibility is used, for instance, by the Flavor Lattice Averaging Group (FLAG) \cite{FlavourLatticeAveragingGroupFLAG:2021npn}.} These two parameters have to be determined using non-perturbative methods such as lattice QCD \cite{FermilabLattice:2016ipl,Dowdall:2019bea} or QCD sum rules \cite{Grozin:2016uqy,Kirk:2017juj,King:2019lal}. In this sense, we should point out that, for the purpose of our analysis, we find more transparent to follow the lattice results from \cite{FermilabLattice:2016ipl} in order to take a particular value of the previous product in both $B_d$ and $B_s$ systems together with their correlation. We note that the semileptonic asymmetries are not particularly sensitive to these parameters and, as such, choosing any other set would not impact significantly our results and final conclusions. We also refer to \cite{Tsang:2023nay} for an updated discussion on the various calculations of $B_q$ meson parameters from the lattice and future prospects. Finally, additional details on the computation of $M_{12}^{q,\mathrm{SM}}$ can be found in App.~\ref{App:DetailsSM}.

The mixing phenomenon in $B_q$ meson systems is triggered dominantly by the mass difference between the heavy and light eigenstates, that is, $\Delta M_q$. The oscillations are considerably faster for $B_s$ mesons (around 35 times faster) than for $B_d$ mesons. This is due to the hierarchy $|V_{ts}| \gg |V_{td}|$ between the CKM elements appearing in the result for $M_{12}^{q,\mathrm{SM}}$. The most recent SM prediction for the meson mass differences reads \cite{Albrecht:2024oyn}:
\begin{align}
    &\Delta M_d^\mathrm{SM} = 0.535 \pm 0.021 \, \mathrm{ps}^{-1},\\
    &\Delta M_s^\mathrm{SM} = 18.23 \pm 0.63 \, \mathrm{ps}^{-1}.
\end{align}

The connection with the rephasing invariant angles $\beta$ and $\beta_s$ of the unitarity triangles corresponding to the $B_d$ and $B_s$ systems is most important. They are defined as
\begin{equation}\label{eq:betabetas}
    \beta \equiv \mathrm{arg}\left(-\frac{V_{cd}V_{cb}^{*}}{V_{td}V_{tb}^{*}}\right),\quad \beta_s \equiv \mathrm{arg}\left(-\frac{V_{ts}V_{tb}^{*}}{V_{cs}V_{cb}^{*}}\right),
\end{equation}
and control the time-dependent CP asymmetries, $S_{J/\Psi K_S} \simeq -\eta_{J/\Psi K_S} \sin 2\beta$ and $S_{J/\Psi \Phi} \simeq \eta_{J/\Psi \Phi} \sin 2\beta_s$ ($\eta_f$ is the CP eigenvalue of the final state $f$), in the gold-plated modes\footnote{The gold-plated modes \cite{Carter:1980tk,Bigi:1981qs} correspond to decays into CP eigenstates common to $B_q$ and $\bar{B}_q$ arising at tree-level. In those cases, hadronic effects cancel in the ratio of the decay amplitudes of $B_q \rightarrow f$ and $\bar{B}_q \rightarrow f$.} $B_d\to J/\Psi K_S$ and $B_s\to J/\Psi\Phi$ due to interference between the decay with and without mixing. More precisely, these CP asymmetries have a clear theoretical interpretation in terms of the angles of the unitarity triangles only when tree-level contributions to the decays are considered (and CP violation in kaon mixing is neglected in $B_d\to J/\Psi K_S$). However, gluon penguin exchange diagrams can also arise in the SM giving a contribution of order $1^\circ$ \cite{Fleischer:2024uru,Barel:2020jvf,LHCb:2015esn,Frings:2015eva,DeBruyn:2014oga,LHCb:2014xpr,Jung:2012mp,Ciuchini:2011kd,Faller:2008zc,Ciuchini:2005mg}, which is at the level of the current experimental precision \cite{HFLAV:2022esi,LHCb:2023zcp,LHCb:2023sim}, and will be determinant if one considers the expected experimental sensitivity at LHCb \cite{LHCb:2018roe} and Belle II \cite{Belle-II:2018jsg}. Therefore, this so-called ``penguin pollution" must be taken into account in order not to misidentify potential NP effects. In this sense, we conveniently write the SM prediction for the phases controlling the time-dependent CP asymmetries as:
\begin{align}
    &\phi_d^{\mathrm{SM}} = 2\beta + \phi_d^{\mathrm{SM,peng}},\label{eq:phidSM}\\
    &\phi_s^{\mathrm{SM}} = -2\beta_s + \phi_s^{\mathrm{SM,peng}}\label{eq:phisSM}.
\end{align}
The first angle on the right-hand side of Eqs.~\eqref{eq:phidSM} and \eqref{eq:phisSM} arises from the argument of $M_{12}^{q,{\mathrm{SM}}}$ normalized to the relative phase between the tree-level decays $B_d\to J/\Psi K_S$ or $B_s\to J/\Psi \Phi$ and their corresponding CP-conjugated processes $\bar B_d\to J/\Psi K_S$ or $\bar B_s\to J/\Psi \Phi$. In particular:
\begin{align}
    &2\beta = \mathrm{arg}(M_{12}^{d,\mathrm{SM}}) - 2\,\mathrm{arg}(V_{cb}V_{cd}^{*}),\label{eq:argM12dSM}\\
    -&2\beta_s = \mathrm{arg}(M_{12}^{s,\mathrm{SM}}) - 2\,\mathrm{arg}(V_{cb}V_{cs}^{*}).\label{eq:argM12sSM}
\end{align}
The parameterization of the CKM matrix can be chosen in such a way that the second term on the right-hand side vanishes.

On the other hand, the theoretical determination of $\Gamma_{12}^{q,\mathrm{SM}}$ is much more challenging. It arises from the absorptive part of box diagrams in Fig.~\ref{fig:Box-SM}  with real intermediate up and charm quarks. The way to proceed with the computation consists of performing a first Operator Product Expansion (OPE) where all degrees of freedom heavier than the $b$ quark are integrated out (similarly to the $M_{12}^{q,\mathrm{SM}}$ case). Then, a second OPE in inverse powers of the heavy $b$ quark mass, the so-called Heavy Quark Expansion (HQE) \cite{Lenz:2014jha}, is carried out. Therefore, this procedure yields $\Gamma_{12}^{q,\mathrm{SM}}$ as a power series in $1/m_b$,\footnote{In the case of $\Gamma_{12}^q$, the HQE starts at order $1/m_b^3$.} being $m_b$ the mass of the $b$ quark, where each term is the product of a perturbative Wilson coefficient and a non-perturbative matrix element between the $B_q$ and $\bar{B}_q$ meson states. In turn, the Wilson coefficients may receive QCD corrections at higher orders in the strong coupling $\alpha_s$. On that respect, contributions at $\mathcal{O}(\alpha_s)$ \cite{Beneke:1998sy,Beneke:2003az,Ciuchini:2003ww,Lenz:2006hd} and $\mathcal{O}(\alpha_s^2)$ \cite{Asatrian:2017qaz,Asatrian:2020zxa,Gerlach:2021xtb,Gerlach:2022wgb,Gerlach:2022hoj} have been computed for the $1/m_b^3$ power, while only leading order terms in QCD are known for $1/m_b^4$ \cite{Beneke:1996gn} and $1/m_b^5$ \cite{Badin:2007bv} powers. Regarding non-perturbative calculations, the hadronic scale fixed by the matrix elements, $\Lambda$, is naively expected to be of the order of $\Lambda_{\mathrm{QCD}}$, but its specific value must be extracted individually for each term. As previously commented, non-perturbative matrix elements of dimension-6 operators are computed in the framework of lattice QCD \cite{FermilabLattice:2016ipl,Dowdall:2019bea} or QCD sum rules \cite{Grozin:2016uqy,Kirk:2017juj,King:2019lal}. More recently, a novel determination of matrix elements of dimension-7 operators was achieved in \cite{Davies:2019gnp}. In App.~\ref{App:DetailsSM}, details on the computation of $\Gamma_{12}^{q,\mathrm{SM}}$ at lowest order in the HQE and QCD can be found. The latest update of the SM prediction for the decay width differences of $B$ mesons is given by \cite{Albrecht:2024oyn}
\begin{align}
    &\Delta \Gamma_d^\mathrm{SM} = (2.7 \pm 0.4)\times 10^{-3} \, \mathrm{ps}^{-1},\\
    &\Delta \Gamma_s^\mathrm{SM} = (9.1 \pm 1.5)\times 10^{-2} \, \mathrm{ps}^{-1}.
\end{align}

In any case, we are mostly interested in the ratio $\Gamma_{12}^{q,\mathrm{SM}}/ M_{12}^{q,\mathrm{SM}}$, which can be expressed as
\begin{equation}
    \frac{\Gamma_{12}^{q,\mathrm{SM}}}{M_{12}^{q,\mathrm{SM}}} = -\frac{(\lambda_{bq}^c)^2 \Gamma_{12}^{q,cc} + 2\lambda_{bq}^c \lambda_{bq}^u \Gamma_{12}^{q,uc} + (\lambda_{bq}^u)^2 \Gamma_{12}^{q,uu}}{(\lambda_{bq}^t)^2 \tilde{M}_{12}^{q,\mathrm{SM}}},
\end{equation}
where $\tilde{M}_{12}^{q,\mathrm{SM}}$ contains the CKM-independent part of $M_{12}^{q,\mathrm{SM}}$, and $\Gamma_{12}^{q,\mathrm{SM}}$ has been decomposed into different CKM structures coming from $uu$, $uc$ and $cc$ quarks contributions of very similar size $\Gamma_{12}^{q,cc} \simeq \Gamma_{12}^{q,uc} \simeq \Gamma_{12}^{q,uu}$ ~\cite{Bobeth:2014rda}. Defining the coefficients
\begin{align}
    &c_q = -\Gamma_{12}^{q,cc}/\tilde{M}_{12}^{q,\mathrm{SM}},\\
    &a_q = 2(\Gamma_{12}^{q,uc} - \Gamma_{12}^{q,cc})/\tilde{M}_{12}^{q,\mathrm{SM}},\\
    &b_q = (2\Gamma_{12}^{q,uc} - \Gamma_{12}^{q,cc} - \Gamma_{12}^{q,uu})/\tilde{M}_{12}^{q,\mathrm{SM}},
\end{align}
and applying $3\times3$ CKM unitarity, i.e. $\lambda_{bq}^u + \lambda_{bq}^c + \lambda_{bq}^t = 0$, one obtains
\begin{equation}\label{eq:Gamma12/M12_SM}
    \frac{\Gamma_{12}^{q,\mathrm{SM}}}{M_{12}^{q,\mathrm{SM}}} = c_q + \left(\frac{\lambda_{bq}^u}{\lambda_{bq}^t}\right)a_q + \left(\frac{\lambda_{bq}^u}{\lambda_{bq}^t}\right)^2 b_q.
\end{equation}
The numerical values of the coefficients $c_q$, $a_q$ and $b_q$ from \cite{Artuso:2015swg} are
\begin{align}\label{eq:cdadbd}
   &c_d = (-49.5 \pm 8.5)\times 10^{-4},\\
   &a_d = (11.7 \pm 1.3)\times 10^{-4},\\
   &b_d = (0.24 \pm 0.06)\times 10^{-4},
\end{align}
\begin{align}
   &c_s = (-48.0 \pm 8.3)\times 10^{-4},\\
   &a_s = (12.3 \pm 1.4)\times 10^{-4},\\
   &b_s = (0.79 \pm 0.12)\times 10^{-4},\label{eq:csasbs}
\end{align}
for the $B_d$ and $B_s$ systems in accordance with the previous comment concerning $\Gamma_{12}^{q,cc} \simeq \Gamma_{12}^{q,uc} \simeq \Gamma_{12}^{q,uu}$. At this point, it is justified to neglect terms of $\mathcal{O}(|\Gamma_{12}^q / M_{12}^q|^2)$ in the expressions for the relevant mixing observables since $|\Gamma_{12}^q / M_{12}^q| \sim 5\times 10^{-3}$ in the SM.\footnote{More precisely, corrections to the mixing observables are of $\mathcal{O}((1/8)|\Gamma_{12}^q / M_{12}^q|^2 \sin^2{\phi_{12}^q})$, that is, $\mathcal{O}(10^{-8})$ for $B_d$ and $\mathcal{O}(10^{-11})$ for $B_s$ in the SM, thus completely negligible given the current experimental precision.} There is a strong hierarchy in the numerical values, namely $|c_q| > a_q \gg b_q$, due to the GIM \cite{Glashow:1970gm} suppression affecting $a_q$ and $b_q$, which vanish in the limit $m_u \rightarrow m_c$. Furthermore, the terms proportional to $a_q$ and $b_q$ in Eq.\eqref{eq:Gamma12/M12_SM} are CKM-suppressed, with 
\begin{align}
    \frac{\lambda_{bq}^u}{\lambda_{bq}^t} &= \begin{cases}
        1.7\times 10^{-2} - 4.2\times 10^{-1}\, i,\quad q = d,\\
        -8.8\times 10^{-3} + 1.8\times 10^{-2}\, i,\quad q = s,
    \end{cases}\\
    \left(\frac{\lambda_{bq}^u}{\lambda_{bq}^t}\right)^2 &= \begin{cases}
        -1.8\times 10^{-1} - 1.5\times 10^{-2}\, i,\quad q = d,\\
        -2.5\times 10^{-4} - 3.2\times 10^{-4}\, i,\quad q = s.
    \end{cases}
\end{align}
It is then clear that the real part of the previous ratio is dominated by the real $c_q$ contribution, while for the imaginary part only the term proportional to $a_q$ contributes significantly. In particular, given that $a_d$ and $a_s$ are of the same order, the values of the semileptonic asymmetries in each system are determined by the fact that the imaginary part of $\lambda_{bq}^{u}/\lambda_{bq}^t$ is around 20 times larger in the $B_d$ sector than in the $B_s$ sector, having both opposite sign. The latest update from \cite{Albrecht:2024oyn} on the SM prediction for the semileptonic asymmetries reads:
\begin{align}
    A_{\mathrm{SL}}^{d,\mathrm{SM}} &= (-5.1\pm 0.5)\times 10^{-4},\label{eq:ASLd-SM}\\
    A_{\mathrm{SL}}^{s,\mathrm{SM}} &= (0.22\pm 0.02)\times 10^{-4}.\label{eq:ASLs-SM}
\end{align}
All in all, the semileptonic asymmetries in the SM are small, with room for variation at the 10\% level, and fully compatible with the current experimental results. Comparing Eqs.~\eqref{eq:ASLq-Exp} with Eqs.~\eqref{eq:ASLd-SM} and~\eqref{eq:ASLs-SM}, one can readily check that the experimental uncertainty is about 3 times larger than the corresponding SM central value in the $B_d$ system, and 130 times larger in the $B_s$ system. This means that there is a priori ample room to accommodate NP effects affecting the semileptonic asymmetries, thus providing new sources of CP violation in $B$ meson mixing. 

In summary, there are two ingredients that suppress the imaginary part of the ratio $\Gamma_{12}^{q,\mathrm{SM}}/M_{12}^{q,\mathrm{SM}}$ and, in particular, that align the phases of $\Gamma_{12}^{q,\mathrm{SM}}$ and $M_{12}^{q,\mathrm{SM}}$: (i) the top dominated contribution in $M_{12}^{q,\mathrm{SM}}$, and (ii) $3\times3$ CKM unitarity and GIM cancellation. Aiming to analyze how this suppression can be evaded BSM, we explore different frameworks in the following. In Sec.~\ref{SEC:NPM12q} we consider, in a model independent approach, scenarios in which only $M_{12}^q$ is modified. In Sec.~\ref{SEC:VLQ}, the inclusion of vector-like quark singlets is considered. Finally, modifications of $\Gamma_{12}^q$ are addressed in Sec.~\ref{SEC:G12q}.

\section{$A_{\rm SL}^q$ with heavy new physics affecting $M_{12}^q$ \label{SEC:NPM12q}}
In this section, we consider the possibility of heavy new physics affecting $B_q$--$\bar{B}_q$ meson mixing. The effects of short-distance contributions can be addressed in the framework of Effective Field Theory (EFT) through higher dimension operators built out of SM light fields (below the $m_b$ scale). On the one hand, in the context of perturbation theory in the weak interactions, $M_{12}^q$ corresponds to $\Delta B = 2$ transitions through intermediate virtual states. These arise at the one-loop level in the SM and they are CKM-suppressed, so that $M_{12}^q$ could be a sensitive probe to elucidate heavy NP effects which might compete with the corresponding SM contribution. On the other hand, $\Gamma_{12}^q$ corresponds to two $\Delta B = 1$ transitions through an intermediate real state that is common to both $B_q$ and $\bar{B}_q$ mesons. $\Delta B = 1$ transitions arise at tree-level in the SM, so that if NP entered also at tree-level, its effects would be in any case suppressed with respect to the SM by powers of $(M_W/ \Lambda)^2$, being $\Lambda$ the NP scale. Therefore, it seems reasonable to neglect heavy NP contributions in $\Gamma_{12}^q$ at first instance and circumscribe its effects to $M_{12}^q$.

In this context, we consider a general model-independent modification of $M_{12}^q$, such that it could be the benchmark to analyze a great variety of models including, e.g., leptoquarks \cite{Dorsner:2016wpm,Buttazzo:2017ixm,DiLuzio:2019jyq}, heavy $Z'$s \cite{DiLuzio:2019jyq,Allanach:2023uxz,Athron:2023hmz}, two-Higgs-doublet models \cite{Trott:2010iz,Iguro:2018qzf,Crivellin:2019dun,Athron:2021auq}, axion-like particles \cite{Bonilla:2022qgm,Li:2024thq}, supersymmetric models \cite{Alonso-Alvarez:2019fym,Zheng:2022ssr}, dark sector particles \cite{Datta:2022zng}, or $SU(2)_L$ triplets of heavy gauge bosons \cite{Davighi:2023xqn}, among other extensions. We should point out that although previous analyses addressing the impact of NP in $B$ meson mixing have been performed in the literature in a similar fashion, see e.g. \cite{Ligeti:2010ia,Lenz:2011zz,Bobeth:2011st,Lenz:2012az,Bevan:2013kaa,Charles:2013aka,Botella:2014qya,Artuso:2015swg,Charles:2020dfl,DeBruyn:2022zhw}, we rather focus our attention on the question of how large $A_{\mathrm{SL}}^q$ can be. On that respect, modifications of these quantities triggered the attention from the theoretical particle physics community some time ago due to the anomalous value of the like-sign dimuon asymmetry reported by the D0 Collaboration \cite{D0:2010sht,D0:2011hom,D0:2013ohp}. In addition, as previously commented, an specific study of the potential size of the semileptonic asymmetries could be of interest as they can be directly related to baryogenesis \cite{Elor:2018twp}. Finally, it might be convenient to revisit these analyses in light of the latest results presented by LHCb concerning the phases that control CP violation in the interference between mixing and decay of $B$ mesons \cite{LHCb:2023zcp,LHCb:2023sim}.

Following previous studies \cite{Lenz:2012az,Artuso:2015swg}, we can write down
\begin{align}
    &M_{12}^q = M_{12}^{q,\mathrm{SM}} \Delta_q = M_{12}^{q,\mathrm{SM}} |\Delta_q|e^{i\phi_q^\Delta},\label{eq:M12-3x3}\\
    &\Gamma_{12}^q = \Gamma_{12}^{q,\mathrm{SM}},\label{eq:phiM12q}\\
    &\phi_{12}^q = \phi_{12}^{q,\mathrm{SM}} + \phi_q^\Delta,
\end{align}
being $|\Delta_q|$ and $\phi_q^\Delta$ the modulus and the phase of the complex number $\Delta_q$ that parameterizes the modification of $M_{12}^q$ with respect to the SM.\footnote{Other authors \cite{Charles:2013aka,Charles:2020dfl,DeBruyn:2022zhw} prefer to parameterize the deviation of the pure NP contribution with respect to the SM, that is, the ratio $M_{12}^{q,\mathrm{NP}}/M_{12}^{q,\mathrm{SM}}$. It is straightforward to relate both parameterizations as $M_{12}^{q}/M_{12}^{q,\mathrm{SM}} = 1 + M_{12}^{q,\mathrm{NP}}/M_{12}^{q,\mathrm{SM}}$.} Of course, the absence of NP translates into $(|\Delta_q|,\phi_q^\Delta) = (1,0)$. In the context of EFT, $\Delta_q$ can be understood as a modification of the Wilson coefficient accompanying the dimension-6 operator that contributes to $M_{12}^{q,\mathrm{SM}}$. In this way, the mixing observables are modified according to:
\begin{equation}
    \Delta M_q = 2|M_{12}^{q,\mathrm{SM}}| |\Delta_q| = \Delta M_q^\mathrm{SM} |\Delta_q|,\hspace{1.0cm}
\end{equation}
\begin{equation}
    \begin{split}
        \Delta \Gamma_q &= 2|\Gamma_{12}^{q,\mathrm{SM}}| \cos{(\phi_{12}^{q,\mathrm{SM}} + \phi_q^\Delta)}\\
        &= \Delta \Gamma_q^\mathrm{SM} \cos{\phi_q^\Delta} - \Delta M_q^\mathrm{SM} A_{\mathrm{SL}}^{q,\mathrm{SM}} \sin{\phi_q^\Delta},
    \end{split}
\end{equation}
\begin{align}\label{eq:ASLq-heavyNP}
        A_{\mathrm{SL}}^q &= \frac{1}{|\Delta_q|} \bigg | \frac{\Gamma_{12}^{q,\mathrm{SM}}}{M_{12}^{q,\mathrm{SM}}} \bigg | \sin{(\phi_{12}^{q,\mathrm{SM}} + \phi_q^\Delta)} \\
        &= \frac{1}{|\Delta_q|}\left[A_{\mathrm{SL}}^{q,\mathrm{SM}}\cos{\phi_q^\Delta} + \left(\frac{\Delta \Gamma_q^\mathrm{SM}}{\Delta M_q^\mathrm{SM}}\right)\sin{\phi_q^\Delta}\right]. \nonumber
\end{align}
The phase controlling the time-dependent CP asymmetries gets modified as:
\begin{equation}
    \phi_q = \phi_q^{\mathrm{SM}} + \phi_q^\Delta,\label{eq:phiq-heavyNP}
\end{equation}
where $\phi_q^{\mathrm{SM}}$ is given in Eqs.~\eqref{eq:phidSM} and~\eqref{eq:phisSM}. In Eq.~\eqref{eq:phiq-heavyNP}, we are assuming that NP does not enter in tree-level decays.  Eq.~\eqref{eq:ASLq-heavyNP} highlights the strong dependence of $A_{\rm SL}^q$ on $\phi_q^\Delta$.

With the aim of understanding how large the CP asymmetries can be in this scenario, we perform a global fit to constrain all the independent parameters: $\{|\Delta_d|,|\Delta_s|,\phi_d^\Delta,\phi_s^\Delta,f_{B_d}^2 B_{B_d},f_{B_s}^2 B_{B_s},\theta_{12},\theta_{13},\theta_{23},\delta\}$, being $\theta_{ij}$ and $\delta$ the three rotation angles and the complex phase of a $3\times3$ unitary CKM matrix. Without loss of generality, one can restrict the angles to lie in the first quadrant provided that the phase is allowed to be free. In particular, we adopt the Chau-Keung parameterization \cite{Chau:1984fp} used by the PDG. The set of relevant constraints includes the following.
\begin{itemize}
    \item $B$ meson mass differences $\Delta M_d$ (measured by ALEPH, DELPHI, L3, OPAL, CDF, D0, BaBar, Belle, and LHCb collaborations) and $\Delta M_s$ (from CDF, D0, and LHCb collaborations). The combination of all different measurements is summarized in \cite{HFLAV:2022esi}, and given in Eq.~\eqref{eq:Mass-diff-Exp}.
    \item CP violating phases arising in the interference between mixing and decay of $B$ mesons. These can be measured in different decay channels governed by the quark level transition $b \rightarrow c\bar{c}s$. On the one hand, $\sin{\phi_d}$ is obtained from the analysis of final states such that $J/\Psi K_{S/L}$, $\Psi(2S) K_S$, or $\chi_{c1} K_S$. On the other hand, $\phi_s$ is extracted using final states such as $J/\Psi \Phi$, $\Psi(2S)\Phi$, or $D_s^{+}D_s^{-}$. On that respect, LHCb has recently published the most precise single measurements of $\sin{\phi_d}$ \cite{LHCb:2023zcp} and $\phi_s$ \cite{LHCb:2023sim} using the decays $B_d \rightarrow \Psi(\rightarrow \ell^{+}\ell^{-})K_S(\rightarrow \pi^{+}\pi^{-})$ and $B_s \rightarrow J/\Psi(\rightarrow \mu^{+}\mu^{-})K^{+}K^{-}$, respectively. A detailed breakdown of the different determinations of the CP violating phases can be found in \cite{HFLAV:2022esi}. Here, we use the average including the latest results from LHCb \cite{WebHFLAV:2024,LHCbseminar:2023}, namely:
    \begin{align}
        &\sin{\phi_d^\mathrm{Exp}} = 0.708 \pm 0.011,\\
        &\phi_s^\mathrm{Exp} = -0.040 \pm 0.016.
    \end{align}
    Only the most precise measurements of $\sin{\phi_d}$ are considered in the previous average.
    
    As previously commented, one should care about the contributions coming from penguin diagrams when implementing this constraint. The penguin pollution in the SM is estimated to be of order 1$^\circ$ ($\sim$ 0.017 rad), while NP penguins are much less constrained. In this sense, we add in quadrature the potential size of penguin topologies and the experimental uncertainty in the measurement of the previous mixing-induced CP phases, and treat the result as a total uncertainty for this constraint. This roughly implies an increase in the uncertainty of $\phi_s^\mathrm{Exp}$ by a factor of $\sqrt{2}$, with approximately no change in the uncertainty of the observable $\sin{\phi_d^\mathrm{Exp}}$. Our limited knowledge on the penguin effects, and even more on the NP penguin contributions (that depend on the specific model), should be appropriately covered in this way. One may also worry about the channel dependence of the penguin pollution,\footnote{In particular, the penguin pollution encodes non-perturbative hadronic effects involving the initial and final states of the transition amplitude. Other factors such as the polarization of the final state particles can also have an impact on this estimation, as it happens for instance in the gold-plated mode $B_s \rightarrow J/\Psi \Phi$. A more extensive explanation can be found in \cite{Artuso:2015swg}.} since we are actually taking an experimental average over different decay modes where the mixing-induced CP phases have been measured. An analysis of different decay channels besides the gold-plated ones has been carried out, for instance, in \cite{Frings:2015eva,Barel:2020jvf}, where the penguin effects are expected to be of order 1$^\circ$ as well. Hence, these further effects are also accommodated within our approach.
    \item Lattice QCD results for the products $f_{B_d}^2 B_{B_d}$ and $f_{B_s}^2 B_{B_s}$, together with their correlation, are taken from \cite{FermilabLattice:2016ipl}:
    \begin{align}
        &f_{B_d}^2 B_{B_d} (\bar{m}_b) = 0.0342(29)(7)\, \mathrm{GeV}^2,\\
        &f_{B_s}^2 B_{B_s} (\bar{m}_b) = 0.0498(30)(10)\, \mathrm{GeV}^2,\\
        &\rho(f_{B_d}^2 B_{B_d},f_{B_s}^2 B_{B_s}) = 0.968,
    \end{align}
    where the first number in parentheses is the ``total error'' accounting for all statistical and systematic uncertainties in the lattice simulation, and the second number in parentheses corresponds to the ``charm sea error''. We add these two contributions in quadrature.
    \item CKM entries $|V_{ud}|$, $|V_{us}|$, $|V_{ub}|$, $|V_{cb}|$ and the angle $\gamma \equiv \mathrm{arg}(-V_{ud}V_{cb}V_{ub}^{*}V_{cd}^{*})$. As we will highlight later in next section, in the context of vector-like quark extensions it is crucial to further include the $|V_{tb}|$ constraint. The experimental values reported by the PDG \cite{ParticleDataGroup:2022pth} are:
    \begin{align}
        &|V_{ud}| = 0.97367 \pm 0.00032,\\
        &|V_{us}| = 0.22431 \pm 0.00085,\\
        &|V_{ub}| = (3.82 \pm 0.20)\times 10^{-3},\\
        &|V_{cb}| = (41.1 \pm 1.2)\times 10^{-3},\\
        &|V_{tb}| = 1.010 \pm 0.027,\\
        &\gamma = (65.7 \pm 3.0)^\circ.
    \end{align}
    \item $B$ meson decay width differences $\Delta \Gamma_d$ and $\Delta \Gamma_s$ \cite{HFLAV:2022esi}, as given in Eqs.~\eqref{eq:Decay-width-diff-Exp}.
    \item Semileptonic asymmetries $A_{\mathrm{SL}}^d$ and $A_{\mathrm{SL}}^s$ \cite{HFLAV:2022esi} in Eqs.~\eqref{eq:ASLq-Exp}.
    \item Parameters $a_q$, $b_q$, $c_q$ in Eqs.~\eqref{eq:cdadbd}--\eqref{eq:csasbs} entering $\Gamma_{12}^q$. It is worth mentioning that their uncertainties are ``theoretical''. We explore two extreme options: (i) allowing them to vary freely within the range given by the central values $\pm$ the uncertainties, (ii) fixing them to their central values. The results obtained with one or the other option differ at a negligible level. A more realistic approach would allow them to vary in a correlated manner, but since the extreme options make essentially no difference, that more realistic approach cannot neither.
\end{itemize}
Other input parameters used in the analysis are given in Table \ref{tab:input-parameters}.
\begin{table}[H]
\begin{center}
 \begin{tabular}{|c|c|c|c|}
 \hline
  Parameter & Value & Units & Reference \\ 
  \hline\hline
  $G_F$ & $1.1663788(6) \times 10^{-5}$ & GeV$^{-2}$ & \cite{ParticleDataGroup:2022pth} \\ 
  \hline
  $\hat{\eta}_B$ & $0.84$ & - & \cite{Buras:1990fn} \\ 
  \hline
  $M_W$ & $80.377(12)$ & GeV & \cite{ParticleDataGroup:2022pth} \\ 
  \hline
  $\bar{m}_t (\bar{m}_t)$ & $161.98(75)$ & GeV & \cite{Huang:2020hdv} \\ 
  \hline
  $M_{B_d}$ & $5.27963(20)$ & GeV & \cite{ParticleDataGroup:2022pth} \\ 
  \hline
   $M_{B_s}$ & $5.36691(11)$ & GeV & \cite{ParticleDataGroup:2022pth} \\ 
  \hline
  \end{tabular}
 \caption{Input parameters.}
 \label{tab:input-parameters}
\end{center}
\end{table}
The previous constraints are implemented in terms of Gaussian likelihoods or equivalent $\chi^2$ terms (except for the parameters $a_q$, $b_q$, $c_q$ as just commented). The global $\chi^2$ is sampled via Markov chain Monte Carlo techniques in order to represent the relevant regions for the different parameters and observables.

\begin{figure*}[t]
\begin{center}
\includegraphics[width=1.0\textwidth]{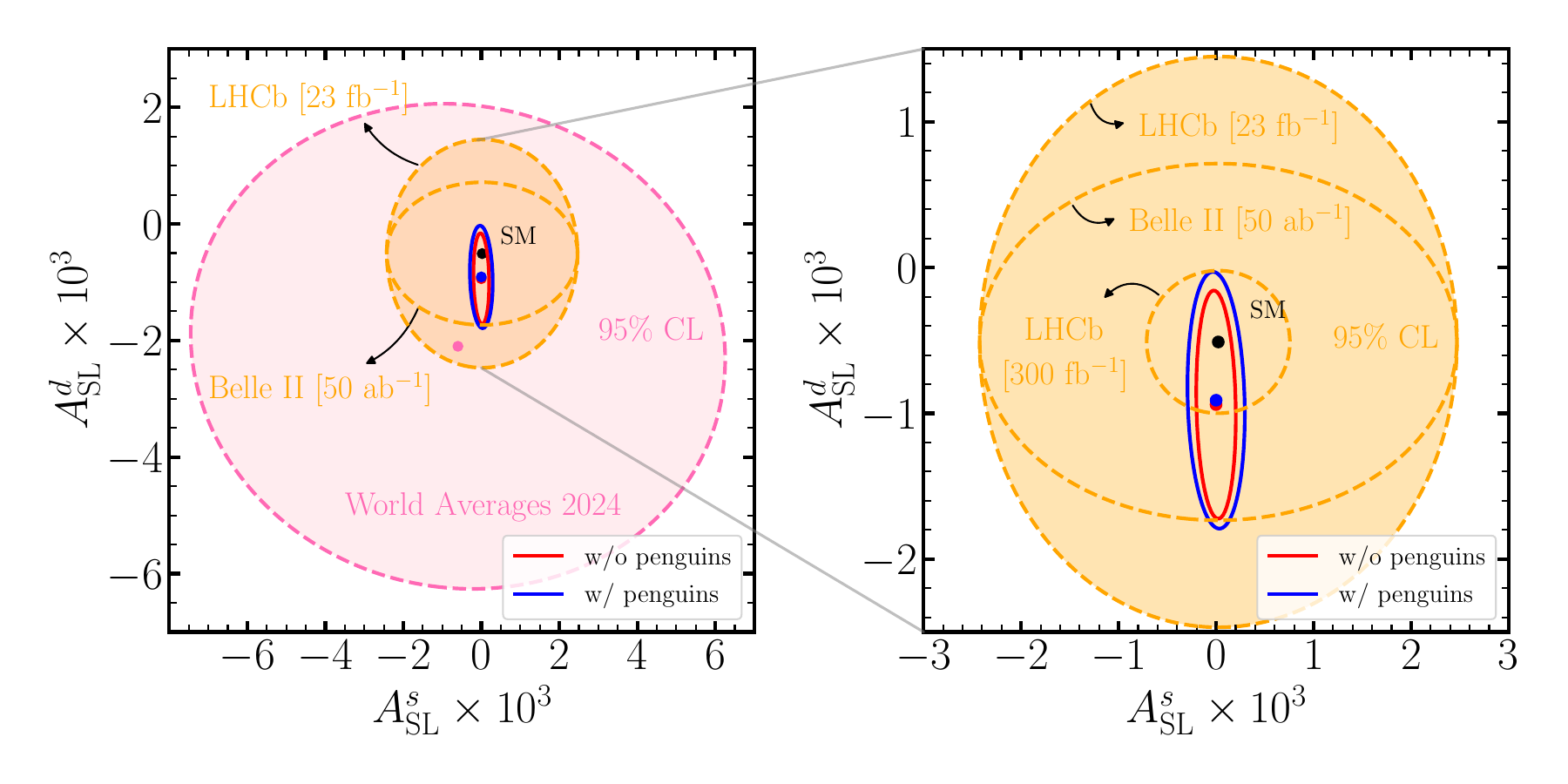}
\caption{\justifying Allowed regions for $A_\mathrm{SL}^d$ and $A_\mathrm{SL}^s$ with heavy new physics solely affecting $M_{12}^q$. All contours are shown at 95\% C.L. (2D-$\Delta \chi^2 = 5.99$), together with the corresponding best fit point. In the left panel, the pink ellipse represents the current experimental world averages for the semileptonic asymmetries measurements as of 2024, while the ellipses in orange project the future expected sensitivity at LHCb Run 3 (23 fb$^{-1}$) and Belle II (50 ab$^{-1}$). The blue and red contours show the results of our analysis within the scenario presented in Sec.~\ref{SEC:NPM12q}, corresponding to the case where SM penguin contributions have been included or neglected, respectively. The point in black is the SM prediction, whose uncertainties are not visible in these axes. In the right panel, we provide the detail of our results along with the projected sensitivities at LHCb Run 3 and Belle II, and, in addition, at LHCb Run 5 (300 fb$^{-1}$).}
\label{fig:NP33}
\end{center}
\end{figure*}

Our results are summarized in Fig.~\ref{fig:NP33}, where we present the allowed regions in the $A_\mathrm{SL}^s$--$A_\mathrm{SL}^d$ plane. In particular, results at 68\% C.L. read:
\begin{subequations}\label{eq:ASL_resultspeng}
\begin{align}
    &A_\mathrm{SL}^d |_\mathrm{w/\, peng} = (-9.1\pm 3.6)\times 10^{-4},\\
    &A_\mathrm{SL}^s |_\mathrm{w/\, peng} = (-0.04 \pm 1.21)\times 10^{-4},\\
    &\rho(A_\mathrm{SL}^d,A_\mathrm{SL}^s) |_\mathrm{w/\, peng} = -0.113,
\end{align}
\end{subequations}
taking into account the penguin pollution, and 
\begin{subequations}
\begin{align}
    &A_\mathrm{SL}^d |_\mathrm{w/o\, peng} = (-9.4\pm 3.2)\times 10^{-4},\\
    &A_\mathrm{SL}^s |_\mathrm{w/o\, peng} = (-0.01\pm 0.83)\times 10^{-4},\\
    &\rho(A_\mathrm{SL}^d,A_\mathrm{SL}^s) |_\mathrm{w/o\, peng} = -0.110,
\end{align}
\end{subequations}
if penguins effects are ignored. Both results are fully compatible with the SM prediction at 95\% C.L., and, in particular, compatible with zero in the $B_s$ system. As we can clearly see, the potential size of the semileptonic asymmetries in light of current data is $\mathcal{O}(10^{-3})$ for $A_\mathrm{SL}^d$ and $\mathcal{O}(10^{-4})$ for $A_\mathrm{SL}^s$, which are still below the projected sensitivity at LHCb Run 3 (23 fb$^{-1}$) and Belle II (50 ab$^{-1}$). Only LHCb Run 5 (300 fb$^{-1}$) data is expected to effectively constrain the allowed range for $A_\mathrm{SL}^d$, as it is shown in Fig.~\ref{fig:NP33}.

The vanishing correlation coefficient between the semileptonic asymmetries in our analyses traces back to the absence of correlation in the measurements of the mixing-induced CP phases that constrain the parameters $\phi_q^\Delta$ in $B_d$ and $B_s$ meson systems. Likewise, the strong correlation between $A_\mathrm{SL}^q$ and $\phi_q^\Delta$ also explains the increase of the uncertainty in $A_\mathrm{SL}^s |_\mathrm{w/\, peng}$ with respect to $A_\mathrm{SL}^s |_\mathrm{w/o\, peng}$ by a factor of $\sqrt{2}$, as commented when analyzing the constraint coming from the mixing-induced CP phases.

For completeness, we provide in the following the values of the parameters $|\Delta_q|$ and $\phi_q^\Delta$ that modify the modulus and the phase of $M_{12}^{q,\mathrm{SM}}$ in this scenario:
\begin{subequations}\label{eq:3x3Globalanalysis}
\begin{align}
    &| \Delta_d |_\mathrm{w/\, peng} = 0.98_{-0.07}^{+0.10},\quad \phi_d^\Delta |_\mathrm{w/\, peng} = -0.071_{-0.057}^{+0.058},\\
    &| \Delta_s |_\mathrm{w/\, peng} = 1.00_{-0.04}^{+0.06}, \quad \phi_s^\Delta |_\mathrm{w/\, peng} = -0.004_{-0.027}^{+0.025},
\end{align}
\end{subequations}
and
\begin{subequations}
\begin{align}
    &| \Delta_d |_\mathrm{w/o\, peng} = 0.90_{-0.07}^{+0.13},\quad \phi_d^\Delta |_\mathrm{w/o\, peng} = -0.055_{-0.059}^{+0.048},\\
    &| \Delta_s |_\mathrm{w/o\, peng} = 0.92_{-0.05}^{+0.10}, \quad \phi_s^\Delta |_\mathrm{w/o\, peng} = -0.006_{-0.019}^{+0.018}.
\end{align}
\end{subequations}
On the one hand, the parameters $|\Delta_q|$ are close to their SM value with uncertainties of $\mathcal{O}(10\%)$, as one could have expected from the agreement between the measurements and the SM prediction of the meson mass differences $\Delta M_q$, and thus they do not contribute to substantially enhance the values of the semileptonic asymmetries with respect to the SM. On the other hand, despite the small values of $\phi_s^\Delta$, they can enhance $A_{\mathrm{SL}}^s$ at the $10^{-4}$ level. In the case of $A_\mathrm{SL}^d$, the phases $\phi_d^\Delta$ could be large enough to saturate the experimental lower bound expected from Belle II (50 ab$^{-1}$). As a final comment, we devote App.~\ref{app:G12M12analysis} to a detailed numerical analysis of how the modifications of $M_{12}^q$ actually achieve the enhancement of $A_{\mathrm{SL}}^d$ and $A_{\mathrm{SL}}^s$ through their parametric effect not only on $M_{12}^q$ but on $\Gamma_{12}^q$ as well. 

\section{$A_{\rm SL}^q$ with non-unitary CKM mixing \label{SEC:VLQ}}
\begin{figure}[t]
\begin{center}
\subfloat[\label{sfig:DVLQ-tree-channel-s}]{\includegraphics[width=0.225\textwidth]{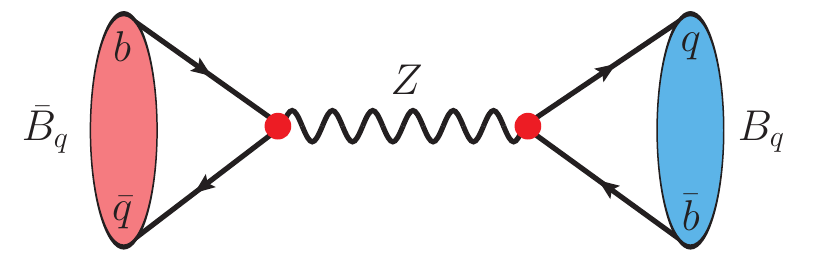}}\
\subfloat[\label{sfig:DVLQ-tree-channel-t}]{\includegraphics[width=0.225\textwidth]{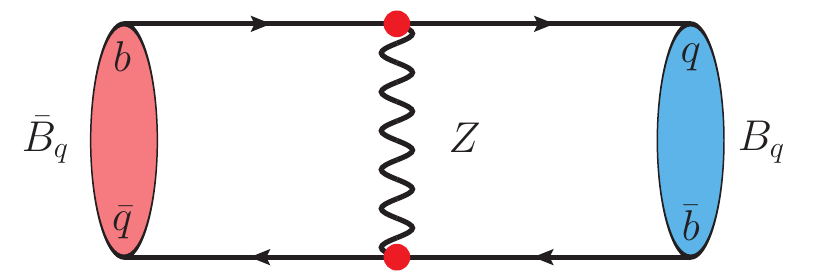}}\\
\subfloat[\label{sfig:DVLQ-blob-left}]{\includegraphics[width=0.225\textwidth]{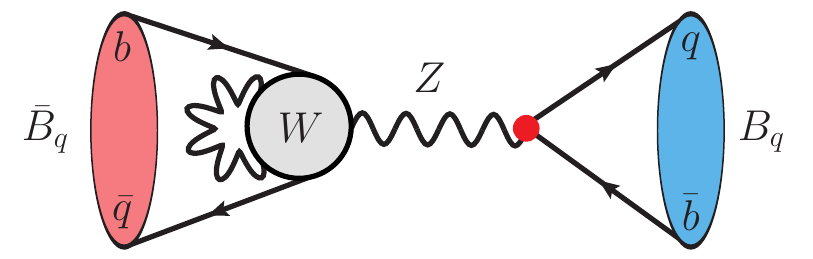}}\
\subfloat[\label{sfig:DVLQ-blob-up}]{\includegraphics[width=0.225\textwidth]{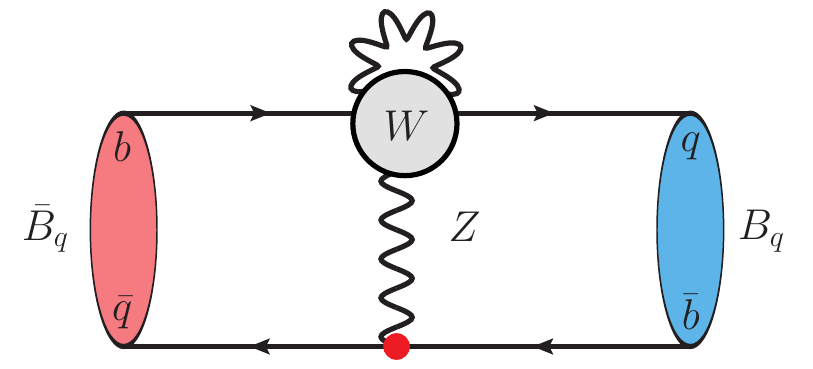}}\\
\caption{\justifying Tree-level and one-loop diagrams contributing to $B_q$--$\bar{B}_q$ mixing in DVLQ models. The red vertex corresponds to the flavor-changing coupling $Z d_i \bar{d}_j$ obtained in this type of models. The blob in grey represents all possible flavor-changing one-loop topologies arising from the exchange of one or two $W$ bosons, including those where the $Z$ couples directly to one of the external lines. In this case, the other topologies where the red vertex and the grey blob exchange positions are not shown, but appropriately included in the calculations.}\label{fig:DVLQ-tree-and-peng}
\end{center}
\end{figure}

We consider in this section models where the SM quark content is extended through the inclusion of ``vector-like'' quarks.\footnote{``Vector-like'' meaning that new left-handed and right-handed fields are added together, with the same $SU(2)_L$ assignment, and thus they are \emph{harmless} from the point of view of anomaly cancellations}
These ``exotic fermions'' naturally arise in a number of theoretically top-down motivated scenarios such as Grand Unified models \cite{delAguila:1982fs,Barger:1995dd}, extra-dimensional models \cite{Carena:2007tn}, or composite Higgs models \cite{Matsedonskyi:2012ym}. From a bottom-up perspective, models with vector-like quarks provide an interestingly rich phenomenology \cite{Branco:1986my,Langacker:1988ur,Nir:1990yq,Barenboim:1997pf,Barenboim:1997qx,delAguila:2000rc,Aguilar-Saavedra:2013qpa,Ishiwata:2015cga,Alves:2023ufm}. 
We will focus on the simplest cases within that large class of models, that is, models with either one up-type or one down-type quark singlet (to be referred in the following as UVLQ and DVLQ, respectively), since they already introduce modifications of interest. In these models,
\begin{itemize}
\item The CKM matrix is not anymore $3\times 3$ unitary, but part of larger unitary matrix: in the UVLQ case, CKM is $4\times 3$ and is embedded in a $4\times 4$ unitary matrix; in the DVLQ case, CKM is $3\times 4$ and is embedded in a $4\times 4$ unitary matrix; just for illustration, although we do not consider such a scenario, with both one up-type and one down-type vector-like quark singlets, CKM is $4\times 4$ and is embedded in a $5\times 5$ unitary matrix.
\item As discussed in detail in App.~\ref{App:ModelsVLQ}, in the UVLQ case there are tree-level $Z$ Flavor-Changing Neutral Currents (FCNC) in the up quark sector but not in the down quark sector, while in the DVLQ case, there are tree-level $Z$-FCNC in the down quark sector, but not in the up quark sector. For illustration again, with both one up-type and one down-type vector-like quark singlets, tree-level $Z$-FCNC are present in both the up and down quark sectors.
\end{itemize}
Focusing on $B_d$ and $B_s$ neutral meson mixings,
\begin{itemize}
\item There are new contributions to $M_{12}^q$. In the UVLQ case, they correspond to additional SM-like diagrams as in Fig.~\ref{fig:Box-SM}, including the new up-type quark $T$ in the loop; these contributions depend on its mass $m_T$ and the combinations of elements $V_{Tb}V_{Tq}^\ast$ of the enlarged CKM matrix. In the DVLQ case, the new contributions involve tree-level flavor changing neutral vertices, as illustrated in Fig.~\ref{fig:DVLQ-tree-and-peng}; they depend on the $3\times 3$ unitarity violations $(D_L)_{qb}=V_{ub}V_{uq}^\ast+V_{cb}V_{cq}^\ast+V_{tb}V_{tq}^\ast\neq 0$ (see Eqs.~\eqref{eq:DL-matrix} and \eqref{eq:DVLQ-unit}), but do not depend on the mass of the new down-type quark.
\item Concerning $\Gamma_{12}^q$, there are no new contributions in the UVLQ case, contrary to the DVLQ case, where new contributions involving one or two $Z$ flavor changing vertices are present. These new contributions in the DVLQ case are nevertheless negligible, as discussed later.
\end{itemize}
The inclusion of vector-like quarks provides two ingredients --a non $3\times 3$ unitary CKM matrix together with new contributions to $M_{12}^q$-- that have the potential to induce a significant misalignment of $\Gamma_{12}^q/M_{12}^q$ with respect to the SM. We scrutinize how these ingredients actually operate in our analyses in App.~\ref{app:G12M12analysis}.

The analyses follow the same approach as those of Sec.~\ref{SEC:NPM12q}, with the following changes. Since the CKM matrix is not $3\times 3$ unitary but part of a larger $4\times 4$ unitary matrix in both analyses (UVLQ and DVLQ scenarios), rather than the 4 parameters required in Sec.~\ref{SEC:NPM12q} for CKM, a further 5 parameters are needed, i.e., 3 new ``mixing angles'' $\theta_{14}$, $\theta_{24}$, $\theta_{34}$, and 2 new phases $\delta_{14}$, $\delta_{24}$ in a PDG-like parameterization, in order to describe the extended CKM matrix ($4\times 3$ in the UVLQ case, $3\times 4$ in the DVLQ case) and the $Z$-FCNC tree-level couplings.

\begin{figure}[htbp]
\begin{center}
\includegraphics[width=0.45\textwidth]{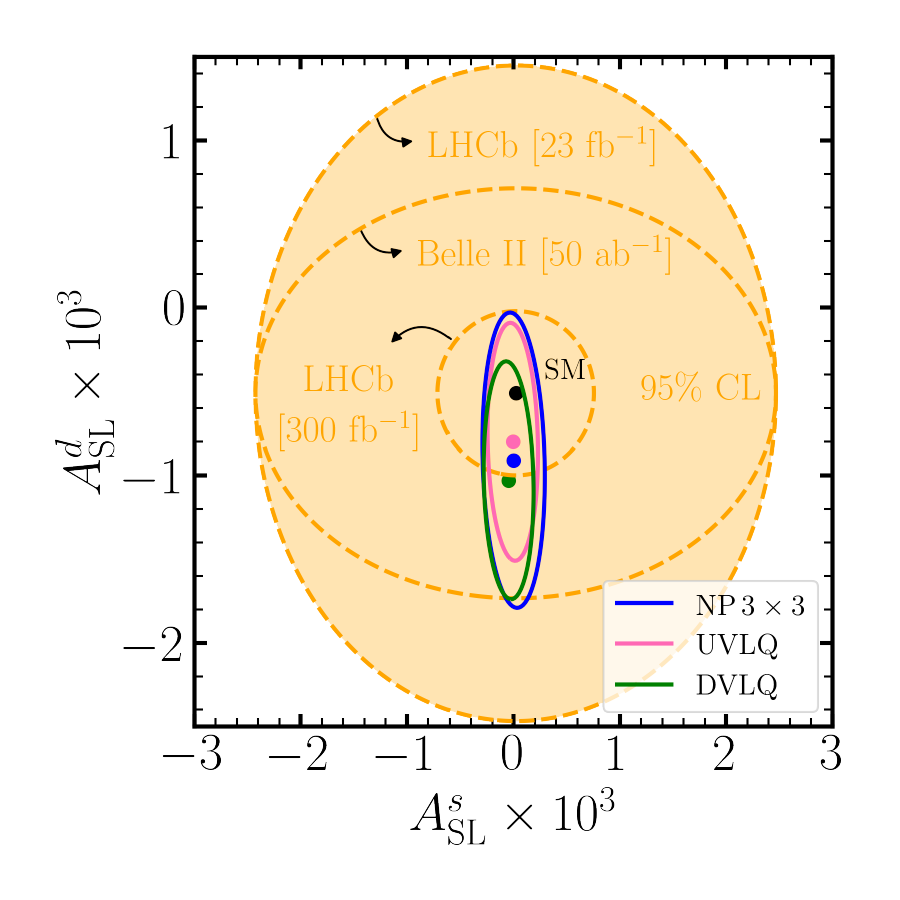}
\vspace{-0.4cm}
\caption{\justifying \justifying Allowed regions for $A_\mathrm{SL}^d$ and $A_\mathrm{SL}^s$ within vector-like quark singlet extensions. All contours are shown at 95\% C.L. (2D-$\Delta \chi^2 = 5.99$), together with the corresponding best fit point. The pink and green contours correspond to UVLQ and DVLQ models, respectively. In blue, we further include the result of our previous analysis with heavy NP affecting $M_{12}^q$ for comparision. All results are obtained including the penguin pollution as discussed in Sec.~\ref{SEC:NPM12q}.}
\label{fig:NP33-UVLQ-DVLQ-scaled}
\end{center}
\end{figure}

In the DVLQ scenario, we ignore the mass of the new down-type quark since it does not enter the quantities of interest $\Gamma_{12}^q$ and $M_{12}^q$. In the UVLQ case, the mass of the new up-type quark $m_T$ is, on the contrary, very relevant. The results shown in Figs.~\ref{fig:NP33-UVLQ-DVLQ-scaled} and \ref{fig:argM12-vs-argG12-normalized} correspond to $m_T=1.6$ TeV. This value is chosen to avoid direct lower bounds~\cite{CMS:2022fck,Banerjee:2024zvg}.\footnote{It is to be mentioned, however, that such direct bounds typically assume patterns of decays, including dominance of decays into third generation quarks, that do not necessarily hold, and thus such bounds are not absolute and could be evaded; in any case, we play safe setting $m_T=1.6$ TeV.} We have also checked that for values of $m_T$ in $[1.6;5]$ TeV, the range of potential enhancement of the asymmetries $A^q_{\mathrm{SL}}$ is unchanged. Concerning the penguin pollution discussed in Sec.~\ref{SEC:NPM12q}, we only consider in this section analyses which include it, without further mention.

There is an important aspect of our analyses that deserves clarification. For both the UVLQ and DVLQ scenarios, we consider the same set of constraints discussed in Sec.~\ref{SEC:NPM12q}, and used in the model independent analyses presented there. In principle, this set could be extended to include other relevant constraints (e.g., $B_q$ rare decays, electroweak precision data/oblique parameters, kaon mixing observables, rare kaon decays, to name a few), if we were to perform detailed phenomenological analyses of these scenarios. We stick to the constraints of Sec.~\ref{SEC:NPM12q} since that is not our goal. As a consequence, the results obtained in this section, in particular the ranges in $A_{\mathrm{SL}}^d$ and $A_{\mathrm{SL}}^s$, come with a qualification. It is not guaranteed that the full range of variation of $A^d_{\mathrm{SL}}$ and $A^s_{\mathrm{SL}}$ would be allowed if additional constraints were to be included. What is guaranteed, however, is that values of $A^d_{\mathrm{SL}}$ and $A^s_{\mathrm{SL}}$ outside the allowed regions emerging from the analyses cannot be accommodated in any case in these scenarios. We have explored the size of the $3\times 3$ unitarity deviations that our analyses allow for, finding, in the DVLQ case $|(D_L)_{db}|<5\times 10^{-4}$ and $|(D_L)_{sb}|<2\times 10^{-3}$, while in the UVLQ case $|V_{Td}^{*}V_{Tb}|<10^{-3}$ and $|V_{Ts}^{*}V_{Tb}|<5\times 10^{-3}$. Since part of these ranges can indeed be in conflict with other constraints not included, we have also explored what are the minimal sizes of these $3\times 3$ unitarity deviations that are able to produce the full range of variation of $A^d_{\mathrm{SL}}$ and $A^s_{\mathrm{SL}}$, finding, in the DVLQ case $|(D_L)_{db}|\sim 2\times 10^{-4}$ and $|(D_L)_{sb}|\sim 5\times 10^{-4}$, while in the UVLQ case $|V_{Td}^{*}V_{Tb}|\sim 4\times 10^{-4}$ and $|V_{Ts}^{*}V_{Tb}|\sim 8\times 10^{-4}$. These values appear to be sufficiently safe from the point of view of other $B_d$ and $B_s$ related constraints \cite{Alves:2023ufm} not considered here. As a bonus, it is also clear from the values of these $3\times 3$ unitarity violations, that the new contributions to $\Gamma_{12}^q$ in the DVLQ case mentioned previously are much suppressed with respect to the usual ones, since $|(D_L)_{db}|\ll |\lambda^u_{bd}|,|\lambda^c_{bd}|\sim\mathcal O(\lambda^3)$ and $|(D_L)_{sb}|\leq |\lambda^u_{bs}|\sim\mathcal O(\lambda^4)\ll |\lambda^c_{bs}|\sim\mathcal O(\lambda^2)$ (with $\lambda\simeq 0.22$ in the Wolfenstein parameterization of CKM).

As a last comment concerning the constraints, the one on $|V_{tb}|$ deserves attention. When CKM is assumed to be $3\times 3$ unitary (as in Sec.~\ref{SEC:NPM12q}), the constraints on $|V_{cb}|$ and $|V_{ub}|$ already force $|V_{tb}|= 1-\mathcal O(\lambda^4)$, while the uncertainty in the experimental determination of $|V_{tb}|$ is (numerically) in the $\mathcal O(\lambda^2)$ to $\mathcal O(\lambda^3)$ ballpark. This means that the role of this constraint when $3\times 3$ unitarity is assumed is irrelevant. That is not the case anymore when CKM is not $3\times 3$ but part of a larger unitary matrix, since it bounds the moduli of elements of that extended matrix beyond the third row and column. 

The main result of this section is shown in Fig.~\ref{fig:NP33-UVLQ-DVLQ-scaled}. As in Fig.~\ref{fig:NP33} in Sec.~\ref{SEC:NPM12q}, the allowed region of $A^d_{\mathrm{SL}}$ vs. $A^s_{\mathrm{SL}}$ is shown, together with experimental expected sensitivities. For comparison, the allowed region obtained in the analysis of Sec.~\ref{SEC:NPM12q}, the NP $3\times 3$ case, is also displayed. One can observe that the enhancements of $A^d_{\mathrm{SL}}$ and $A^s_{\mathrm{SL}}$ in these simple scenarios are similar to the ones achieved in Sec.~\ref{SEC:NPM12q}, with values of $A^d_{\mathrm{SL}}$ at the $10^{-3}$ level, values of $A^s_{\mathrm{SL}}$ at the $10^{-4}$ level, and small correlation at the $-0.1$ level. More precisely:
\begin{itemize}
\item UVLQ case,
\begin{align}
 &A^d_{\mathrm{SL}}=(-8.0\pm 2.9)\times 10^{-4},\\
 &A^s_{\mathrm{SL}}=(-0.08\pm 0.97)\times 10^{-4},\\ 
 &\rho(A^d_{\mathrm{SL}},A^s_{\mathrm{SL}}) = -0.094.
\end{align}
\item DVLQ case,
\begin{align}
 &A^d_{\mathrm{SL}}=(-10.3\pm 2.9)\times 10^{-4},\\
 &A^s_{\mathrm{SL}}=(-0.47\pm 0.96)\times 10^{-4},\\ 
 &\rho(A^d_{\mathrm{SL}},A^s_{\mathrm{SL}}) = -0.101.
\end{align}
\end{itemize}
At a finer level of detail, it is to be noticed that these regions, although similar, are both slightly smaller than in the NP $3\times 3$ case, covering the same region in $A^s_{\mathrm{SL}}$ but a smaller one in $A^d_{\mathrm{SL}}$, with, in addition, a small shift of around $\sim 2\times 10^{-4}$ in the central value of $A^d_{\mathrm{SL}}$ in the UVLQ case with respect to the DVLQ scenario.

\section{$A_{\rm SL}^q$ with modifications to $\Gamma_{12}^q$ \label{SEC:G12q}}

In this section we explore scenarios which could lead to contributions to $\Gamma_{12}^q$ and its phase and explore its impact on $A_{\rm SL}^q$. This is something that was explored before in a range of models, in particular after the anomalously large asymmetries reported by D0 more than a decade ago, see~\cite{D0:2010sht,D0:2011hom,D0:2013ohp} for the measurements and~\cite{Bauer:2010dga,Dobrescu:2010rh,Trott:2010iz,Bai:2010kf,Botella:2014qya} for several references discussing phenomenology.

Trying to enhance the semileptonic asymmetries by modifying $\Gamma_{12}^q$ requires a priori light new physics. Why? Since $M_{12}^q$ is a transition amplitude dominated by virtual particles, while $\Gamma_{12}^q$ is a decay rate, they typically scale as $|\Gamma_{12}^q/M_{12}^q| \sim \mathcal{O}(m_b^2/m_{\rm virtual}^2)$. For instance, in the Standard Model $m_{\rm virtual} = m_t$ and therefore $|\Gamma_{12}^q| \ll |M_{12}^q|$ and consequently $A_{\rm SL}^q$ is also small. Beyond the Standard Model it becomes a model dependent task to understand how large $\Gamma_{12}^q$ can be and its impact on $A_{\rm SL}^q$. To address this question we opt for two avenues. First, we review existing studies in the literature that explore various possible common final states for the $B_q$ and $\bar{B}_q$ mesons. Second, we explicitly calculate how large the semileptonic asymmetry can be within the minimal realization of the $B$-Mesogenesis paradigm, which contains a color-triplet scalar that can mediate $B_q$ and $\bar{B}_q$ transitions and in particular modify $\Gamma_{12}^q$.

\subsection{Possible channels contributing to $\Gamma_{12}^q$\label{SEC:G12q_channels}}

It has been known for a while that there are not many possible channels that can substantially modify $\Gamma_{12}^q$. In particular, the viable options as of 2010 were~\cite{Bauer:2010dga}: 1) $b\to u_i \bar{u}_j q$ decays, where $u_i = u$ or $c$, 2) $b\to \tau \tau s$ decays, and 3) decays into modes involving invisible particles. Overall, they were viable because for case 1) substantial BSM effects could hide within large hadronic uncertainties, and cases 2) (taus) and 3) (invisibles) involve final state modes that are hard to detect, so that large branching fractions could still be allowed. The current status of each of these options is different and here we provide a summary and reinterpretation of recent works for options 1) and 2), and study one specific example for cases 1) and 3).

\begin{center}
    \underline{\textit{Modifying $\Gamma_{12}^q$ with $b\to u_i \bar{u}_j q$ decays at the EFT level}}
\end{center}
Recent global analyses of effective operators that can modify $b\to u_i \bar{u}_j q$ decays include~\cite{Bobeth:2014rda,Jager:2017gal,Lenz:2019lvd,Jager:2019bgk,Lenz:2022pgw}. In full generality, 20 combinations of $\bar{u}_i b \bar{u}_j q$ operators can be constructed with the quarks in different color and Lorentz structures. Out of those, most of them cannot lead to significant contributions to $\Gamma_{12}^q$ because, rather generically, they also substantially contribute radiatively to rare processes in the Standard Model, such as $b\to q \gamma$ and $b\to q \ell \ell$ --see e.g.~\cite{Jager:2017gal} for their relationship, and~\cite{Altmannshofer:2021qrr,Greljo:2022jac,Bause:2022rrs} for recent global analyses of these modes. Out of the operators investigated in~\cite{Jager:2017gal,Lenz:2019lvd,Jager:2019bgk}, color-rearranged structures of the type $Q_{1}^{d,\,cc} = (\bar{c}^\beta_L \gamma_\mu b_L^\alpha) (\bar{d}^\alpha_L \gamma^\mu c_L^\beta) $, where $\alpha, \,\beta$ are color indices, can lead to effects that saturate the experimental limit on $A_{\rm SL}^d$, namely, $|A_{\rm SL}^d| \lesssim 2\times 10^{-3}$. $\Gamma_{12}^s$ is strongly dominated by $b\to c\bar{c}s$ decays, but in this scenario operators of the type $Q_{1}^{s,\,cc}$ are severely constrained by $b\to s\gamma$, while color-singlet operators of the type $Q_2^{s, cc} = (\bar{c}^\alpha_L \gamma_\mu b_L^\alpha) (\bar{s}^\beta_L \gamma^\mu c_L^\beta) $ can lead to the largest modifications allowing for $|A_{\rm SL}^s| \lesssim 2\times 10^{-3}$ as well. 

\begin{center}
    \underline{\textit{Modifying $\Gamma_{12}^q$ with $b\to \tau \tau s$ at the EFT level}}
\end{center}
The recent global analysis of~\cite{Bordone:2023ybl} studied the current status of all possible effective operators that can significantly contribute to $\Gamma^{s}_{12}$ via $b\to \tau \tau s$ decays (see also~\cite{Bobeth:2011st}). Although in~\cite{Bordone:2023ybl} the authors restricted themselves to real Wilson coefficients, their results can be extended to show that in the presence of complex Wilson coefficients $|A_{\rm SL}^s| \lesssim 10^{-3}$, again a rather large number.\footnote{Note that the case of $b\to \tau \tau d$ is significantly more constrained than that of $b\to \tau \tau s$ because ${\rm Br}(B_d \to \tau\tau) < 2.3\times 10^{-3}$ while ${\rm Br}(B_s \to \tau\tau) < 6.8\times 10^{-3}$ at 90\% C.L.~\cite{LHCb:2017myy}. Thus, there appears to be more room for $b\to \tau\tau s$ transitions as compared to $b\to \tau\tau d$ ones (although a dedicated study is not currently available in the literature).}

\begin{center}
    \underline{\textit{Modifying $\Gamma_{12}^q$ within UV complete models}}
\end{center}
The studies mentioned before worked in terms of effective $\Delta B = 1$ operators involving light states (below the $m_b$ scale), whose Wilson coefficients are appropriately constrained by all relevant observables, including, in particular, the meson mass differences $\Delta M_q$ that are generated through $\Delta B = 2$ transitions. On that respect, it is crucial to check that contributions to $\Delta M_q$ controlled by heavier states do not spoil the agreement between the experimental measurement and the Standard Model prediction. To understand this observation, it is illustrative to think, for instance, in the Standard Model case: as $\Delta M_q^\mathrm{SM}$ is top-dominated, in an effective theory such as those proposed in the previous studies this contribution would be missed. Therefore, we include in this section a review of some UV completions capable of generating the aforementioned $\Delta B = 1$ operators in order to check the actual enhancement of $A_\mathrm{SL}^q$ one is able to achieve.
 
Recent examples of UV models featuring $\Delta B = 1$ operators include Ref.~\cite{Crivellin:2023saq} for operators of the type 1), and Ref.~\cite{Cornella:2021sby} for operators of the type 2). Taking the parameter space in~\cite{Crivellin:2023saq} at face value, the authors report that $A_{\rm SL}^s$ could be as large as $A_{\rm SL}^s = -4\times 10^{-5}$ and thus only a factor of two larger than the Standard Model prediction in Eq.~\eqref{eq:ASLs-SM}, and a factor of $\sim \! 50$ times smaller than what appears to be allowed when considering only $\Delta B = 1$ operators. Regarding the type 2), the UV complete model that realizes potentially large $b\to \tau \tau s$ transitions of~\cite{Cornella:2021sby} would allow for contributions to $A_{\rm SL}^s$ that are $\lesssim 10^{-5}$, and thus again rather small and comparable to the effects one can find by modifying the phase of $M_{12}^q$. 

Thus, although this does not constitute an exhaustive list of UV completions, it generically highlights that, indeed, in UV theories that could potentially induce modifications to $\Gamma_{12}^q$ it is not easy to find large values of the CP asymmetries. The reason is rather simple: the contribution of $\Gamma_{12}^q$ appears from tree-level decays and mediators of these interactions are generically constrained to have $M \gtrsim 1\,{\rm TeV}$ from LHC searches, and this in turn leads to very suppressed contributions to $\Gamma_{12}^q$. 

To complete this section, in what follows, we consider at face value the minimal scenario capable of realizing $B$-Mesogenesis with the aim of understanding whether it can lead to substantially large CP asymmetries. This is motivated because it can contribute both to $\Delta B = 2$ transitions as well as to $\Delta B = 1$ operators of the type 1) $b\to u_i \bar{u}_j q$, and those of the type 3) involving a new invisible particle, $b\to \psi\bar{\psi} q$.

\subsection{$\Gamma_{12}^q$ within the minimal $B$-Mesogenesis realization\label{SEC:G12q_BMesogenesis}}
The minimal realization of $B$-Mesogenesis~\cite{Elor:2018twp,Alonso-Alvarez:2021qfd} involves a color-triplet scalar $Y$ with hypercharge $-1/3$, and a dark sector antibaryon $\psi$.\footnote{Not to be confused with the $J/\Psi$ hadronic resonance.} The interaction Lagrangian reads:
\begin{align}
    \!\! \mathcal{L}_{-1/3} =& - \sum_{i,\,j} y_{u_id_j} Y^* \bar{u}_{iR} d_{jR}^c  + \text{h.c.}  \nonumber \\
    & - \sum_k y_{\psi d_k} Y d_{kR}^c \bar{\psi} + \text{h.c.} \,, \label{eq:LflavorUV_13} 
\end{align}
where $y$ represent (complex) coupling constants.\footnote{Note that there is also a version of the model with $Y= 2/3$, but we would expect even smaller $\Gamma_{12}^q$ contributions in that scenario, see~\cite{Alonso-Alvarez:2021qfd} for the phenomenology.} While $Y$ features a coupling to two quarks, proton decay is evaded by imposing baryon number conservation, with $B(\psi) = -1$, $B(Y) = -2/3$, and by kinematics, with $m_\psi > m_p - m_e$.

\begin{figure}[htbp]
\begin{center}
\includegraphics[width=0.45\textwidth]{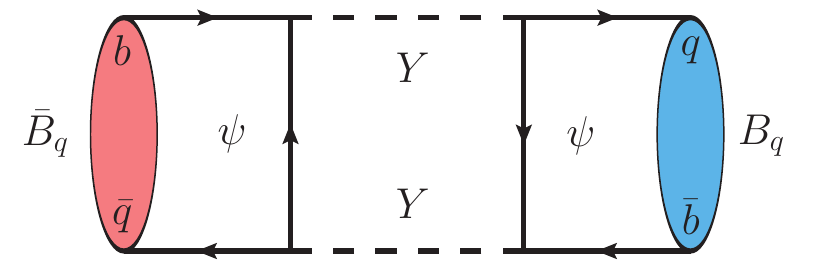} 
\caption{\justifying Box diagram triggering $B_q$--$\bar{B}_q$ mixing from $\psi$--$Y$ mediation. The other topology is not shown, but it is considered in the calculation.}\label{fig:Box-BSM_psi}
\end{center}
\end{figure}

\begin{center}
    \underline{\textit{Modifying $\Gamma_{12}^q$ with semi-invisible $b\to \psi \bar{\psi} q$ decays}}
\end{center}

The $\psi$ fermion and $Y$ boson mediate the transition amplitudes shown in Fig.~\ref{fig:Box-BSM_psi}. They contribute to $M_{12}^q$ as well as $\Gamma_{12}^q$ via $b\to \psi \bar{\psi} q$ decays provided that $m_\psi < m_b/2+m_q$. In this kinematical regime and considering $m_q = 0$, direct computation leads to:
\begin{align}
    \Gamma_{12}^{q,\mathrm{NP}}(\psi) &= -\frac{f_{B_q}^2 M_{B_q}}{256\pi} \frac{(y_{\psi q}y_{\psi b}^{*})^2 m_b^2}{M_Y^4} \\
    &\times \left(1-\frac{2}{3}\frac{m_\psi^2}{m_b^2}\right)\sqrt{1-4\frac{m_\psi^2}{m_b^2}}, \nonumber
\end{align}
and
\begin{equation}\label{eq:M12qPsi}
    M_{12}^{q,\mathrm{NP}}(\psi) = \frac{f_{B_q}^2 M_{B_q}}{384\pi^2} \frac{(y_{\psi q}y_{\psi b}^{*})^2}{M_Y^2} G(x_{\psi Y}),
\end{equation}
with $x_{\psi Y} = m_\psi^2/M_Y^2$, and $G(x_{\psi Y})$ the function reported in App.~\ref{app:formulaeBMesogenesis}, although for the relevant range of parameters $G\simeq 1$. 

We know that any new physics contributions to $M_{12}^q$ are bounded to be at most $\sim 10\,\%$ of the SM one given Eqs.~\eqref{eq:3x3Globalanalysis}. From Eq.~\eqref{eq:M12qPsi}, this leads to a constraint on the couplings:
\begin{subequations}\label{eq:couppsibounds}
    \begin{align}
   |y_{\psi d}y_{\psi b}^{*}| & \lesssim 0.01 \times {M_Y}/{500} \,{\rm GeV} \,,\\
    |y_{\psi s}y_{\psi b}^{*}| & \lesssim 0.05 \times {M_Y}/{500} \,{\rm GeV} \,,
\end{align}
\end{subequations}
where the bound has been normalized considering that $M_Y> 500\,{\rm GeV}$ is the most conservative limit from direct LHC searches on pair produced $Y$  resonances~\cite{ATLAS:2017jnp,CMS:2018mts,CMS:2019zmd,ATLAS:2020syg}.\footnote{More restrictive LHC bounds on the $Y$ mass may apply as it can be produced singly and also decay into jets and missing energy, see~\cite{Alonso-Alvarez:2021qfd} for a detailed assessment of these bounds.}
To understand the potential relevance of $\Gamma_{12}^{q,{\rm NP}}$ into the semileptonic asymmetry we can write:
\begin{align}\label{eq:ASLexpanded}
    A_{\rm SL}^q &= {\rm Im} \left(\frac{\Gamma_{12}^{q,{\rm SM}}+\Gamma_{12}^{q,{\rm NP}}}{M_{12}^{q,{\rm SM}}+M_{12}^{q,{\rm NP}}}\right) \\
    &\simeq {\rm Im} \left(\frac{\Gamma_{12}^{q,{\rm SM}}}{M_{12}^{q,{\rm SM}}+M_{12}^{q,{\rm NP}}}\right)  +  {\rm Im} \left(\frac{\Gamma_{12}^{q,{\rm NP}}}{M_{12}^{q,{\rm SM}}}\right) \,, \nonumber
\end{align}
where in the second term, for simplicity, we have neglected $M_{12}^{q,{\rm NP}}$ as it can be at most 10\% of the SM. Note that here the first term is precisely the one that we explored in a model independent way in Sec.~\ref{SEC:NPM12q}. Taking into account the constraints in Eqs.~\eqref{eq:couppsibounds}, we see that the new physics contribution from the last term in Eq.~\eqref{eq:ASLexpanded} involving $\psi$ exchange is bounded to be
\begin{align}
    |A_{\rm SL}^{q, {\rm NP}} (\psi)| < 4\times 10^{-5} \left(\frac{500\,{\rm GeV}}{M_Y}\right)^2\,.
\end{align}
These contributions to the semileptonic asymmetries are smaller than the one from $M_{12}^{q,{\rm NP}}$ in Eq.~\eqref{eq:ASLexpanded}, see Eq.~\eqref{eq:ASL_resultspeng}. 

\vspace{0.2cm}

\begin{center}
    \underline{\textit{Modifying $\Gamma_{12}^q$ with $b\to u_i \bar{u}_j q$ decays}}
\end{center}

Within the minimal realization of $B$-Mesogenesis it is also possible for $Y$ to mediate a transition amplitude from the quark-quark coupling present in Eq.~\eqref{eq:LflavorUV_13}. Working in the vacuum insertion approximation ($B_{B_q} = 1$) and taking the simplification that $M_{B_q}/(m_b+m_q)\simeq 1$, we find:
\begin{align}\label{eq:Gamma_12_diquark}
    \Gamma_{12}^{q,\mathrm{NP}}(\cancel{\psi}) &= \frac{f_{B_q}^2 M_{B_q}}{384\pi^2} \sum_{i,j=u,c} \frac{\pi \sqrt{\lambda(m_b^2,m_i^2,m_j^2)}}{m_b^2} \\ 
    &\times \left[ (V_{ib}V_{jq}^{*} y_{iq}y_{jb}^{*}) \frac{m_i m_j}{M_W^2 M_Y^2} 8 g^2 \right. \nonumber \\ 
    &\left.\,\,\,\,\,\,+\,(y_{iq}y_{ib}^{*} y_{jq}y_{jb}^{*}) \frac{m_b^2}{12M_Y^4} (8g_{2}^{ij}-5 g_{3}^{ij})\right]\,,\nonumber
\end{align}
where $g$ is the $SU(2)_L$ coupling constant, $\lambda$ is the Källen function, and $g_{2,3}^{ij}$ are functions of the quark mass ratios as given in App.~\ref{app:formulaeBMesogenesis} that numerically are of $\mathcal{O}(1)$.  

The equivalent contribution to $M_{12}^q$ contains terms with all up-type quarks, and we find:
\begin{align}\label{eq:M_12_diquark}
    M_{12}^{q,\mathrm{NP}}(\cancel{\psi}) = &-\frac{f_{B_q}^2 M_{B_q}}{384\pi^2} \\ &\times\sum_{i,j = u,c,t} \left[(V_{ib}V_{jq}^{*} y_{iq}y_{jb}^{*}) \frac{m_i m_j}{M_W^2 M_Y^2} g^2 f_{1}^{ij} \right. \nonumber\\ 
    &\left.\,\,\,\,\,\, \,\,\,\,\,\,\,\,\,\,\,\,\,\,\,\,\,\,\,\, -\,(y_{iq}y_{ib}^{*} y_{jq}y_{jb}^{*}) \frac{1}{M_Y^2} f_{2}^{ij}\right]\,\nonumber,
\end{align}
which agrees with the results within the same model from~\cite{Agrawal:2014aoa,Alonso-Alvarez:2021qfd}, and where the functions $f_1^{ij}$ and $f_2^{ij}$ depend upon the quark and $Y$ masses. They are given in App.~\ref{app:formulaeBMesogenesis}, but  numerically are $\mathcal{O}(1-10)$.

\begin{figure}[!t]
\begin{center}
\subfloat[\label{sfig:Box1-BSM_psi}]{\includegraphics[width=0.225\textwidth]{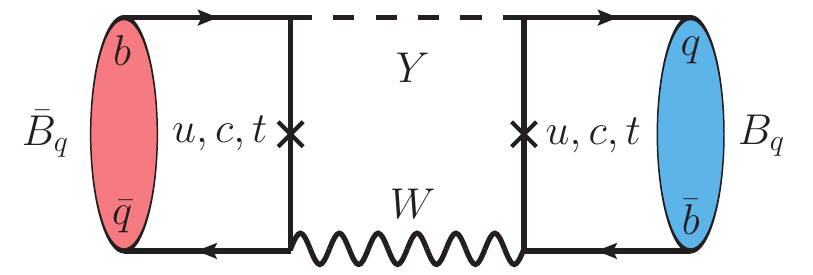}}\quad
\subfloat[\label{sfig:Box1-BSM_psi}]{\includegraphics[width=0.225\textwidth]{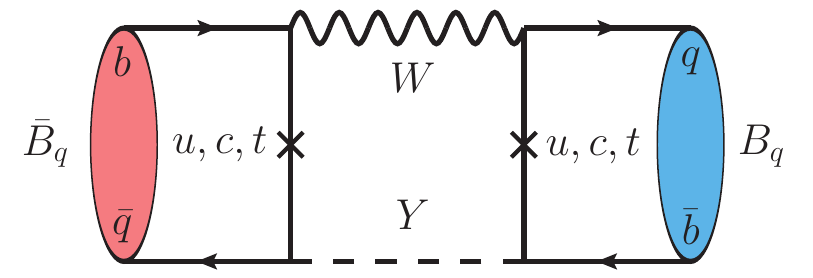}}\\
\subfloat[\label{sfig:Box1-BSM_psi}]{\includegraphics[width=0.225\textwidth]{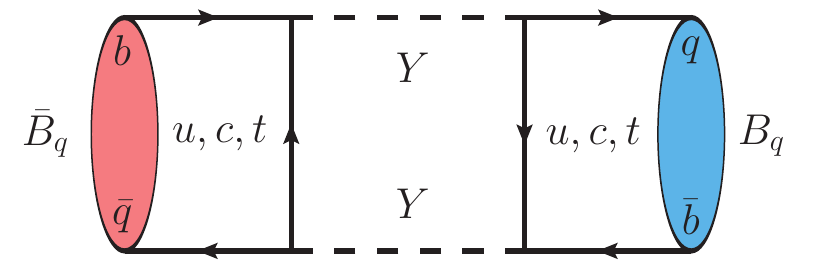}}
\caption{\justifying Box diagrams triggering $B_q$--$\bar{B}_q$ mixing from $u_i$--$Y/W$ mediation.}\label{fig:Box-BSM_diquark}
\end{center}
\end{figure}

From Eq.~\eqref{eq:Gamma_12_diquark} we can clearly see that $\Gamma_{12}^q$ will be dominated by the $b\to c\bar{c} q$ channel. On the other hand, if the $y_{iq}$ coupling constant matrix is homogeneous, then $M_{12}^q$ will be top-dominated, and therefore the contribution to $\Gamma_{12}^q
$ in the model will be strongly suppressed. In order to maximize the size of $\Gamma_{12}^q$ and $A_{\rm SL}^q$, we can consider a favorable (and tuned) scenario where only the $b\to c\bar{c} q$ channel contributes both to $M_{12}^q$ and $\Gamma_{12}^q$. By requiring that $|M_{12}^{q,{\rm NP}}|$ does not exceed more than 10\% of the SM one (see Eqs.~\eqref{eq:3x3Globalanalysis}), one finds:
\begin{subequations}
\begin{align}
    |y_{cd} y_{cb}^{*}| &\lesssim 0.01 \times M_Y/500\,{\rm GeV}  \,,\\
    |y_{cs} y_{cb}^{*}| &\lesssim 0.05 \times M_Y/500\,{\rm GeV}  \,,
\end{align}
\end{subequations}
where the bounds scale linearly with $M_Y$ because the dominant contribution comes from the second term in Eq.~\eqref{eq:M_12_diquark}. Proceeding as before, noting that at least $M_Y > 500\,{\rm GeV}$ from LHC searches, and taking the phase of $y_{cb}^{*}y_{cq}$ to be $\pi/2$ with respect to the SM one (in order to maximize the $A_{\rm SL}^q$ asymmetry), we can use Eq.~\eqref{eq:Gamma_12_diquark} to find that the contribution from $\Gamma_{12}^q$ to the CP asymmetries is:
\begin{align}\label{eq:ASL_Gamma12-BMesogenesis}
    |A_{\rm SL}^{q,{\rm NP}}(\cancel{\psi})| &\lesssim 10^{-4}  \,.
\end{align}
This means that even in the most favorable scenario where only $b\to c\bar{c}q$ decays appear in both $\Gamma_{12}^q$ and $M_{12}^q$, the contributions from $\Gamma_{12}^q$ to $A_{\rm SL}^q$ are small. This results from a combination of the fact that the very same channel contributes to $\Delta M_{q}$, which cannot be modified beyond 10\%, and because we know that at least $M_Y > 500\,{\rm GeV}$ from LHC constraints. We have further checked that using $b\to \bar{u}c q$, $b\to \bar{c}u q$ or $b\to \bar{u}u q$ decays leads to even smaller asymmetries than the ones quoted in Eq.~\eqref{eq:ASL_Gamma12-BMesogenesis}.

We thus conclude that in the absence of fine tuning the maximum values of the semileptonic asymmetries in the minimal realization of the $B$-Mesogenesis mechanism are mainly dominated by the $\sim 1^\circ$ phase that the color-triplet scalar could induce to $M_{12}^q$, and not by new contributions to $\Gamma_{12}^q$. This means that the same limits we found in Sec.~\ref{SEC:NPM12q} for generic heavy new physics scenarios that only modify the phase and magnitude of $M_{12}^q$ apply, i.e, the semileptonic asymmetries cannot be larger than those in Eqs.~\eqref{eq:ASL_resultspeng}. An exception to this conclusion can be reached if there are large contributions to $\Gamma_{12}^q$ from a given channel, say $b\to c\bar{c}s$, but then the contributions from this channel to $M_{12}^q$ are compensated by another one with opposite sign from a different channel, say from the top. In that case, $\Delta M_{q}$ would still be in agreement with the Standard Model prediction. While this is in principle possible, it is finely tuned as it requires very specific phases and hierarchies of coupling constants. In this context, we finally comment on two additional options that have been invoked in the literature to enlarge the CP asymmetries with light new particles. Ref.~\cite{Bai:2010kf} considered a new hidden pseudoscalar that mixes with $B_q$ and $\bar{B}_q$. In order for the CP asymmetries to be enlarged this new state should have a mass $M \simeq M_B$ and very specific decay modes. In addition, Ref.~\cite{Oh:2010vc} considered the contributions from a very light $Z'$ with $M \lesssim M_B$. Significant enhancements to the CP asymmetries can be obtained by considering only $bq$ couplings, but the necessary presence of other types of interactions are likely to strongly constrain this possibility, see~\cite{Crivellin:2022obd}.

\section{Implications for $B$-Mesogenesis and future searches}\label{sec:implications}
\setlength\parskip{0pt}

Having performed global analyses to explore the magnitude of the CP asymmetries in neutral $B$ meson mixing within heavy new physics scenarios modifying mass mixing in Sec.~\ref{SEC:NPM12q}, scenarios featuring vector-like quarks leading to a non-unitary CKM in Sec.~\ref{SEC:VLQ}, and scenarios modifying decay width mixing $\Gamma_{12}^q$ in Sec.~\ref{SEC:G12q}, we are now in good position to compare our findings with 1) the expected experimental sensitivity for these quantities from LHCb and Belle II (see Eqs.~\eqref{eq:ASL_sensitivities_future}), and 2) with their minimum required value for a successful $B$-Mesogenesis.

As introduced in Sec.~\ref{SEC:Introduction}, within the $B$-Mesogenesis framework~\cite{Elor:2018twp,Nelson:2019fln,Alonso-Alvarez:2021qfd}, the amount of CP violation in neutral $B$ meson mixing is directly proportional to the baryon asymmetry of the Universe. In particular, a minimal requirement is that~\cite{Alonso-Alvarez:2021qfd}:\footnote{See also footnote~\ref{footnote1}.}
\begin{align}
    A_{\rm SL}^q > +10^{-4} \,.
\end{align}
Importantly, at least one of the CP asymmetries needs to be positive as we live in a Universe where structures are made of matter. The prediction for the baryon asymmetry does depend upon the sign and magnitude of $A_{\rm SL}^d$ and $A_{\rm SL}^s$ (as each meson system oscillates at a different rate and they are produced in different numbers in the early Universe), but also upon the branching ratio of the new decay mode of a $B$ meson into a dark sector antibaryon, a visible baryon, and any number of light mesons, that is, ${\rm Br}(B\to {\rm baryon} + \psi +{\rm mesons})$. These two parameters are strongly correlated, with the latter constrained by direct searches at $B$-factories~\cite{Belle:2021gmc,BaBar:2023rer,BaBar:2023dtq} and from a recast of old searches at LEP~\cite{Alonso-Alvarez:2021qfd}, see also~\cite{Khodjamirian:2022vta,Elor:2022jxy,Boushmelev:2023huu,Shi:2023riy} for refined theoretical predictions. While these bounds depend upon the $\psi$ mass, a rather global conservative constraint is ${\rm Br} \lesssim 1\%$ for $m_\psi \sim 1\,{\rm GeV}$. For heavier $\psi$ masses and depending upon the particular flavor combination that dominates in Eq.~\eqref{eq:LflavorUV_13}, the bound can actually be more restrictive, namely ${\rm Br} <(0.01-1)\%$. The smaller this branching ratio is, the larger in magnitude the CP asymmetries should be to provide the observed baryon asymmetry of the Universe. 

\begin{figure*}[ht]
\begin{center}
\includegraphics[width=0.7\textwidth]{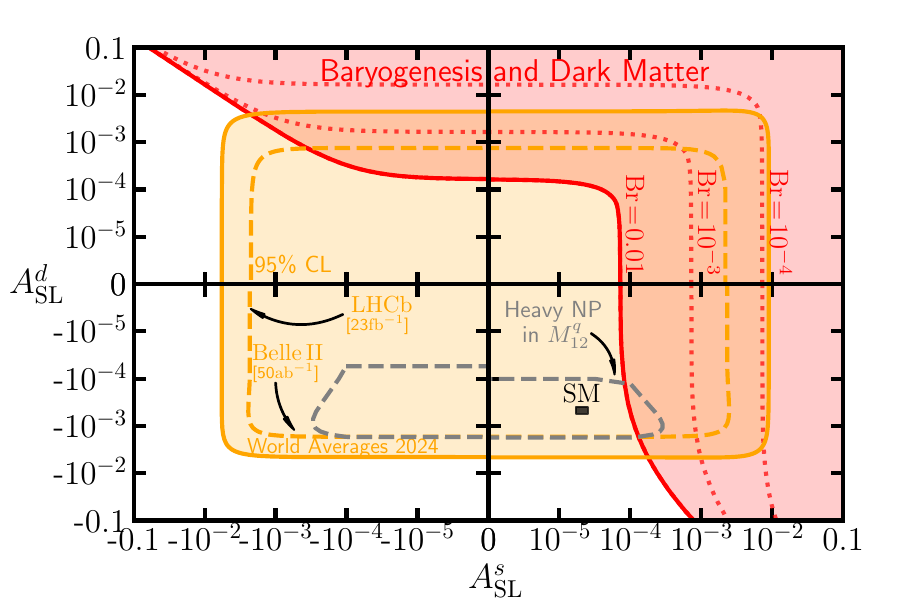}
\caption{\justifying Semileptonic asymmetries of the $B_d$ and $B_s$ systems in logarithmic scale. In orange we show the experimentally allowed regions at 95\% C.L. (2D-$\Delta \chi^2 = 5.99$, Eq.~\eqref{eq:ASLq-Exp}), and in dashed orange the expected sensitivity from LHCb and Belle II assuming their measurements are centered around the SM prediction, see Eqs.~\eqref{eq:ASL_sensitivities_future}. In red we highlight the region of parameter space identified in~\cite{Alonso-Alvarez:2021qfd} in which the baryon asymmetry of the Universe can be explained via the $B$-Mesogenesis mechanism~\cite{Elor:2018twp,Nelson:2019fln,Alonso-Alvarez:2021qfd}. The dashed red lines correspond to isocontours of ${\rm Br}(B \to {\rm baryon} + \psi + {\rm mesons})$. Only the region of parameter space with ${\rm Br} < 0.01$ is shown since larger branching fractions are conservatively excluded. Depending upon other parameters, the constraints can reach up to ${\rm Br} < 10^{-4}$. The dashed grey line is one of the main results of our study and highlights the values of the semileptonic asymmetries that heavy new physics models contributing to $M_{12}^q$ can reach, see Eq.~\eqref{eq:ASL_resultspeng}. We see that the overlap between the red and the grey dashed regions is small, and this puts the $B$-Mesogenesis mechanism in tension as for it to be successful $B$ mesons should posses an inclusive $\sim \mathcal{O}(1\%)$ branching ratio into a baryon, any number of light mesons, and missing energy. We finally note that semileptonic asymmetries larger than those contained in the grey contour could be obtained in tuned scenarios where $\Gamma_{12}^q$ is modified, see Sec.~\ref{SEC:G12q} for a discussion.}
\label{fig:ASL_global_BMesogenesis}
\end{center}
\end{figure*}

We depict the parameter space studied in~\cite{Alonso-Alvarez:2021qfd} in Fig.~\ref{fig:ASL_global_BMesogenesis}, where we show the allowed region for the $A_{\rm SL}^d$ and $A_{\rm SL}^s$ asymmetries, and highlight in red the region of parameter space where the mechanism could lead to the observed baryon asymmetry of the Universe according to~\cite{Alonso-Alvarez:2021qfd}. In addition, we show in orange the current  experimental world averages, the expected sensitivity from LHCb and Belle II, and most importantly our results for the scenarios where $M_{12}^q$ is modified by heavy new physics. We show this case because, as we have checked in the previous sections, scenarios with a non-unitary CKM or UV complete models that could modify $\Gamma_{12}^q$ cannot lead to larger CP asymmetries. In light of this, Fig.~\ref{fig:ASL_global_BMesogenesis} allows us to state two of our key results:
\begin{enumerate}
    \item \textbf{New measurements of the semileptonic asymmetries by LHCb and Belle II are not expected to test the most generic new physics scenarios}. All the models we have considered feature smaller asymmetries than their future sensitivity. This includes any BSM scenario that modifies mass mixing $M_{12}^q$, scenarios featuring vector-like quarks leading to a non-unitary CKM, as well as non-tuned UV complete models which could also contribute to decay width mixing $\Gamma_{12}^q$. The only scenarios that could lead to large asymmetries are those which introduce substantial modifications to $\Gamma_{12}^q$ but also tuned contributions to $M_{12}^q$ in order to evade the strong constraints from $\Delta M_q$.
    \item \textbf{The small CP asymmetries that can be obtained BSM put the $B$-Mesogenesis mechanism in theoretical tension}. As shown in red, the mechanism requires positive and rather large CP asymmetries in order to explain the observed baryon asymmetry of the Universe. However, only the region with $A_{\rm SL}^s \simeq (1-5) \times 10^{-4}$ and $A_{\rm SL}^d \simeq A_{\rm SL}^{d,\rm SM}$ can be theoretically achieved from a generic contribution to $M_{12}^q$. In the rest of the parameter space, and in the absence of tuning, there is no BSM scenario explored in our study which can lead to the required CP asymmetries to have successful baryogenesis. We note that while ${\rm Br } < 1\%$ is a rather conservative constraint on the new decay mode $B\to {\rm baryon} + \psi + {\rm mesons}$, in many regions of the parameter space the bound is stronger and therefore the tension is even larger.
\end{enumerate}

\section{Summary, Conclusions and Outlook}\label{sec:conclusions}

\setlength\parskip{1pt}

In this study we have explored in detail how large CP violation in neutral $B$ meson mixing could be beyond the Standard Model. This is motivated by the fact that 1) $B$-factories and the LHC have significantly constrained the $B_d$ and $B_s$ mixing parameters, and 2) we know that this type of CP violation could be at the origin of the matter-antimatter asymmetry of the Universe through the $B$-Mesogenesis mechanism, but 3) the SM CP violation in $B$ mesons falls short by $\sim \! 1$ order of magnitude to explain the observed baryon asymmetry of the Universe.

In this context, we have first considered in Sec.~\ref{SEC:NPM12q} a very general scenario where heavy new physics affects meson mass mixings $M_{12}^q$. We have followed a model independent approach allowing the modulus and phase of $M_{12}^d$ and $M_{12}^s$ to vary. Then, we have performed a global analysis of these quantities and the CKM matrix by using relevant flavor data, including the meson mass differences and CP violation measurements in the interference between mixing and decay. Our 68\% C.L. results projected in terms of the semileptonic asymmetries are:
\begin{subequations}\label{eq:ASL_resultspeng_conclusions}
\begin{align}
    &A_\mathrm{SL}^d |_\mathrm{w/\, peng} = (-9.1\pm 3.6)\times 10^{-4},\\
    &A_\mathrm{SL}^s |_\mathrm{w/\, peng} = (-0.04 \pm 1.21)\times 10^{-4},\\
    &\rho(A_\mathrm{SL}^d,A_\mathrm{SL}^s) |_\mathrm{w/\, peng} = -0.113.
\end{align}
\end{subequations}
Effectively, these results are partly driven by the $\sim 1^\circ$ precision measurement of the phase of $M_{12}^d$ and $M_{12}^s$ from measurements of CP violation in the interference between mixing and decay, and partly from the allowed freedom in the CKM parameters.

In Sec.~\ref{SEC:VLQ} we have considered the case of an additional generation of up-type and down-type vector-like quarks that lead to a non-unitary $3\times 3$ CKM matrix. This is motivated because one of the reasons the CP violation in $B$ meson mixing in the Standard Model is so small is precisely the unitarity of the CKM matrix. Our results from an enlarged global CKM fit show that these scenarios can lead to similar CP asymmetries to the ones in Eq.~\eqref{eq:ASL_resultspeng_conclusions}, but not larger in magnitude. The main 
reasons why the CP asymmetries are not substantially larger stem from the fact that 1) the allowed deviations of CKM $3\times 3$ unitarity are small, and 2) up-type vector-like quarks are constrained by direct searches to be rather heavy and thus cannot lead to large effects on the relation between $\Gamma_{12}^q$ and $M_{12}^q$.

Finally, in Sec.~\ref{SEC:G12q} we have considered scenarios that can directly alter decay width mixings $\Gamma_{12}^q$. While at first sight it appears that in an EFT where only operators that contribute to $\Gamma_{12}^q$ the CP asymmetries could be large, the actual results in UV complete models show that the resulting contribution to the CP asymmetries from $\Gamma_{12}^q$ ends up being significantly smaller than those in Eq.~\eqref{eq:ASL_resultspeng_conclusions}. Our results include the first full calculation of the CP asymmetries within the minimal realization of $B$-Mesogenesis in Sec.~\ref{SEC:G12q_BMesogenesis}. The overall difficulty in obtaining large contributions to $A_{\rm SL}^q$ from $\Gamma_{12}^q$ is due to the fact that when a full UV complete model is considered 1) the contributions to $\Gamma_{12}^q$ are typically suppressed by $m_b^2/M_{\rm heavy}^2$ with respect to those for $M_{12}^q$, 2) the modulus of $M_{12}^q$ should be within $\sim 10\%$ of the SM value and its phase within $1^\circ$, and 3) the LHC sets typically very strong constraints on $M_{\rm heavy}$, hence diminishing the effects on $\Gamma_{12}^q$. 

To conclude, in Sec.~\ref{sec:implications} we have put our results in context by comparing these findings with the sensitivity of LHCb and Belle II on the CP asymmetries quantifying CP violation in $B$ meson mixing. Our main results are summarized in Fig.~\ref{fig:ASL_global_BMesogenesis} and highlight that 1) we do not expect upcoming measurements of the CP asymmetries to test any of the many new physics models we have considered in this study (unless significant tunings are present), and 2) we clearly see that the $B$-Mesogenesis mechanism is in theoretical tension. This is because the rather large and positive CP asymmetries needed for the mechanism to be successful are difficult to feature beyond the Standard Model. The only region of parameter space that could be covered is the one with $A_{\rm SL}^s\simeq (1-5)\times 10^{-4}$ and $A_{\rm SL}^d \simeq A_{\rm SL}^{d,{\rm SM}}$. This is interesting because in this region of parameter space $B$ mesons should have a rather large branching ratio ($0.2-1\%$) into a visible baryon and missing energy. While this represents a rather strong requirement, it makes the mechanism even more predictive and we note that its minimal realization could actually incorporate the two ingredients: the amount of CP violation needed (from the triplet scalar contributions to $M_{12}^s$), as well as the required branching ratio for the new decay mode. 

Overall, we find that it appears challenging to significantly enlarge the CP violation in the neutral $B$ meson mixing beyond the Standard Model. This has implications both for the interpretation of future LHCb and Belle II measurements of the semileptonic asymmetries as well as for the $B$-Mesogenesis mechanism, as the latter is put in theoretical tension. Despite the rich ongoing flavor physics program at LHC and Belle II, we expect our results not to be significantly altered for the next 5--10 years. The reason is that the ranges of the CP asymmetries found in Eq.~\eqref{eq:ASL_resultspeng_conclusions} are primarily driven by the measurements of the phase of $M_{12}^q$ from time-dependent CP asymmetries arising from the interference between the decays with and without oscillations. The relevant phases $\phi_d$ and $\phi_s$ have been measured with $\sim \! 1^\circ$ precision; this is precisely the expected size of SM penguin pollution and hence, at first sight, contain irreducible uncertainties. 

The quest for beyond the Standard Model physics continues, but our study highlights that it is likely not to be found for the most generic models through upcoming measurements of CP violation in $B$ meson mixing. Very high precision measurements of these asymmetries at a future Tera-Z factory such as FCC-ee~\cite{Monteil:2021ith,FCC:2018evy} or finely tuned contributions to $B$ meson mixing may change this picture.

\section*{Acknowledgments}
We would like to thank Alexander Lenz for very useful comments on the manuscript. We are grateful to Tobias Tsang for useful clarifications on lattice results on $B$ decay constants and bag parameters, to Marzia Bordone for useful discussions on the $b\to \tau \tau s$ decay contributions to $A_{\rm SL}^s$, and to Peter Stangl for useful discussions on $b\to q \ell \ell$ and  $b\to q\gamma$ bounds and their connection with $b\to u \bar{u} q$ decays. CM is funded by \textit{Conselleria de Innovación, Universidades, Ciencia y Sociedad Digital} from \textit{Generalitat Valenciana} and by \textit{Fondo Social Europeo} under grants ACIF/2021/284, CIBEFP/2022/92, and CIBEFP/2023/96. CM specially thanks CERN Theoretical Physics Department for the hospitality during his visit. MN acknowledges support from \textit{Generalitat Valenciana} through project CIDEGENT/2019/024, and Spanish MICIU/AEI/10.13039/501100011033/ through grants PID2020-113334GB-I00, PID2023-151418NB-I00, and the \emph{Severo Ochoa} project CEX2023-001292-S.

\newpage

\bibliography{biblio}

\begin{thebibliography}{151}%
\makeatletter
\providecommand \@ifxundefined [1]{%
 \@ifx{#1\undefined}
}%
\providecommand \@ifnum [1]{%
 \ifnum #1\expandafter \@firstoftwo
 \else \expandafter \@secondoftwo
 \fi
}%
\providecommand \@ifx [1]{%
 \ifx #1\expandafter \@firstoftwo
 \else \expandafter \@secondoftwo
 \fi
}%
\providecommand \natexlab [1]{#1}%
\providecommand \enquote  [1]{``#1''}%
\providecommand \bibnamefont  [1]{#1}%
\providecommand \bibfnamefont [1]{#1}%
\providecommand \citenamefont [1]{#1}%
\providecommand \href@noop [0]{\@secondoftwo}%
\providecommand \href [0]{\begingroup \@sanitize@url \@href}%
\providecommand \@href[1]{\@@startlink{#1}\@@href}%
\providecommand \@@href[1]{\endgroup#1\@@endlink}%
\providecommand \@sanitize@url [0]{\catcode `\\12\catcode `\$12\catcode
  `\&12\catcode `\#12\catcode `\^12\catcode `\_12\catcode `\%12\relax}%
\providecommand \@@startlink[1]{}%
\providecommand \@@endlink[0]{}%
\providecommand \url  [0]{\begingroup\@sanitize@url \@url }%
\providecommand \@url [1]{\endgroup\@href {#1}{\urlprefix }}%
\providecommand \urlprefix  [0]{URL }%
\providecommand \Eprint [0]{\href }%
\providecommand \doibase [0]{http://dx.doi.org/}%
\providecommand \selectlanguage [0]{\@gobble}%
\providecommand \bibinfo  [0]{\@secondoftwo}%
\providecommand \bibfield  [0]{\@secondoftwo}%
\providecommand \translation [1]{[#1]}%
\providecommand \BibitemOpen [0]{}%
\providecommand \bibitemStop [0]{}%
\providecommand \bibitemNoStop [0]{.\EOS\space}%
\providecommand \EOS [0]{\spacefactor3000\relax}%
\providecommand \BibitemShut  [1]{\csname bibitem#1\endcsname}%
\let\auto@bib@innerbib\@empty
\bibitem [{\citenamefont {Kobayashi}\ and\ \citenamefont
  {Maskawa}(1973)}]{Kobayashi:1973fv}%
  \BibitemOpen
  \bibfield  {author} {\bibinfo {author} {\bibfnamefont {M.}~\bibnamefont
  {Kobayashi}}\ and\ \bibinfo {author} {\bibfnamefont {T.}~\bibnamefont
  {Maskawa}},\ }\href {\doibase 10.1143/PTP.49.652} {\bibfield  {journal}
  {\bibinfo  {journal} {Prog. Theor. Phys.}\ }\textbf {\bibinfo {volume}
  {49}},\ \bibinfo {pages} {652} (\bibinfo {year} {1973})}\BibitemShut
  {NoStop}%
\bibitem [{\citenamefont {Workman}\ and\ \citenamefont
  {Others}(2022)}]{Workman:2022ynf}%
  \BibitemOpen
  \bibfield  {author} {\bibinfo {author} {\bibfnamefont {R.~L.}\ \bibnamefont
  {Workman}}\ and\ \bibinfo {author} {\bibnamefont {Others}} (\bibinfo
  {collaboration} {Particle Data Group}),\ }\href {\doibase
  10.1093/ptep/ptac097} {\bibfield  {journal} {\bibinfo  {journal} {PTEP}\
  }\textbf {\bibinfo {volume} {2022}},\ \bibinfo {pages} {083C01} (\bibinfo
  {year} {2022})}\BibitemShut {NoStop}%
\bibitem [{\citenamefont {Artuso}\ \emph {et~al.}(2016)\citenamefont {Artuso},
  \citenamefont {Borissov},\ and\ \citenamefont {Lenz}}]{Artuso:2015swg}%
  \BibitemOpen
  \bibfield  {author} {\bibinfo {author} {\bibfnamefont {M.}~\bibnamefont
  {Artuso}}, \bibinfo {author} {\bibfnamefont {G.}~\bibnamefont {Borissov}}, \
  and\ \bibinfo {author} {\bibfnamefont {A.}~\bibnamefont {Lenz}},\ }\href
  {\doibase 10.1103/RevModPhys.88.045002} {\bibfield  {journal} {\bibinfo
  {journal} {Rev. Mod. Phys.}\ }\textbf {\bibinfo {volume} {88}},\ \bibinfo
  {pages} {045002} (\bibinfo {year} {2016})},\ \bibinfo {note} {[Addendum:
  Rev.Mod.Phys. 91, 049901 (2019)]},\ \Eprint {http://arxiv.org/abs/1511.09466}
  {arXiv:1511.09466 [hep-ph]} \BibitemShut {NoStop}%
\bibitem [{\citenamefont {Albrecht}\ \emph {et~al.}(2024)\citenamefont
  {Albrecht}, \citenamefont {Bernlochner}, \citenamefont {Lenz},\ and\
  \citenamefont {Rusov}}]{Albrecht:2024oyn}%
  \BibitemOpen
  \bibfield  {author} {\bibinfo {author} {\bibfnamefont {J.}~\bibnamefont
  {Albrecht}}, \bibinfo {author} {\bibfnamefont {F.}~\bibnamefont
  {Bernlochner}}, \bibinfo {author} {\bibfnamefont {A.}~\bibnamefont {Lenz}}, \
  and\ \bibinfo {author} {\bibfnamefont {A.}~\bibnamefont {Rusov}},\
  }\href@noop {} {\  (\bibinfo {year} {2024})},\ \Eprint
  {http://arxiv.org/abs/2402.04224} {arXiv:2402.04224 [hep-ph]} \BibitemShut
  {NoStop}%
\bibitem [{\citenamefont {Elor}\ \emph {et~al.}(2019)\citenamefont {Elor},
  \citenamefont {Escudero},\ and\ \citenamefont {Nelson}}]{Elor:2018twp}%
  \BibitemOpen
  \bibfield  {author} {\bibinfo {author} {\bibfnamefont {G.}~\bibnamefont
  {Elor}}, \bibinfo {author} {\bibfnamefont {M.}~\bibnamefont {Escudero}}, \
  and\ \bibinfo {author} {\bibfnamefont {A.}~\bibnamefont {Nelson}},\ }\href
  {\doibase 10.1103/PhysRevD.99.035031} {\bibfield  {journal} {\bibinfo
  {journal} {Phys. Rev. D}\ }\textbf {\bibinfo {volume} {99}},\ \bibinfo
  {pages} {035031} (\bibinfo {year} {2019})},\ \Eprint
  {http://arxiv.org/abs/1810.00880} {arXiv:1810.00880 [hep-ph]} \BibitemShut
  {NoStop}%
\bibitem [{\citenamefont {Nelson}\ and\ \citenamefont
  {Xiao}(2019)}]{Nelson:2019fln}%
  \BibitemOpen
  \bibfield  {author} {\bibinfo {author} {\bibfnamefont {A.~E.}\ \bibnamefont
  {Nelson}}\ and\ \bibinfo {author} {\bibfnamefont {H.}~\bibnamefont {Xiao}},\
  }\href {\doibase 10.1103/PhysRevD.100.075002} {\bibfield  {journal} {\bibinfo
   {journal} {Phys. Rev. D}\ }\textbf {\bibinfo {volume} {100}},\ \bibinfo
  {pages} {075002} (\bibinfo {year} {2019})},\ \Eprint
  {http://arxiv.org/abs/1901.08141} {arXiv:1901.08141 [hep-ph]} \BibitemShut
  {NoStop}%
\bibitem [{\citenamefont {Alonso-\'Alvarez}\ \emph {et~al.}(2021)\citenamefont
  {Alonso-\'Alvarez}, \citenamefont {Elor},\ and\ \citenamefont
  {Escudero}}]{Alonso-Alvarez:2021qfd}%
  \BibitemOpen
  \bibfield  {author} {\bibinfo {author} {\bibfnamefont {G.}~\bibnamefont
  {Alonso-\'Alvarez}}, \bibinfo {author} {\bibfnamefont {G.}~\bibnamefont
  {Elor}}, \ and\ \bibinfo {author} {\bibfnamefont {M.}~\bibnamefont
  {Escudero}},\ }\href {\doibase 10.1103/PhysRevD.104.035028} {\bibfield
  {journal} {\bibinfo  {journal} {Phys. Rev. D}\ }\textbf {\bibinfo {volume}
  {104}},\ \bibinfo {pages} {035028} (\bibinfo {year} {2021})},\ \Eprint
  {http://arxiv.org/abs/2101.02706} {arXiv:2101.02706 [hep-ph]} \BibitemShut
  {NoStop}%
\bibitem [{\citenamefont {Ghalsasi}\ \emph {et~al.}(2015)\citenamefont
  {Ghalsasi}, \citenamefont {McKeen},\ and\ \citenamefont
  {Nelson}}]{Ghalsasi:2015mxa}%
  \BibitemOpen
  \bibfield  {author} {\bibinfo {author} {\bibfnamefont {A.}~\bibnamefont
  {Ghalsasi}}, \bibinfo {author} {\bibfnamefont {D.}~\bibnamefont {McKeen}}, \
  and\ \bibinfo {author} {\bibfnamefont {A.~E.}\ \bibnamefont {Nelson}},\
  }\href {\doibase 10.1103/PhysRevD.92.076014} {\bibfield  {journal} {\bibinfo
  {journal} {Phys. Rev. D}\ }\textbf {\bibinfo {volume} {92}},\ \bibinfo
  {pages} {076014} (\bibinfo {year} {2015})},\ \Eprint
  {http://arxiv.org/abs/1508.05392} {arXiv:1508.05392 [hep-ph]} \BibitemShut
  {NoStop}%
\bibitem [{\citenamefont {McKeen}\ and\ \citenamefont
  {Nelson}(2016)}]{McKeen:2015cuz}%
  \BibitemOpen
  \bibfield  {author} {\bibinfo {author} {\bibfnamefont {D.}~\bibnamefont
  {McKeen}}\ and\ \bibinfo {author} {\bibfnamefont {A.~E.}\ \bibnamefont
  {Nelson}},\ }\href {\doibase 10.1103/PhysRevD.94.076002} {\bibfield
  {journal} {\bibinfo  {journal} {Phys. Rev. D}\ }\textbf {\bibinfo {volume}
  {94}},\ \bibinfo {pages} {076002} (\bibinfo {year} {2016})},\ \Eprint
  {http://arxiv.org/abs/1512.05359} {arXiv:1512.05359 [hep-ph]} \BibitemShut
  {NoStop}%
\bibitem [{\citenamefont {Aitken}\ \emph {et~al.}(2017)\citenamefont {Aitken},
  \citenamefont {McKeen}, \citenamefont {Neder},\ and\ \citenamefont
  {Nelson}}]{Aitken:2017wie}%
  \BibitemOpen
  \bibfield  {author} {\bibinfo {author} {\bibfnamefont {K.}~\bibnamefont
  {Aitken}}, \bibinfo {author} {\bibfnamefont {D.}~\bibnamefont {McKeen}},
  \bibinfo {author} {\bibfnamefont {T.}~\bibnamefont {Neder}}, \ and\ \bibinfo
  {author} {\bibfnamefont {A.~E.}\ \bibnamefont {Nelson}},\ }\href {\doibase
  10.1103/PhysRevD.96.075009} {\bibfield  {journal} {\bibinfo  {journal} {Phys.
  Rev. D}\ }\textbf {\bibinfo {volume} {96}},\ \bibinfo {pages} {075009}
  (\bibinfo {year} {2017})},\ \Eprint {http://arxiv.org/abs/1708.01259}
  {arXiv:1708.01259 [hep-ph]} \BibitemShut {NoStop}%
\bibitem [{\citenamefont {Alonso-\'Alvarez}\ \emph {et~al.}(2020)\citenamefont
  {Alonso-\'Alvarez}, \citenamefont {Elor}, \citenamefont {Nelson},\ and\
  \citenamefont {Xiao}}]{Alonso-Alvarez:2019fym}%
  \BibitemOpen
  \bibfield  {author} {\bibinfo {author} {\bibfnamefont {G.}~\bibnamefont
  {Alonso-\'Alvarez}}, \bibinfo {author} {\bibfnamefont {G.}~\bibnamefont
  {Elor}}, \bibinfo {author} {\bibfnamefont {A.~E.}\ \bibnamefont {Nelson}}, \
  and\ \bibinfo {author} {\bibfnamefont {H.}~\bibnamefont {Xiao}},\ }\href
  {\doibase 10.1007/JHEP03(2020)046} {\bibfield  {journal} {\bibinfo  {journal}
  {JHEP}\ }\textbf {\bibinfo {volume} {03}},\ \bibinfo {pages} {046} (\bibinfo
  {year} {2020})},\ \Eprint {http://arxiv.org/abs/1907.10612} {arXiv:1907.10612
  [hep-ph]} \BibitemShut {NoStop}%
\bibitem [{\citenamefont {Elor}\ and\ \citenamefont
  {McGehee}(2021)}]{Elor:2020tkc}%
  \BibitemOpen
  \bibfield  {author} {\bibinfo {author} {\bibfnamefont {G.}~\bibnamefont
  {Elor}}\ and\ \bibinfo {author} {\bibfnamefont {R.}~\bibnamefont {McGehee}},\
  }\href {\doibase 10.1103/PhysRevD.103.035005} {\bibfield  {journal} {\bibinfo
   {journal} {Phys. Rev. D}\ }\textbf {\bibinfo {volume} {103}},\ \bibinfo
  {pages} {035005} (\bibinfo {year} {2021})},\ \Eprint
  {http://arxiv.org/abs/2011.06115} {arXiv:2011.06115 [hep-ph]} \BibitemShut
  {NoStop}%
\bibitem [{\citenamefont {Elahi}\ \emph {et~al.}(2022)\citenamefont {Elahi},
  \citenamefont {Elor},\ and\ \citenamefont {McGehee}}]{Elahi:2021jia}%
  \BibitemOpen
  \bibfield  {author} {\bibinfo {author} {\bibfnamefont {F.}~\bibnamefont
  {Elahi}}, \bibinfo {author} {\bibfnamefont {G.}~\bibnamefont {Elor}}, \ and\
  \bibinfo {author} {\bibfnamefont {R.}~\bibnamefont {McGehee}},\ }\href
  {\doibase 10.1103/PhysRevD.105.055024} {\bibfield  {journal} {\bibinfo
  {journal} {Phys. Rev. D}\ }\textbf {\bibinfo {volume} {105}},\ \bibinfo
  {pages} {055024} (\bibinfo {year} {2022})},\ \Eprint
  {http://arxiv.org/abs/2109.09751} {arXiv:2109.09751 [hep-ph]} \BibitemShut
  {NoStop}%
\bibitem [{\citenamefont {Elor}\ \emph {et~al.}(2024)\citenamefont {Elor},
  \citenamefont {Houtz}, \citenamefont {Ipek},\ and\ \citenamefont
  {Ulloa}}]{Elor:2024cea}%
  \BibitemOpen
  \bibfield  {author} {\bibinfo {author} {\bibfnamefont {G.}~\bibnamefont
  {Elor}}, \bibinfo {author} {\bibfnamefont {R.}~\bibnamefont {Houtz}},
  \bibinfo {author} {\bibfnamefont {S.}~\bibnamefont {Ipek}}, \ and\ \bibinfo
  {author} {\bibfnamefont {M.}~\bibnamefont {Ulloa}},\ }\href@noop {} {\
  (\bibinfo {year} {2024})},\ \Eprint {http://arxiv.org/abs/2408.12647}
  {arXiv:2408.12647 [hep-ph]} \BibitemShut {NoStop}%
\bibitem [{\citenamefont {Alonso-\'Alvarez}\ \emph {et~al.}(2024)\citenamefont
  {Alonso-\'Alvarez}, \citenamefont {Escudero}, \citenamefont {Grinstein},
  \citenamefont {Martin-Camalich},\ and\ \citenamefont
  {Murgui}}]{Alonso-Alvarez:2024inprep}%
  \BibitemOpen
  \bibfield  {author} {\bibinfo {author} {\bibfnamefont {G.}~\bibnamefont
  {Alonso-\'Alvarez}}, \bibinfo {author} {\bibfnamefont {M.}~\bibnamefont
  {Escudero}}, \bibinfo {author} {\bibfnamefont {B.}~\bibnamefont {Grinstein}},
  \bibinfo {author} {\bibfnamefont {J.}~\bibnamefont {Martin-Camalich}}, \ and\
  \bibinfo {author} {\bibfnamefont {C.}~\bibnamefont {Murgui}},\ }\href@noop {}
  {\bibfield  {journal} {\bibinfo  {journal} {In preparation}\ } (\bibinfo
  {year} {2024})}\BibitemShut {NoStop}%
\bibitem [{\citenamefont {Bauer}\ and\ \citenamefont
  {Dunn}(2011)}]{Bauer:2010dga}%
  \BibitemOpen
  \bibfield  {author} {\bibinfo {author} {\bibfnamefont {C.~W.}\ \bibnamefont
  {Bauer}}\ and\ \bibinfo {author} {\bibfnamefont {N.~D.}\ \bibnamefont
  {Dunn}},\ }\href {\doibase 10.1016/j.physletb.2010.12.039} {\bibfield
  {journal} {\bibinfo  {journal} {Phys. Lett. B}\ }\textbf {\bibinfo {volume}
  {696}},\ \bibinfo {pages} {362} (\bibinfo {year} {2011})},\ \Eprint
  {http://arxiv.org/abs/1006.1629} {arXiv:1006.1629 [hep-ph]} \BibitemShut
  {NoStop}%
\bibitem [{\citenamefont {Ligeti}\ \emph {et~al.}(2010)\citenamefont {Ligeti},
  \citenamefont {Papucci}, \citenamefont {Perez},\ and\ \citenamefont
  {Zupan}}]{Ligeti:2010ia}%
  \BibitemOpen
  \bibfield  {author} {\bibinfo {author} {\bibfnamefont {Z.}~\bibnamefont
  {Ligeti}}, \bibinfo {author} {\bibfnamefont {M.}~\bibnamefont {Papucci}},
  \bibinfo {author} {\bibfnamefont {G.}~\bibnamefont {Perez}}, \ and\ \bibinfo
  {author} {\bibfnamefont {J.}~\bibnamefont {Zupan}},\ }\href {\doibase
  10.1103/PhysRevLett.105.131601} {\bibfield  {journal} {\bibinfo  {journal}
  {Phys. Rev. Lett.}\ }\textbf {\bibinfo {volume} {105}},\ \bibinfo {pages}
  {131601} (\bibinfo {year} {2010})},\ \Eprint {http://arxiv.org/abs/1006.0432}
  {arXiv:1006.0432 [hep-ph]} \BibitemShut {NoStop}%
\bibitem [{\citenamefont {Lenz}(2011)}]{Lenz:2011zz}%
  \BibitemOpen
  \bibfield  {author} {\bibinfo {author} {\bibfnamefont {A.~J.}\ \bibnamefont
  {Lenz}},\ }\href {\doibase 10.1103/PhysRevD.84.031501} {\bibfield  {journal}
  {\bibinfo  {journal} {Phys. Rev. D}\ }\textbf {\bibinfo {volume} {84}},\
  \bibinfo {pages} {031501} (\bibinfo {year} {2011})},\ \Eprint
  {http://arxiv.org/abs/1106.3200} {arXiv:1106.3200 [hep-ph]} \BibitemShut
  {NoStop}%
\bibitem [{\citenamefont {Bobeth}\ and\ \citenamefont
  {Haisch}(2013)}]{Bobeth:2011st}%
  \BibitemOpen
  \bibfield  {author} {\bibinfo {author} {\bibfnamefont {C.}~\bibnamefont
  {Bobeth}}\ and\ \bibinfo {author} {\bibfnamefont {U.}~\bibnamefont
  {Haisch}},\ }\href {\doibase 10.5506/APhysPolB.44.127} {\bibfield  {journal}
  {\bibinfo  {journal} {Acta Phys. Polon. B}\ }\textbf {\bibinfo {volume}
  {44}},\ \bibinfo {pages} {127} (\bibinfo {year} {2013})},\ \Eprint
  {http://arxiv.org/abs/1109.1826} {arXiv:1109.1826 [hep-ph]} \BibitemShut
  {NoStop}%
\bibitem [{\citenamefont {Lenz}\ \emph {et~al.}(2012)\citenamefont {Lenz},
  \citenamefont {Nierste}, \citenamefont {Charles}, \citenamefont
  {Descotes-Genon}, \citenamefont {Lacker}, \citenamefont {Monteil},
  \citenamefont {Niess},\ and\ \citenamefont {T'Jampens}}]{Lenz:2012az}%
  \BibitemOpen
  \bibfield  {author} {\bibinfo {author} {\bibfnamefont {A.}~\bibnamefont
  {Lenz}}, \bibinfo {author} {\bibfnamefont {U.}~\bibnamefont {Nierste}},
  \bibinfo {author} {\bibfnamefont {J.}~\bibnamefont {Charles}}, \bibinfo
  {author} {\bibfnamefont {S.}~\bibnamefont {Descotes-Genon}}, \bibinfo
  {author} {\bibfnamefont {H.}~\bibnamefont {Lacker}}, \bibinfo {author}
  {\bibfnamefont {S.}~\bibnamefont {Monteil}}, \bibinfo {author} {\bibfnamefont
  {V.}~\bibnamefont {Niess}}, \ and\ \bibinfo {author} {\bibfnamefont
  {S.}~\bibnamefont {T'Jampens}},\ }\href {\doibase 10.1103/PhysRevD.86.033008}
  {\bibfield  {journal} {\bibinfo  {journal} {Phys. Rev. D}\ }\textbf {\bibinfo
  {volume} {86}},\ \bibinfo {pages} {033008} (\bibinfo {year} {2012})},\
  \Eprint {http://arxiv.org/abs/1203.0238} {arXiv:1203.0238 [hep-ph]}
  \BibitemShut {NoStop}%
\bibitem [{\citenamefont {Bevan}\ \emph {et~al.}(2013)\citenamefont {Bevan}
  \emph {et~al.}}]{Bevan:2013kaa}%
  \BibitemOpen
  \bibfield  {author} {\bibinfo {author} {\bibfnamefont {A.}~\bibnamefont
  {Bevan}} \emph {et~al.},\ }\href {\doibase 10.1016/j.nuclphysbps.2013.06.015}
  {\bibfield  {journal} {\bibinfo  {journal} {Nucl. Phys. B Proc. Suppl.}\
  }\textbf {\bibinfo {volume} {241-242}},\ \bibinfo {pages} {89} (\bibinfo
  {year} {2013})}\BibitemShut {NoStop}%
\bibitem [{\citenamefont {Charles}\ \emph {et~al.}(2014)\citenamefont
  {Charles}, \citenamefont {Descotes-Genon}, \citenamefont {Ligeti},
  \citenamefont {Monteil}, \citenamefont {Papucci},\ and\ \citenamefont
  {Trabelsi}}]{Charles:2013aka}%
  \BibitemOpen
  \bibfield  {author} {\bibinfo {author} {\bibfnamefont {J.}~\bibnamefont
  {Charles}}, \bibinfo {author} {\bibfnamefont {S.}~\bibnamefont
  {Descotes-Genon}}, \bibinfo {author} {\bibfnamefont {Z.}~\bibnamefont
  {Ligeti}}, \bibinfo {author} {\bibfnamefont {S.}~\bibnamefont {Monteil}},
  \bibinfo {author} {\bibfnamefont {M.}~\bibnamefont {Papucci}}, \ and\
  \bibinfo {author} {\bibfnamefont {K.}~\bibnamefont {Trabelsi}},\ }\href
  {\doibase 10.1103/PhysRevD.89.033016} {\bibfield  {journal} {\bibinfo
  {journal} {Phys. Rev. D}\ }\textbf {\bibinfo {volume} {89}},\ \bibinfo
  {pages} {033016} (\bibinfo {year} {2014})},\ \Eprint
  {http://arxiv.org/abs/1309.2293} {arXiv:1309.2293 [hep-ph]} \BibitemShut
  {NoStop}%
\bibitem [{\citenamefont {Charles}\ \emph {et~al.}(2020)\citenamefont
  {Charles}, \citenamefont {Descotes-Genon}, \citenamefont {Ligeti},
  \citenamefont {Monteil}, \citenamefont {Papucci}, \citenamefont {Trabelsi},\
  and\ \citenamefont {Vale~Silva}}]{Charles:2020dfl}%
  \BibitemOpen
  \bibfield  {author} {\bibinfo {author} {\bibfnamefont {J.}~\bibnamefont
  {Charles}}, \bibinfo {author} {\bibfnamefont {S.}~\bibnamefont
  {Descotes-Genon}}, \bibinfo {author} {\bibfnamefont {Z.}~\bibnamefont
  {Ligeti}}, \bibinfo {author} {\bibfnamefont {S.}~\bibnamefont {Monteil}},
  \bibinfo {author} {\bibfnamefont {M.}~\bibnamefont {Papucci}}, \bibinfo
  {author} {\bibfnamefont {K.}~\bibnamefont {Trabelsi}}, \ and\ \bibinfo
  {author} {\bibfnamefont {L.}~\bibnamefont {Vale~Silva}},\ }\href {\doibase
  10.1103/PhysRevD.102.056023} {\bibfield  {journal} {\bibinfo  {journal}
  {Phys. Rev. D}\ }\textbf {\bibinfo {volume} {102}},\ \bibinfo {pages}
  {056023} (\bibinfo {year} {2020})},\ \Eprint
  {http://arxiv.org/abs/2006.04824} {arXiv:2006.04824 [hep-ph]} \BibitemShut
  {NoStop}%
\bibitem [{\citenamefont {De~Bruyn}\ \emph {et~al.}(2023)\citenamefont
  {De~Bruyn}, \citenamefont {Fleischer}, \citenamefont {Malami},\ and\
  \citenamefont {van Vliet}}]{DeBruyn:2022zhw}%
  \BibitemOpen
  \bibfield  {author} {\bibinfo {author} {\bibfnamefont {K.}~\bibnamefont
  {De~Bruyn}}, \bibinfo {author} {\bibfnamefont {R.}~\bibnamefont {Fleischer}},
  \bibinfo {author} {\bibfnamefont {E.}~\bibnamefont {Malami}}, \ and\ \bibinfo
  {author} {\bibfnamefont {P.}~\bibnamefont {van Vliet}},\ }\href {\doibase
  10.1088/1361-6471/acab1d} {\bibfield  {journal} {\bibinfo  {journal} {J.
  Phys. G}\ }\textbf {\bibinfo {volume} {50}},\ \bibinfo {pages} {045003}
  (\bibinfo {year} {2023})},\ \Eprint {http://arxiv.org/abs/2208.14910}
  {arXiv:2208.14910 [hep-ph]} \BibitemShut {NoStop}%
\bibitem [{\citenamefont {Dor\v{s}ner}\ \emph {et~al.}(2016)\citenamefont
  {Dor\v{s}ner}, \citenamefont {Fajfer}, \citenamefont {Greljo}, \citenamefont
  {Kamenik},\ and\ \citenamefont {Ko\v{s}nik}}]{Dorsner:2016wpm}%
  \BibitemOpen
  \bibfield  {author} {\bibinfo {author} {\bibfnamefont {I.}~\bibnamefont
  {Dor\v{s}ner}}, \bibinfo {author} {\bibfnamefont {S.}~\bibnamefont {Fajfer}},
  \bibinfo {author} {\bibfnamefont {A.}~\bibnamefont {Greljo}}, \bibinfo
  {author} {\bibfnamefont {J.~F.}\ \bibnamefont {Kamenik}}, \ and\ \bibinfo
  {author} {\bibfnamefont {N.}~\bibnamefont {Ko\v{s}nik}},\ }\href {\doibase
  10.1016/j.physrep.2016.06.001} {\bibfield  {journal} {\bibinfo  {journal}
  {Phys. Rept.}\ }\textbf {\bibinfo {volume} {641}},\ \bibinfo {pages} {1}
  (\bibinfo {year} {2016})},\ \Eprint {http://arxiv.org/abs/1603.04993}
  {arXiv:1603.04993 [hep-ph]} \BibitemShut {NoStop}%
\bibitem [{\citenamefont {Buttazzo}\ \emph {et~al.}(2017)\citenamefont
  {Buttazzo}, \citenamefont {Greljo}, \citenamefont {Isidori},\ and\
  \citenamefont {Marzocca}}]{Buttazzo:2017ixm}%
  \BibitemOpen
  \bibfield  {author} {\bibinfo {author} {\bibfnamefont {D.}~\bibnamefont
  {Buttazzo}}, \bibinfo {author} {\bibfnamefont {A.}~\bibnamefont {Greljo}},
  \bibinfo {author} {\bibfnamefont {G.}~\bibnamefont {Isidori}}, \ and\
  \bibinfo {author} {\bibfnamefont {D.}~\bibnamefont {Marzocca}},\ }\href
  {\doibase 10.1007/JHEP11(2017)044} {\bibfield  {journal} {\bibinfo  {journal}
  {JHEP}\ }\textbf {\bibinfo {volume} {11}},\ \bibinfo {pages} {044} (\bibinfo
  {year} {2017})},\ \Eprint {http://arxiv.org/abs/1706.07808} {arXiv:1706.07808
  [hep-ph]} \BibitemShut {NoStop}%
\bibitem [{\citenamefont {Di~Luzio}\ \emph {et~al.}(2019)\citenamefont
  {Di~Luzio}, \citenamefont {Kirk}, \citenamefont {Lenz},\ and\ \citenamefont
  {Rauh}}]{DiLuzio:2019jyq}%
  \BibitemOpen
  \bibfield  {author} {\bibinfo {author} {\bibfnamefont {L.}~\bibnamefont
  {Di~Luzio}}, \bibinfo {author} {\bibfnamefont {M.}~\bibnamefont {Kirk}},
  \bibinfo {author} {\bibfnamefont {A.}~\bibnamefont {Lenz}}, \ and\ \bibinfo
  {author} {\bibfnamefont {T.}~\bibnamefont {Rauh}},\ }\href {\doibase
  10.1007/JHEP12(2019)009} {\bibfield  {journal} {\bibinfo  {journal} {JHEP}\
  }\textbf {\bibinfo {volume} {12}},\ \bibinfo {pages} {009} (\bibinfo {year}
  {2019})},\ \Eprint {http://arxiv.org/abs/1909.11087} {arXiv:1909.11087
  [hep-ph]} \BibitemShut {NoStop}%
\bibitem [{\citenamefont {Allanach}\ and\ \citenamefont
  {Mullin}(2023)}]{Allanach:2023uxz}%
  \BibitemOpen
  \bibfield  {author} {\bibinfo {author} {\bibfnamefont {B.}~\bibnamefont
  {Allanach}}\ and\ \bibinfo {author} {\bibfnamefont {A.}~\bibnamefont
  {Mullin}},\ }\href {\doibase 10.1007/JHEP09(2023)173} {\bibfield  {journal}
  {\bibinfo  {journal} {JHEP}\ }\textbf {\bibinfo {volume} {09}},\ \bibinfo
  {pages} {173} (\bibinfo {year} {2023})},\ \Eprint
  {http://arxiv.org/abs/2306.08669} {arXiv:2306.08669 [hep-ph]} \BibitemShut
  {NoStop}%
\bibitem [{\citenamefont {Athron}\ \emph {et~al.}(2024)\citenamefont {Athron},
  \citenamefont {Martinez},\ and\ \citenamefont {Sierra}}]{Athron:2023hmz}%
  \BibitemOpen
  \bibfield  {author} {\bibinfo {author} {\bibfnamefont {P.}~\bibnamefont
  {Athron}}, \bibinfo {author} {\bibfnamefont {R.}~\bibnamefont {Martinez}}, \
  and\ \bibinfo {author} {\bibfnamefont {C.}~\bibnamefont {Sierra}},\ }\href
  {\doibase 10.1007/JHEP02(2024)121} {\bibfield  {journal} {\bibinfo  {journal}
  {JHEP}\ }\textbf {\bibinfo {volume} {02}},\ \bibinfo {pages} {121} (\bibinfo
  {year} {2024})},\ \Eprint {http://arxiv.org/abs/2308.13426} {arXiv:2308.13426
  [hep-ph]} \BibitemShut {NoStop}%
\bibitem [{\citenamefont {Dobrescu}\ \emph {et~al.}(2010)\citenamefont
  {Dobrescu}, \citenamefont {Fox},\ and\ \citenamefont
  {Martin}}]{Dobrescu:2010rh}%
  \BibitemOpen
  \bibfield  {author} {\bibinfo {author} {\bibfnamefont {B.~A.}\ \bibnamefont
  {Dobrescu}}, \bibinfo {author} {\bibfnamefont {P.~J.}\ \bibnamefont {Fox}}, \
  and\ \bibinfo {author} {\bibfnamefont {A.}~\bibnamefont {Martin}},\ }\href
  {\doibase 10.1103/PhysRevLett.105.041801} {\bibfield  {journal} {\bibinfo
  {journal} {Phys. Rev. Lett.}\ }\textbf {\bibinfo {volume} {105}},\ \bibinfo
  {pages} {041801} (\bibinfo {year} {2010})},\ \Eprint
  {http://arxiv.org/abs/1005.4238} {arXiv:1005.4238 [hep-ph]} \BibitemShut
  {NoStop}%
\bibitem [{\citenamefont {Trott}\ and\ \citenamefont
  {Wise}(2010)}]{Trott:2010iz}%
  \BibitemOpen
  \bibfield  {author} {\bibinfo {author} {\bibfnamefont {M.}~\bibnamefont
  {Trott}}\ and\ \bibinfo {author} {\bibfnamefont {M.~B.}\ \bibnamefont
  {Wise}},\ }\href {\doibase 10.1007/JHEP11(2010)157} {\bibfield  {journal}
  {\bibinfo  {journal} {JHEP}\ }\textbf {\bibinfo {volume} {11}},\ \bibinfo
  {pages} {157} (\bibinfo {year} {2010})},\ \Eprint
  {http://arxiv.org/abs/1009.2813} {arXiv:1009.2813 [hep-ph]} \BibitemShut
  {NoStop}%
\bibitem [{\citenamefont {Bai}\ and\ \citenamefont
  {Nelson}(2010)}]{Bai:2010kf}%
  \BibitemOpen
  \bibfield  {author} {\bibinfo {author} {\bibfnamefont {Y.}~\bibnamefont
  {Bai}}\ and\ \bibinfo {author} {\bibfnamefont {A.~E.}\ \bibnamefont
  {Nelson}},\ }\href {\doibase 10.1103/PhysRevD.82.114027} {\bibfield
  {journal} {\bibinfo  {journal} {Phys. Rev. D}\ }\textbf {\bibinfo {volume}
  {82}},\ \bibinfo {pages} {114027} (\bibinfo {year} {2010})},\ \Eprint
  {http://arxiv.org/abs/1007.0596} {arXiv:1007.0596 [hep-ph]} \BibitemShut
  {NoStop}%
\bibitem [{\citenamefont {Botella}\ \emph {et~al.}(2015)\citenamefont
  {Botella}, \citenamefont {Branco}, \citenamefont {Nebot},\ and\ \citenamefont
  {S\'anchez}}]{Botella:2014qya}%
  \BibitemOpen
  \bibfield  {author} {\bibinfo {author} {\bibfnamefont {F.~J.}\ \bibnamefont
  {Botella}}, \bibinfo {author} {\bibfnamefont {G.~C.}\ \bibnamefont {Branco}},
  \bibinfo {author} {\bibfnamefont {M.}~\bibnamefont {Nebot}}, \ and\ \bibinfo
  {author} {\bibfnamefont {A.}~\bibnamefont {S\'anchez}},\ }\href {\doibase
  10.1103/PhysRevD.91.035013} {\bibfield  {journal} {\bibinfo  {journal} {Phys.
  Rev. D}\ }\textbf {\bibinfo {volume} {91}},\ \bibinfo {pages} {035013}
  (\bibinfo {year} {2015})},\ \Eprint {http://arxiv.org/abs/1402.1181}
  {arXiv:1402.1181 [hep-ph]} \BibitemShut {NoStop}%
\bibitem [{\citenamefont {Iguro}\ and\ \citenamefont
  {Omura}(2018)}]{Iguro:2018qzf}%
  \BibitemOpen
  \bibfield  {author} {\bibinfo {author} {\bibfnamefont {S.}~\bibnamefont
  {Iguro}}\ and\ \bibinfo {author} {\bibfnamefont {Y.}~\bibnamefont {Omura}},\
  }\href {\doibase 10.1007/JHEP05(2018)173} {\bibfield  {journal} {\bibinfo
  {journal} {JHEP}\ }\textbf {\bibinfo {volume} {05}},\ \bibinfo {pages} {173}
  (\bibinfo {year} {2018})},\ \Eprint {http://arxiv.org/abs/1802.01732}
  {arXiv:1802.01732 [hep-ph]} \BibitemShut {NoStop}%
\bibitem [{\citenamefont {Crivellin}\ \emph {et~al.}(2019)\citenamefont
  {Crivellin}, \citenamefont {M\"uller},\ and\ \citenamefont
  {Wiegand}}]{Crivellin:2019dun}%
  \BibitemOpen
  \bibfield  {author} {\bibinfo {author} {\bibfnamefont {A.}~\bibnamefont
  {Crivellin}}, \bibinfo {author} {\bibfnamefont {D.}~\bibnamefont {M\"uller}},
  \ and\ \bibinfo {author} {\bibfnamefont {C.}~\bibnamefont {Wiegand}},\ }\href
  {\doibase 10.1007/JHEP06(2019)119} {\bibfield  {journal} {\bibinfo  {journal}
  {JHEP}\ }\textbf {\bibinfo {volume} {06}},\ \bibinfo {pages} {119} (\bibinfo
  {year} {2019})},\ \Eprint {http://arxiv.org/abs/1903.10440} {arXiv:1903.10440
  [hep-ph]} \BibitemShut {NoStop}%
\bibitem [{\citenamefont {Athron}\ \emph {et~al.}(2022)\citenamefont {Athron},
  \citenamefont {Balazs}, \citenamefont {Gonzalo}, \citenamefont {Jacob},
  \citenamefont {Mahmoudi},\ and\ \citenamefont {Sierra}}]{Athron:2021auq}%
  \BibitemOpen
  \bibfield  {author} {\bibinfo {author} {\bibfnamefont {P.}~\bibnamefont
  {Athron}}, \bibinfo {author} {\bibfnamefont {C.}~\bibnamefont {Balazs}},
  \bibinfo {author} {\bibfnamefont {T.~E.}\ \bibnamefont {Gonzalo}}, \bibinfo
  {author} {\bibfnamefont {D.}~\bibnamefont {Jacob}}, \bibinfo {author}
  {\bibfnamefont {F.}~\bibnamefont {Mahmoudi}}, \ and\ \bibinfo {author}
  {\bibfnamefont {C.}~\bibnamefont {Sierra}},\ }\href {\doibase
  10.1007/JHEP01(2022)037} {\bibfield  {journal} {\bibinfo  {journal} {JHEP}\
  }\textbf {\bibinfo {volume} {01}},\ \bibinfo {pages} {037} (\bibinfo {year}
  {2022})},\ \Eprint {http://arxiv.org/abs/2111.10464} {arXiv:2111.10464
  [hep-ph]} \BibitemShut {NoStop}%
\bibitem [{\citenamefont {Bonilla}\ \emph {et~al.}(2023)\citenamefont
  {Bonilla}, \citenamefont {de~Giorgi}, \citenamefont {Gavela}, \citenamefont
  {Merlo},\ and\ \citenamefont {Ramos}}]{Bonilla:2022qgm}%
  \BibitemOpen
  \bibfield  {author} {\bibinfo {author} {\bibfnamefont {J.}~\bibnamefont
  {Bonilla}}, \bibinfo {author} {\bibfnamefont {A.}~\bibnamefont {de~Giorgi}},
  \bibinfo {author} {\bibfnamefont {B.}~\bibnamefont {Gavela}}, \bibinfo
  {author} {\bibfnamefont {L.}~\bibnamefont {Merlo}}, \ and\ \bibinfo {author}
  {\bibfnamefont {M.}~\bibnamefont {Ramos}},\ }\href {\doibase
  10.1007/JHEP02(2023)138} {\bibfield  {journal} {\bibinfo  {journal} {JHEP}\
  }\textbf {\bibinfo {volume} {02}},\ \bibinfo {pages} {138} (\bibinfo {year}
  {2023})},\ \Eprint {http://arxiv.org/abs/2209.11247} {arXiv:2209.11247
  [hep-ph]} \BibitemShut {NoStop}%
\bibitem [{\citenamefont {Li}\ \emph {et~al.}(2024)\citenamefont {Li},
  \citenamefont {Qian}, \citenamefont {Schmidt},\ and\ \citenamefont
  {Yuan}}]{Li:2024thq}%
  \BibitemOpen
  \bibfield  {author} {\bibinfo {author} {\bibfnamefont {T.}~\bibnamefont
  {Li}}, \bibinfo {author} {\bibfnamefont {Z.}~\bibnamefont {Qian}}, \bibinfo
  {author} {\bibfnamefont {M.~A.}\ \bibnamefont {Schmidt}}, \ and\ \bibinfo
  {author} {\bibfnamefont {M.}~\bibnamefont {Yuan}},\ }\href@noop {} {\
  (\bibinfo {year} {2024})},\ \Eprint {http://arxiv.org/abs/2402.14232}
  {arXiv:2402.14232 [hep-ph]} \BibitemShut {NoStop}%
\bibitem [{\citenamefont {Zheng}\ \emph {et~al.}(2022)\citenamefont {Zheng},
  \citenamefont {Chen},\ and\ \citenamefont {Zhang}}]{Zheng:2022ssr}%
  \BibitemOpen
  \bibfield  {author} {\bibinfo {author} {\bibfnamefont {M.-D.}\ \bibnamefont
  {Zheng}}, \bibinfo {author} {\bibfnamefont {F.-Z.}\ \bibnamefont {Chen}}, \
  and\ \bibinfo {author} {\bibfnamefont {H.-H.}\ \bibnamefont {Zhang}},\ }\href
  {\doibase 10.1140/epjc/s10052-022-10822-y} {\bibfield  {journal} {\bibinfo
  {journal} {Eur. Phys. J. C}\ }\textbf {\bibinfo {volume} {82}},\ \bibinfo
  {pages} {895} (\bibinfo {year} {2022})},\ \Eprint
  {http://arxiv.org/abs/2207.07636} {arXiv:2207.07636 [hep-ph]} \BibitemShut
  {NoStop}%
\bibitem [{\citenamefont {Datta}\ \emph {et~al.}(2023)\citenamefont {Datta},
  \citenamefont {Hammad}, \citenamefont {Marfatia}, \citenamefont {Mukherjee},\
  and\ \citenamefont {Rashed}}]{Datta:2022zng}%
  \BibitemOpen
  \bibfield  {author} {\bibinfo {author} {\bibfnamefont {A.}~\bibnamefont
  {Datta}}, \bibinfo {author} {\bibfnamefont {A.}~\bibnamefont {Hammad}},
  \bibinfo {author} {\bibfnamefont {D.}~\bibnamefont {Marfatia}}, \bibinfo
  {author} {\bibfnamefont {L.}~\bibnamefont {Mukherjee}}, \ and\ \bibinfo
  {author} {\bibfnamefont {A.}~\bibnamefont {Rashed}},\ }\href {\doibase
  10.1007/JHEP03(2023)108} {\bibfield  {journal} {\bibinfo  {journal} {JHEP}\
  }\textbf {\bibinfo {volume} {03}},\ \bibinfo {pages} {108} (\bibinfo {year}
  {2023})},\ \Eprint {http://arxiv.org/abs/2210.15662} {arXiv:2210.15662
  [hep-ph]} \BibitemShut {NoStop}%
\bibitem [{\citenamefont {Davighi}\ \emph {et~al.}(2023)\citenamefont
  {Davighi}, \citenamefont {Gosnay}, \citenamefont {Miller},\ and\
  \citenamefont {Renner}}]{Davighi:2023xqn}%
  \BibitemOpen
  \bibfield  {author} {\bibinfo {author} {\bibfnamefont {J.}~\bibnamefont
  {Davighi}}, \bibinfo {author} {\bibfnamefont {A.}~\bibnamefont {Gosnay}},
  \bibinfo {author} {\bibfnamefont {D.~J.}\ \bibnamefont {Miller}}, \ and\
  \bibinfo {author} {\bibfnamefont {S.}~\bibnamefont {Renner}},\ }\href@noop {}
  {\  (\bibinfo {year} {2023})},\ \Eprint {http://arxiv.org/abs/2312.13346}
  {arXiv:2312.13346 [hep-ph]} \BibitemShut {NoStop}%
\bibitem [{\citenamefont {Abazov}\ \emph {et~al.}(2010)\citenamefont {Abazov}
  \emph {et~al.}}]{D0:2010sht}%
  \BibitemOpen
  \bibfield  {author} {\bibinfo {author} {\bibfnamefont {V.~M.}\ \bibnamefont
  {Abazov}} \emph {et~al.} (\bibinfo {collaboration} {D0}),\ }\href {\doibase
  10.1103/PhysRevD.82.032001} {\bibfield  {journal} {\bibinfo  {journal} {Phys.
  Rev. D}\ }\textbf {\bibinfo {volume} {82}},\ \bibinfo {pages} {032001}
  (\bibinfo {year} {2010})},\ \Eprint {http://arxiv.org/abs/1005.2757}
  {arXiv:1005.2757 [hep-ex]} \BibitemShut {NoStop}%
\bibitem [{\citenamefont {Abazov}\ \emph {et~al.}(2011)\citenamefont {Abazov}
  \emph {et~al.}}]{D0:2011hom}%
  \BibitemOpen
  \bibfield  {author} {\bibinfo {author} {\bibfnamefont {V.~M.}\ \bibnamefont
  {Abazov}} \emph {et~al.} (\bibinfo {collaboration} {D0}),\ }\href {\doibase
  10.1103/PhysRevD.84.052007} {\bibfield  {journal} {\bibinfo  {journal} {Phys.
  Rev. D}\ }\textbf {\bibinfo {volume} {84}},\ \bibinfo {pages} {052007}
  (\bibinfo {year} {2011})},\ \Eprint {http://arxiv.org/abs/1106.6308}
  {arXiv:1106.6308 [hep-ex]} \BibitemShut {NoStop}%
\bibitem [{\citenamefont {Abazov}\ \emph {et~al.}(2014)\citenamefont {Abazov}
  \emph {et~al.}}]{D0:2013ohp}%
  \BibitemOpen
  \bibfield  {author} {\bibinfo {author} {\bibfnamefont {V.~M.}\ \bibnamefont
  {Abazov}} \emph {et~al.} (\bibinfo {collaboration} {D0}),\ }\href {\doibase
  10.1103/PhysRevD.89.012002} {\bibfield  {journal} {\bibinfo  {journal} {Phys.
  Rev. D}\ }\textbf {\bibinfo {volume} {89}},\ \bibinfo {pages} {012002}
  (\bibinfo {year} {2014})},\ \Eprint {http://arxiv.org/abs/1310.0447}
  {arXiv:1310.0447 [hep-ex]} \BibitemShut {NoStop}%
\bibitem [{\citenamefont {Amhis}\ \emph {et~al.}(2023)\citenamefont {Amhis}
  \emph {et~al.}}]{HFLAV:2022esi}%
  \BibitemOpen
  \bibfield  {author} {\bibinfo {author} {\bibfnamefont {Y.~S.}\ \bibnamefont
  {Amhis}} \emph {et~al.} (\bibinfo {collaboration} {HFLAV}),\ }\href {\doibase
  10.1103/PhysRevD.107.052008} {\bibfield  {journal} {\bibinfo  {journal}
  {Phys. Rev. D}\ }\textbf {\bibinfo {volume} {107}},\ \bibinfo {pages}
  {052008} (\bibinfo {year} {2023})},\ \Eprint
  {http://arxiv.org/abs/2206.07501} {arXiv:2206.07501 [hep-ex]} \BibitemShut
  {NoStop}%
\bibitem [{\citenamefont {Weisskopf}\ and\ \citenamefont
  {Wigner}(1930{\natexlab{a}})}]{Weisskopf:1930ps}%
  \BibitemOpen
  \bibfield  {author} {\bibinfo {author} {\bibfnamefont {V.}~\bibnamefont
  {Weisskopf}}\ and\ \bibinfo {author} {\bibfnamefont {E.}~\bibnamefont
  {Wigner}},\ }\href {\doibase 10.1007/BF01397406} {\bibfield  {journal}
  {\bibinfo  {journal} {Z. Phys.}\ }\textbf {\bibinfo {volume} {65}},\ \bibinfo
  {pages} {18} (\bibinfo {year} {1930}{\natexlab{a}})}\BibitemShut {NoStop}%
\bibitem [{\citenamefont {Weisskopf}\ and\ \citenamefont
  {Wigner}(1930{\natexlab{b}})}]{Weisskopf:1930au}%
  \BibitemOpen
  \bibfield  {author} {\bibinfo {author} {\bibfnamefont {V.}~\bibnamefont
  {Weisskopf}}\ and\ \bibinfo {author} {\bibfnamefont {E.~P.}\ \bibnamefont
  {Wigner}},\ }\href {\doibase 10.1007/BF01336768} {\bibfield  {journal}
  {\bibinfo  {journal} {Z. Phys.}\ }\textbf {\bibinfo {volume} {63}},\ \bibinfo
  {pages} {54} (\bibinfo {year} {1930}{\natexlab{b}})}\BibitemShut {NoStop}%
\bibitem [{\citenamefont {Aaij}\ \emph {et~al.}(2018)\citenamefont {Aaij} \emph
  {et~al.}}]{LHCb:2018roe}%
  \BibitemOpen
  \bibfield  {author} {\bibinfo {author} {\bibfnamefont {R.}~\bibnamefont
  {Aaij}} \emph {et~al.} (\bibinfo {collaboration} {LHCb}),\ }\href@noop {} {\
  (\bibinfo {year} {2018})},\ \Eprint {http://arxiv.org/abs/1808.08865}
  {arXiv:1808.08865 [hep-ex]} \BibitemShut {NoStop}%
\bibitem [{\citenamefont {Altmannshofer}\ \emph {et~al.}(2019)\citenamefont
  {Altmannshofer} \emph {et~al.}}]{Belle-II:2018jsg}%
  \BibitemOpen
  \bibfield  {author} {\bibinfo {author} {\bibfnamefont {W.}~\bibnamefont
  {Altmannshofer}} \emph {et~al.} (\bibinfo {collaboration} {Belle-II}),\
  }\href {\doibase 10.1093/ptep/ptz106} {\bibfield  {journal} {\bibinfo
  {journal} {PTEP}\ }\textbf {\bibinfo {volume} {2019}},\ \bibinfo {pages}
  {123C01} (\bibinfo {year} {2019})},\ \bibinfo {note} {[Erratum: PTEP 2020,
  029201 (2020)]},\ \Eprint {http://arxiv.org/abs/1808.10567} {arXiv:1808.10567
  [hep-ex]} \BibitemShut {NoStop}%
\bibitem [{\citenamefont {Aihara}\ \emph {et~al.}(2024)\citenamefont {Aihara}
  \emph {et~al.}}]{Aihara:2024zds}%
  \BibitemOpen
  \bibfield  {author} {\bibinfo {author} {\bibfnamefont {H.}~\bibnamefont
  {Aihara}} \emph {et~al.},\ }\href@noop {} {\  (\bibinfo {year} {2024})},\
  \Eprint {http://arxiv.org/abs/2406.19421} {arXiv:2406.19421 [hep-ex]}
  \BibitemShut {NoStop}%
\bibitem [{\citenamefont {Abada}\ \emph {et~al.}(2019)\citenamefont {Abada}
  \emph {et~al.}}]{FCC:2018evy}%
  \BibitemOpen
  \bibfield  {author} {\bibinfo {author} {\bibfnamefont {A.}~\bibnamefont
  {Abada}} \emph {et~al.} (\bibinfo {collaboration} {FCC}),\ }\href {\doibase
  10.1140/epjst/e2019-900045-4} {\bibfield  {journal} {\bibinfo  {journal}
  {Eur. Phys. J. ST}\ }\textbf {\bibinfo {volume} {228}},\ \bibinfo {pages}
  {261} (\bibinfo {year} {2019})}\BibitemShut {NoStop}%
\bibitem [{\citenamefont {Monteil}\ and\ \citenamefont
  {Wilkinson}(2021)}]{Monteil:2021ith}%
  \BibitemOpen
  \bibfield  {author} {\bibinfo {author} {\bibfnamefont {S.}~\bibnamefont
  {Monteil}}\ and\ \bibinfo {author} {\bibfnamefont {G.}~\bibnamefont
  {Wilkinson}},\ }\href {\doibase 10.1140/epjp/s13360-021-01814-0} {\bibfield
  {journal} {\bibinfo  {journal} {Eur. Phys. J. Plus}\ }\textbf {\bibinfo
  {volume} {136}},\ \bibinfo {pages} {837} (\bibinfo {year} {2021})},\ \Eprint
  {http://arxiv.org/abs/2106.01259} {arXiv:2106.01259 [hep-ex]} \BibitemShut
  {NoStop}%
\bibitem [{\citenamefont {Branco}\ \emph {et~al.}(1999)\citenamefont {Branco},
  \citenamefont {Lavoura},\ and\ \citenamefont {Silva}}]{Branco:1999fs}%
  \BibitemOpen
  \bibfield  {author} {\bibinfo {author} {\bibfnamefont {G.~C.}\ \bibnamefont
  {Branco}}, \bibinfo {author} {\bibfnamefont {L.}~\bibnamefont {Lavoura}}, \
  and\ \bibinfo {author} {\bibfnamefont {J.~P.}\ \bibnamefont {Silva}},\
  }\href@noop {} {\emph {\bibinfo {title} {{CP Violation}}}},\ Vol.\ \bibinfo
  {volume} {103}\ (\bibinfo {year} {1999})\BibitemShut {NoStop}%
\bibitem [{\citenamefont {Buras}(2020)}]{Buras:2020xsm}%
  \BibitemOpen
  \bibfield  {author} {\bibinfo {author} {\bibfnamefont {A.}~\bibnamefont
  {Buras}},\ }\href {\doibase 10.1017/9781139524100} {\emph {\bibinfo {title}
  {{Gauge Theory of Weak Decays}}}}\ (\bibinfo  {publisher} {Cambridge
  University Press},\ \bibinfo {year} {2020})\BibitemShut {NoStop}%
\bibitem [{\citenamefont {Inami}\ and\ \citenamefont
  {Lim}(1981)}]{Inami:1980fz}%
  \BibitemOpen
  \bibfield  {author} {\bibinfo {author} {\bibfnamefont {T.}~\bibnamefont
  {Inami}}\ and\ \bibinfo {author} {\bibfnamefont {C.~S.}\ \bibnamefont
  {Lim}},\ }\href {\doibase 10.1143/PTP.65.297} {\bibfield  {journal} {\bibinfo
   {journal} {Prog. Theor. Phys.}\ }\textbf {\bibinfo {volume} {65}},\ \bibinfo
  {pages} {297} (\bibinfo {year} {1981})},\ \bibinfo {note} {[Erratum:
  Prog.Theor.Phys. 65, 1772 (1981)]}\BibitemShut {NoStop}%
\bibitem [{\citenamefont {Bardeen}\ \emph {et~al.}(1978)\citenamefont
  {Bardeen}, \citenamefont {Buras}, \citenamefont {Duke},\ and\ \citenamefont
  {Muta}}]{Bardeen:1978yd}%
  \BibitemOpen
  \bibfield  {author} {\bibinfo {author} {\bibfnamefont {W.~A.}\ \bibnamefont
  {Bardeen}}, \bibinfo {author} {\bibfnamefont {A.~J.}\ \bibnamefont {Buras}},
  \bibinfo {author} {\bibfnamefont {D.~W.}\ \bibnamefont {Duke}}, \ and\
  \bibinfo {author} {\bibfnamefont {T.}~\bibnamefont {Muta}},\ }\href {\doibase
  10.1103/PhysRevD.18.3998} {\bibfield  {journal} {\bibinfo  {journal} {Phys.
  Rev. D}\ }\textbf {\bibinfo {volume} {18}},\ \bibinfo {pages} {3998}
  (\bibinfo {year} {1978})}\BibitemShut {NoStop}%
\bibitem [{\citenamefont {Buras}\ \emph {et~al.}(1990)\citenamefont {Buras},
  \citenamefont {Jamin},\ and\ \citenamefont {Weisz}}]{Buras:1990fn}%
  \BibitemOpen
  \bibfield  {author} {\bibinfo {author} {\bibfnamefont {A.~J.}\ \bibnamefont
  {Buras}}, \bibinfo {author} {\bibfnamefont {M.}~\bibnamefont {Jamin}}, \ and\
  \bibinfo {author} {\bibfnamefont {P.~H.}\ \bibnamefont {Weisz}},\ }\href
  {\doibase 10.1016/0550-3213(90)90373-L} {\bibfield  {journal} {\bibinfo
  {journal} {Nucl. Phys. B}\ }\textbf {\bibinfo {volume} {347}},\ \bibinfo
  {pages} {491} (\bibinfo {year} {1990})}\BibitemShut {NoStop}%
\bibitem [{\citenamefont {Aoki}\ \emph {et~al.}(2022)\citenamefont {Aoki} \emph
  {et~al.}}]{FlavourLatticeAveragingGroupFLAG:2021npn}%
  \BibitemOpen
  \bibfield  {author} {\bibinfo {author} {\bibfnamefont {Y.}~\bibnamefont
  {Aoki}} \emph {et~al.} (\bibinfo {collaboration} {Flavour Lattice Averaging
  Group (FLAG)}),\ }\href {\doibase 10.1140/epjc/s10052-022-10536-1} {\bibfield
   {journal} {\bibinfo  {journal} {Eur. Phys. J. C}\ }\textbf {\bibinfo
  {volume} {82}},\ \bibinfo {pages} {869} (\bibinfo {year} {2022})},\ \Eprint
  {http://arxiv.org/abs/2111.09849} {arXiv:2111.09849 [hep-lat]} \BibitemShut
  {NoStop}%
\bibitem [{\citenamefont {Bazavov}\ \emph {et~al.}(2016)\citenamefont {Bazavov}
  \emph {et~al.}}]{FermilabLattice:2016ipl}%
  \BibitemOpen
  \bibfield  {author} {\bibinfo {author} {\bibfnamefont {A.}~\bibnamefont
  {Bazavov}} \emph {et~al.} (\bibinfo {collaboration} {Fermilab Lattice,
  MILC}),\ }\href {\doibase 10.1103/PhysRevD.93.113016} {\bibfield  {journal}
  {\bibinfo  {journal} {Phys. Rev. D}\ }\textbf {\bibinfo {volume} {93}},\
  \bibinfo {pages} {113016} (\bibinfo {year} {2016})},\ \Eprint
  {http://arxiv.org/abs/1602.03560} {arXiv:1602.03560 [hep-lat]} \BibitemShut
  {NoStop}%
\bibitem [{\citenamefont {Dowdall}\ \emph {et~al.}(2019)\citenamefont
  {Dowdall}, \citenamefont {Davies}, \citenamefont {Horgan}, \citenamefont
  {Lepage}, \citenamefont {Monahan}, \citenamefont {Shigemitsu},\ and\
  \citenamefont {Wingate}}]{Dowdall:2019bea}%
  \BibitemOpen
  \bibfield  {author} {\bibinfo {author} {\bibfnamefont {R.~J.}\ \bibnamefont
  {Dowdall}}, \bibinfo {author} {\bibfnamefont {C.~T.~H.}\ \bibnamefont
  {Davies}}, \bibinfo {author} {\bibfnamefont {R.~R.}\ \bibnamefont {Horgan}},
  \bibinfo {author} {\bibfnamefont {G.~P.}\ \bibnamefont {Lepage}}, \bibinfo
  {author} {\bibfnamefont {C.~J.}\ \bibnamefont {Monahan}}, \bibinfo {author}
  {\bibfnamefont {J.}~\bibnamefont {Shigemitsu}}, \ and\ \bibinfo {author}
  {\bibfnamefont {M.}~\bibnamefont {Wingate}},\ }\href {\doibase
  10.1103/PhysRevD.100.094508} {\bibfield  {journal} {\bibinfo  {journal}
  {Phys. Rev. D}\ }\textbf {\bibinfo {volume} {100}},\ \bibinfo {pages}
  {094508} (\bibinfo {year} {2019})},\ \Eprint
  {http://arxiv.org/abs/1907.01025} {arXiv:1907.01025 [hep-lat]} \BibitemShut
  {NoStop}%
\bibitem [{\citenamefont {Grozin}\ \emph {et~al.}(2016)\citenamefont {Grozin},
  \citenamefont {Klein}, \citenamefont {Mannel},\ and\ \citenamefont
  {Pivovarov}}]{Grozin:2016uqy}%
  \BibitemOpen
  \bibfield  {author} {\bibinfo {author} {\bibfnamefont {A.~G.}\ \bibnamefont
  {Grozin}}, \bibinfo {author} {\bibfnamefont {R.}~\bibnamefont {Klein}},
  \bibinfo {author} {\bibfnamefont {T.}~\bibnamefont {Mannel}}, \ and\ \bibinfo
  {author} {\bibfnamefont {A.~A.}\ \bibnamefont {Pivovarov}},\ }\href {\doibase
  10.1103/PhysRevD.94.034024} {\bibfield  {journal} {\bibinfo  {journal} {Phys.
  Rev. D}\ }\textbf {\bibinfo {volume} {94}},\ \bibinfo {pages} {034024}
  (\bibinfo {year} {2016})},\ \Eprint {http://arxiv.org/abs/1606.06054}
  {arXiv:1606.06054 [hep-ph]} \BibitemShut {NoStop}%
\bibitem [{\citenamefont {Kirk}\ \emph {et~al.}(2017)\citenamefont {Kirk},
  \citenamefont {Lenz},\ and\ \citenamefont {Rauh}}]{Kirk:2017juj}%
  \BibitemOpen
  \bibfield  {author} {\bibinfo {author} {\bibfnamefont {M.}~\bibnamefont
  {Kirk}}, \bibinfo {author} {\bibfnamefont {A.}~\bibnamefont {Lenz}}, \ and\
  \bibinfo {author} {\bibfnamefont {T.}~\bibnamefont {Rauh}},\ }\href {\doibase
  10.1007/JHEP12(2017)068} {\bibfield  {journal} {\bibinfo  {journal} {JHEP}\
  }\textbf {\bibinfo {volume} {12}},\ \bibinfo {pages} {068} (\bibinfo {year}
  {2017})},\ \bibinfo {note} {[Erratum: JHEP 06, 162 (2020)]},\ \Eprint
  {http://arxiv.org/abs/1711.02100} {arXiv:1711.02100 [hep-ph]} \BibitemShut
  {NoStop}%
\bibitem [{\citenamefont {King}\ \emph {et~al.}(2019)\citenamefont {King},
  \citenamefont {Lenz},\ and\ \citenamefont {Rauh}}]{King:2019lal}%
  \BibitemOpen
  \bibfield  {author} {\bibinfo {author} {\bibfnamefont {D.}~\bibnamefont
  {King}}, \bibinfo {author} {\bibfnamefont {A.}~\bibnamefont {Lenz}}, \ and\
  \bibinfo {author} {\bibfnamefont {T.}~\bibnamefont {Rauh}},\ }\href {\doibase
  10.1007/JHEP05(2019)034} {\bibfield  {journal} {\bibinfo  {journal} {JHEP}\
  }\textbf {\bibinfo {volume} {05}},\ \bibinfo {pages} {034} (\bibinfo {year}
  {2019})},\ \Eprint {http://arxiv.org/abs/1904.00940} {arXiv:1904.00940
  [hep-ph]} \BibitemShut {NoStop}%
\bibitem [{\citenamefont {Tsang}\ and\ \citenamefont
  {Della~Morte}(2023)}]{Tsang:2023nay}%
  \BibitemOpen
  \bibfield  {author} {\bibinfo {author} {\bibfnamefont {J.~T.}\ \bibnamefont
  {Tsang}}\ and\ \bibinfo {author} {\bibfnamefont {M.}~\bibnamefont
  {Della~Morte}},\ }\href@noop {} {\  (\bibinfo {year} {2023})},\ \Eprint
  {http://arxiv.org/abs/2310.02705} {arXiv:2310.02705 [hep-lat]} \BibitemShut
  {NoStop}%
\bibitem [{\citenamefont {Carter}\ and\ \citenamefont
  {Sanda}(1981)}]{Carter:1980tk}%
  \BibitemOpen
  \bibfield  {author} {\bibinfo {author} {\bibfnamefont {A.~B.}\ \bibnamefont
  {Carter}}\ and\ \bibinfo {author} {\bibfnamefont {A.~I.}\ \bibnamefont
  {Sanda}},\ }\href {\doibase 10.1103/PhysRevD.23.1567} {\bibfield  {journal}
  {\bibinfo  {journal} {Phys. Rev. D}\ }\textbf {\bibinfo {volume} {23}},\
  \bibinfo {pages} {1567} (\bibinfo {year} {1981})}\BibitemShut {NoStop}%
\bibitem [{\citenamefont {Bigi}\ and\ \citenamefont
  {Sanda}(1981)}]{Bigi:1981qs}%
  \BibitemOpen
  \bibfield  {author} {\bibinfo {author} {\bibfnamefont {I.~I.~Y.}\
  \bibnamefont {Bigi}}\ and\ \bibinfo {author} {\bibfnamefont {A.~I.}\
  \bibnamefont {Sanda}},\ }\href {\doibase 10.1016/0550-3213(81)90519-8}
  {\bibfield  {journal} {\bibinfo  {journal} {Nucl. Phys. B}\ }\textbf
  {\bibinfo {volume} {193}},\ \bibinfo {pages} {85} (\bibinfo {year}
  {1981})}\BibitemShut {NoStop}%
\bibitem [{\citenamefont {Fleischer}(2024)}]{Fleischer:2024uru}%
  \BibitemOpen
  \bibfield  {author} {\bibinfo {author} {\bibfnamefont {R.}~\bibnamefont
  {Fleischer}}\ }(\bibinfo {year} {2024})\ \Eprint
  {http://arxiv.org/abs/2402.00710} {arXiv:2402.00710 [hep-ph]} \BibitemShut
  {NoStop}%
\bibitem [{\citenamefont {Barel}\ \emph {et~al.}(2021)\citenamefont {Barel},
  \citenamefont {De~Bruyn}, \citenamefont {Fleischer},\ and\ \citenamefont
  {Malami}}]{Barel:2020jvf}%
  \BibitemOpen
  \bibfield  {author} {\bibinfo {author} {\bibfnamefont {M.~Z.}\ \bibnamefont
  {Barel}}, \bibinfo {author} {\bibfnamefont {K.}~\bibnamefont {De~Bruyn}},
  \bibinfo {author} {\bibfnamefont {R.}~\bibnamefont {Fleischer}}, \ and\
  \bibinfo {author} {\bibfnamefont {E.}~\bibnamefont {Malami}},\ }\href
  {\doibase 10.1088/1361-6471/abf2a2} {\bibfield  {journal} {\bibinfo
  {journal} {J. Phys. G}\ }\textbf {\bibinfo {volume} {48}},\ \bibinfo {pages}
  {065002} (\bibinfo {year} {2021})},\ \Eprint
  {http://arxiv.org/abs/2010.14423} {arXiv:2010.14423 [hep-ph]} \BibitemShut
  {NoStop}%
\bibitem [{\citenamefont {Aaij}\ \emph
  {et~al.}(2015{\natexlab{a}})\citenamefont {Aaij} \emph
  {et~al.}}]{LHCb:2015esn}%
  \BibitemOpen
  \bibfield  {author} {\bibinfo {author} {\bibfnamefont {R.}~\bibnamefont
  {Aaij}} \emph {et~al.} (\bibinfo {collaboration} {LHCb}),\ }\href {\doibase
  10.1007/JHEP11(2015)082} {\bibfield  {journal} {\bibinfo  {journal} {JHEP}\
  }\textbf {\bibinfo {volume} {11}},\ \bibinfo {pages} {082} (\bibinfo {year}
  {2015}{\natexlab{a}})},\ \Eprint {http://arxiv.org/abs/1509.00400}
  {arXiv:1509.00400 [hep-ex]} \BibitemShut {NoStop}%
\bibitem [{\citenamefont {Frings}\ \emph {et~al.}(2015)\citenamefont {Frings},
  \citenamefont {Nierste},\ and\ \citenamefont {Wiebusch}}]{Frings:2015eva}%
  \BibitemOpen
  \bibfield  {author} {\bibinfo {author} {\bibfnamefont {P.}~\bibnamefont
  {Frings}}, \bibinfo {author} {\bibfnamefont {U.}~\bibnamefont {Nierste}}, \
  and\ \bibinfo {author} {\bibfnamefont {M.}~\bibnamefont {Wiebusch}},\ }\href
  {\doibase 10.1103/PhysRevLett.115.061802} {\bibfield  {journal} {\bibinfo
  {journal} {Phys. Rev. Lett.}\ }\textbf {\bibinfo {volume} {115}},\ \bibinfo
  {pages} {061802} (\bibinfo {year} {2015})},\ \Eprint
  {http://arxiv.org/abs/1503.00859} {arXiv:1503.00859 [hep-ph]} \BibitemShut
  {NoStop}%
\bibitem [{\citenamefont {De~Bruyn}\ and\ \citenamefont
  {Fleischer}(2015)}]{DeBruyn:2014oga}%
  \BibitemOpen
  \bibfield  {author} {\bibinfo {author} {\bibfnamefont {K.}~\bibnamefont
  {De~Bruyn}}\ and\ \bibinfo {author} {\bibfnamefont {R.}~\bibnamefont
  {Fleischer}},\ }\href {\doibase 10.1007/JHEP03(2015)145} {\bibfield
  {journal} {\bibinfo  {journal} {JHEP}\ }\textbf {\bibinfo {volume} {03}},\
  \bibinfo {pages} {145} (\bibinfo {year} {2015})},\ \Eprint
  {http://arxiv.org/abs/1412.6834} {arXiv:1412.6834 [hep-ph]} \BibitemShut
  {NoStop}%
\bibitem [{\citenamefont {Aaij}\ \emph
  {et~al.}(2015{\natexlab{b}})\citenamefont {Aaij} \emph
  {et~al.}}]{LHCb:2014xpr}%
  \BibitemOpen
  \bibfield  {author} {\bibinfo {author} {\bibfnamefont {R.}~\bibnamefont
  {Aaij}} \emph {et~al.} (\bibinfo {collaboration} {LHCb}),\ }\href {\doibase
  10.1016/j.physletb.2015.01.008} {\bibfield  {journal} {\bibinfo  {journal}
  {Phys. Lett. B}\ }\textbf {\bibinfo {volume} {742}},\ \bibinfo {pages} {38}
  (\bibinfo {year} {2015}{\natexlab{b}})},\ \Eprint
  {http://arxiv.org/abs/1411.1634} {arXiv:1411.1634 [hep-ex]} \BibitemShut
  {NoStop}%
\bibitem [{\citenamefont {Jung}(2012)}]{Jung:2012mp}%
  \BibitemOpen
  \bibfield  {author} {\bibinfo {author} {\bibfnamefont {M.}~\bibnamefont
  {Jung}},\ }\href {\doibase 10.1103/PhysRevD.86.053008} {\bibfield  {journal}
  {\bibinfo  {journal} {Phys. Rev. D}\ }\textbf {\bibinfo {volume} {86}},\
  \bibinfo {pages} {053008} (\bibinfo {year} {2012})},\ \Eprint
  {http://arxiv.org/abs/1206.2050} {arXiv:1206.2050 [hep-ph]} \BibitemShut
  {NoStop}%
\bibitem [{\citenamefont {Ciuchini}\ \emph {et~al.}(2011)\citenamefont
  {Ciuchini}, \citenamefont {Pierini},\ and\ \citenamefont
  {Silvestrini}}]{Ciuchini:2011kd}%
  \BibitemOpen
  \bibfield  {author} {\bibinfo {author} {\bibfnamefont {M.}~\bibnamefont
  {Ciuchini}}, \bibinfo {author} {\bibfnamefont {M.}~\bibnamefont {Pierini}}, \
  and\ \bibinfo {author} {\bibfnamefont {L.}~\bibnamefont {Silvestrini}},\ }in\
  \href@noop {} {\emph {\bibinfo {booktitle} {{6th International Workshop on
  the CKM Unitarity Triangle}}}}\ (\bibinfo {year} {2011})\ \Eprint
  {http://arxiv.org/abs/1102.0392} {arXiv:1102.0392 [hep-ph]} \BibitemShut
  {NoStop}%
\bibitem [{\citenamefont {Faller}\ \emph {et~al.}(2009)\citenamefont {Faller},
  \citenamefont {Jung}, \citenamefont {Fleischer},\ and\ \citenamefont
  {Mannel}}]{Faller:2008zc}%
  \BibitemOpen
  \bibfield  {author} {\bibinfo {author} {\bibfnamefont {S.}~\bibnamefont
  {Faller}}, \bibinfo {author} {\bibfnamefont {M.}~\bibnamefont {Jung}},
  \bibinfo {author} {\bibfnamefont {R.}~\bibnamefont {Fleischer}}, \ and\
  \bibinfo {author} {\bibfnamefont {T.}~\bibnamefont {Mannel}},\ }\href
  {\doibase 10.1103/PhysRevD.79.014030} {\bibfield  {journal} {\bibinfo
  {journal} {Phys. Rev. D}\ }\textbf {\bibinfo {volume} {79}},\ \bibinfo
  {pages} {014030} (\bibinfo {year} {2009})},\ \Eprint
  {http://arxiv.org/abs/0809.0842} {arXiv:0809.0842 [hep-ph]} \BibitemShut
  {NoStop}%
\bibitem [{\citenamefont {Ciuchini}\ \emph {et~al.}(2005)\citenamefont
  {Ciuchini}, \citenamefont {Pierini},\ and\ \citenamefont
  {Silvestrini}}]{Ciuchini:2005mg}%
  \BibitemOpen
  \bibfield  {author} {\bibinfo {author} {\bibfnamefont {M.}~\bibnamefont
  {Ciuchini}}, \bibinfo {author} {\bibfnamefont {M.}~\bibnamefont {Pierini}}, \
  and\ \bibinfo {author} {\bibfnamefont {L.}~\bibnamefont {Silvestrini}},\
  }\href {\doibase 10.1103/PhysRevLett.95.221804} {\bibfield  {journal}
  {\bibinfo  {journal} {Phys. Rev. Lett.}\ }\textbf {\bibinfo {volume} {95}},\
  \bibinfo {pages} {221804} (\bibinfo {year} {2005})},\ \Eprint
  {http://arxiv.org/abs/hep-ph/0507290} {arXiv:hep-ph/0507290} \BibitemShut
  {NoStop}%
\bibitem [{\citenamefont {Aaij}\ \emph {et~al.}(2024)\citenamefont {Aaij} \emph
  {et~al.}}]{LHCb:2023zcp}%
  \BibitemOpen
  \bibfield  {author} {\bibinfo {author} {\bibfnamefont {R.}~\bibnamefont
  {Aaij}} \emph {et~al.} (\bibinfo {collaboration} {LHCb}),\ }\href {\doibase
  10.1103/PhysRevLett.132.021801} {\bibfield  {journal} {\bibinfo  {journal}
  {Phys. Rev. Lett.}\ }\textbf {\bibinfo {volume} {132}},\ \bibinfo {pages}
  {021801} (\bibinfo {year} {2024})},\ \Eprint
  {http://arxiv.org/abs/2309.09728} {arXiv:2309.09728 [hep-ex]} \BibitemShut
  {NoStop}%
\bibitem [{\citenamefont {Bezshyiko}\ \emph {et~al.}(2024)\citenamefont
  {Bezshyiko} \emph {et~al.}}]{LHCb:2023sim}%
  \BibitemOpen
  \bibfield  {author} {\bibinfo {author} {\bibfnamefont {I.}~\bibnamefont
  {Bezshyiko}} \emph {et~al.} (\bibinfo {collaboration} {LHCb}),\ }\href
  {\doibase 10.1103/PhysRevLett.132.051802} {\bibfield  {journal} {\bibinfo
  {journal} {Phys. Rev. Lett.}\ }\textbf {\bibinfo {volume} {132}},\ \bibinfo
  {pages} {051802} (\bibinfo {year} {2024})},\ \Eprint
  {http://arxiv.org/abs/2308.01468} {arXiv:2308.01468 [hep-ex]} \BibitemShut
  {NoStop}%
\bibitem [{\citenamefont {Lenz}(2015)}]{Lenz:2014jha}%
  \BibitemOpen
  \bibfield  {author} {\bibinfo {author} {\bibfnamefont {A.}~\bibnamefont
  {Lenz}},\ }\href {\doibase 10.1142/S0217751X15430058} {\bibfield  {journal}
  {\bibinfo  {journal} {Int. J. Mod. Phys. A}\ }\textbf {\bibinfo {volume}
  {30}},\ \bibinfo {pages} {1543005} (\bibinfo {year} {2015})},\ \Eprint
  {http://arxiv.org/abs/1405.3601} {arXiv:1405.3601 [hep-ph]} \BibitemShut
  {NoStop}%
\bibitem [{\citenamefont {Beneke}\ \emph {et~al.}(1999)\citenamefont {Beneke},
  \citenamefont {Buchalla}, \citenamefont {Greub}, \citenamefont {Lenz},\ and\
  \citenamefont {Nierste}}]{Beneke:1998sy}%
  \BibitemOpen
  \bibfield  {author} {\bibinfo {author} {\bibfnamefont {M.}~\bibnamefont
  {Beneke}}, \bibinfo {author} {\bibfnamefont {G.}~\bibnamefont {Buchalla}},
  \bibinfo {author} {\bibfnamefont {C.}~\bibnamefont {Greub}}, \bibinfo
  {author} {\bibfnamefont {A.}~\bibnamefont {Lenz}}, \ and\ \bibinfo {author}
  {\bibfnamefont {U.}~\bibnamefont {Nierste}},\ }\href {\doibase
  10.1016/S0370-2693(99)00684-X} {\bibfield  {journal} {\bibinfo  {journal}
  {Phys. Lett. B}\ }\textbf {\bibinfo {volume} {459}},\ \bibinfo {pages} {631}
  (\bibinfo {year} {1999})},\ \Eprint {http://arxiv.org/abs/hep-ph/9808385}
  {arXiv:hep-ph/9808385} \BibitemShut {NoStop}%
\bibitem [{\citenamefont {Beneke}\ \emph {et~al.}(2003)\citenamefont {Beneke},
  \citenamefont {Buchalla}, \citenamefont {Lenz},\ and\ \citenamefont
  {Nierste}}]{Beneke:2003az}%
  \BibitemOpen
  \bibfield  {author} {\bibinfo {author} {\bibfnamefont {M.}~\bibnamefont
  {Beneke}}, \bibinfo {author} {\bibfnamefont {G.}~\bibnamefont {Buchalla}},
  \bibinfo {author} {\bibfnamefont {A.}~\bibnamefont {Lenz}}, \ and\ \bibinfo
  {author} {\bibfnamefont {U.}~\bibnamefont {Nierste}},\ }\href {\doibase
  10.1016/j.physletb.2003.09.089} {\bibfield  {journal} {\bibinfo  {journal}
  {Phys. Lett. B}\ }\textbf {\bibinfo {volume} {576}},\ \bibinfo {pages} {173}
  (\bibinfo {year} {2003})},\ \Eprint {http://arxiv.org/abs/hep-ph/0307344}
  {arXiv:hep-ph/0307344} \BibitemShut {NoStop}%
\bibitem [{\citenamefont {Ciuchini}\ \emph {et~al.}(2003)\citenamefont
  {Ciuchini}, \citenamefont {Franco}, \citenamefont {Lubicz}, \citenamefont
  {Mescia},\ and\ \citenamefont {Tarantino}}]{Ciuchini:2003ww}%
  \BibitemOpen
  \bibfield  {author} {\bibinfo {author} {\bibfnamefont {M.}~\bibnamefont
  {Ciuchini}}, \bibinfo {author} {\bibfnamefont {E.}~\bibnamefont {Franco}},
  \bibinfo {author} {\bibfnamefont {V.}~\bibnamefont {Lubicz}}, \bibinfo
  {author} {\bibfnamefont {F.}~\bibnamefont {Mescia}}, \ and\ \bibinfo {author}
  {\bibfnamefont {C.}~\bibnamefont {Tarantino}},\ }\href {\doibase
  10.1088/1126-6708/2003/08/031} {\bibfield  {journal} {\bibinfo  {journal}
  {JHEP}\ }\textbf {\bibinfo {volume} {08}},\ \bibinfo {pages} {031} (\bibinfo
  {year} {2003})},\ \Eprint {http://arxiv.org/abs/hep-ph/0308029}
  {arXiv:hep-ph/0308029} \BibitemShut {NoStop}%
\bibitem [{\citenamefont {Lenz}\ and\ \citenamefont
  {Nierste}(2007)}]{Lenz:2006hd}%
  \BibitemOpen
  \bibfield  {author} {\bibinfo {author} {\bibfnamefont {A.}~\bibnamefont
  {Lenz}}\ and\ \bibinfo {author} {\bibfnamefont {U.}~\bibnamefont {Nierste}},\
  }\href {\doibase 10.1088/1126-6708/2007/06/072} {\bibfield  {journal}
  {\bibinfo  {journal} {JHEP}\ }\textbf {\bibinfo {volume} {06}},\ \bibinfo
  {pages} {072} (\bibinfo {year} {2007})},\ \Eprint
  {http://arxiv.org/abs/hep-ph/0612167} {arXiv:hep-ph/0612167} \BibitemShut
  {NoStop}%
\bibitem [{\citenamefont {Asatrian}\ \emph {et~al.}(2017)\citenamefont
  {Asatrian}, \citenamefont {Hovhannisyan}, \citenamefont {Nierste},\ and\
  \citenamefont {Yeghiazaryan}}]{Asatrian:2017qaz}%
  \BibitemOpen
  \bibfield  {author} {\bibinfo {author} {\bibfnamefont {H.~M.}\ \bibnamefont
  {Asatrian}}, \bibinfo {author} {\bibfnamefont {A.}~\bibnamefont
  {Hovhannisyan}}, \bibinfo {author} {\bibfnamefont {U.}~\bibnamefont
  {Nierste}}, \ and\ \bibinfo {author} {\bibfnamefont {A.}~\bibnamefont
  {Yeghiazaryan}},\ }\href {\doibase 10.1007/JHEP10(2017)191} {\bibfield
  {journal} {\bibinfo  {journal} {JHEP}\ }\textbf {\bibinfo {volume} {10}},\
  \bibinfo {pages} {191} (\bibinfo {year} {2017})},\ \Eprint
  {http://arxiv.org/abs/1709.02160} {arXiv:1709.02160 [hep-ph]} \BibitemShut
  {NoStop}%
\bibitem [{\citenamefont {Asatrian}\ \emph {et~al.}(2020)\citenamefont
  {Asatrian}, \citenamefont {Asatryan}, \citenamefont {Hovhannisyan},
  \citenamefont {Nierste}, \citenamefont {Tumasyan},\ and\ \citenamefont
  {Yeghiazaryan}}]{Asatrian:2020zxa}%
  \BibitemOpen
  \bibfield  {author} {\bibinfo {author} {\bibfnamefont {H.~M.}\ \bibnamefont
  {Asatrian}}, \bibinfo {author} {\bibfnamefont {H.~H.}\ \bibnamefont
  {Asatryan}}, \bibinfo {author} {\bibfnamefont {A.}~\bibnamefont
  {Hovhannisyan}}, \bibinfo {author} {\bibfnamefont {U.}~\bibnamefont
  {Nierste}}, \bibinfo {author} {\bibfnamefont {S.}~\bibnamefont {Tumasyan}}, \
  and\ \bibinfo {author} {\bibfnamefont {A.}~\bibnamefont {Yeghiazaryan}},\
  }\href {\doibase 10.1103/PhysRevD.102.033007} {\bibfield  {journal} {\bibinfo
   {journal} {Phys. Rev. D}\ }\textbf {\bibinfo {volume} {102}},\ \bibinfo
  {pages} {033007} (\bibinfo {year} {2020})},\ \Eprint
  {http://arxiv.org/abs/2006.13227} {arXiv:2006.13227 [hep-ph]} \BibitemShut
  {NoStop}%
\bibitem [{\citenamefont {Gerlach}\ \emph {et~al.}(2021)\citenamefont
  {Gerlach}, \citenamefont {Nierste}, \citenamefont {Shtabovenko},\ and\
  \citenamefont {Steinhauser}}]{Gerlach:2021xtb}%
  \BibitemOpen
  \bibfield  {author} {\bibinfo {author} {\bibfnamefont {M.}~\bibnamefont
  {Gerlach}}, \bibinfo {author} {\bibfnamefont {U.}~\bibnamefont {Nierste}},
  \bibinfo {author} {\bibfnamefont {V.}~\bibnamefont {Shtabovenko}}, \ and\
  \bibinfo {author} {\bibfnamefont {M.}~\bibnamefont {Steinhauser}},\ }\href
  {\doibase 10.1007/JHEP07(2021)043} {\bibfield  {journal} {\bibinfo  {journal}
  {JHEP}\ }\textbf {\bibinfo {volume} {07}},\ \bibinfo {pages} {043} (\bibinfo
  {year} {2021})},\ \Eprint {http://arxiv.org/abs/2106.05979} {arXiv:2106.05979
  [hep-ph]} \BibitemShut {NoStop}%
\bibitem [{\citenamefont {Gerlach}\ \emph
  {et~al.}(2022{\natexlab{a}})\citenamefont {Gerlach}, \citenamefont {Nierste},
  \citenamefont {Shtabovenko},\ and\ \citenamefont
  {Steinhauser}}]{Gerlach:2022wgb}%
  \BibitemOpen
  \bibfield  {author} {\bibinfo {author} {\bibfnamefont {M.}~\bibnamefont
  {Gerlach}}, \bibinfo {author} {\bibfnamefont {U.}~\bibnamefont {Nierste}},
  \bibinfo {author} {\bibfnamefont {V.}~\bibnamefont {Shtabovenko}}, \ and\
  \bibinfo {author} {\bibfnamefont {M.}~\bibnamefont {Steinhauser}},\ }\href
  {\doibase 10.1007/JHEP04(2022)006} {\bibfield  {journal} {\bibinfo  {journal}
  {JHEP}\ }\textbf {\bibinfo {volume} {04}},\ \bibinfo {pages} {006} (\bibinfo
  {year} {2022}{\natexlab{a}})},\ \Eprint {http://arxiv.org/abs/2202.12305}
  {arXiv:2202.12305 [hep-ph]} \BibitemShut {NoStop}%
\bibitem [{\citenamefont {Gerlach}\ \emph
  {et~al.}(2022{\natexlab{b}})\citenamefont {Gerlach}, \citenamefont {Nierste},
  \citenamefont {Shtabovenko},\ and\ \citenamefont
  {Steinhauser}}]{Gerlach:2022hoj}%
  \BibitemOpen
  \bibfield  {author} {\bibinfo {author} {\bibfnamefont {M.}~\bibnamefont
  {Gerlach}}, \bibinfo {author} {\bibfnamefont {U.}~\bibnamefont {Nierste}},
  \bibinfo {author} {\bibfnamefont {V.}~\bibnamefont {Shtabovenko}}, \ and\
  \bibinfo {author} {\bibfnamefont {M.}~\bibnamefont {Steinhauser}},\ }\href
  {\doibase 10.1103/PhysRevLett.129.102001} {\bibfield  {journal} {\bibinfo
  {journal} {Phys. Rev. Lett.}\ }\textbf {\bibinfo {volume} {129}},\ \bibinfo
  {pages} {102001} (\bibinfo {year} {2022}{\natexlab{b}})},\ \Eprint
  {http://arxiv.org/abs/2205.07907} {arXiv:2205.07907 [hep-ph]} \BibitemShut
  {NoStop}%
\bibitem [{\citenamefont {Beneke}\ \emph {et~al.}(1996)\citenamefont {Beneke},
  \citenamefont {Buchalla},\ and\ \citenamefont {Dunietz}}]{Beneke:1996gn}%
  \BibitemOpen
  \bibfield  {author} {\bibinfo {author} {\bibfnamefont {M.}~\bibnamefont
  {Beneke}}, \bibinfo {author} {\bibfnamefont {G.}~\bibnamefont {Buchalla}}, \
  and\ \bibinfo {author} {\bibfnamefont {I.}~\bibnamefont {Dunietz}},\ }\href
  {\doibase 10.1103/PhysRevD.54.4419} {\bibfield  {journal} {\bibinfo
  {journal} {Phys. Rev. D}\ }\textbf {\bibinfo {volume} {54}},\ \bibinfo
  {pages} {4419} (\bibinfo {year} {1996})},\ \bibinfo {note} {[Erratum:
  Phys.Rev.D 83, 119902 (2011)]},\ \Eprint
  {http://arxiv.org/abs/hep-ph/9605259} {arXiv:hep-ph/9605259} \BibitemShut
  {NoStop}%
\bibitem [{\citenamefont {Badin}\ \emph {et~al.}(2007)\citenamefont {Badin},
  \citenamefont {Gabbiani},\ and\ \citenamefont {Petrov}}]{Badin:2007bv}%
  \BibitemOpen
  \bibfield  {author} {\bibinfo {author} {\bibfnamefont {A.}~\bibnamefont
  {Badin}}, \bibinfo {author} {\bibfnamefont {F.}~\bibnamefont {Gabbiani}}, \
  and\ \bibinfo {author} {\bibfnamefont {A.~A.}\ \bibnamefont {Petrov}},\
  }\href {\doibase 10.1016/j.physletb.2007.07.049} {\bibfield  {journal}
  {\bibinfo  {journal} {Phys. Lett. B}\ }\textbf {\bibinfo {volume} {653}},\
  \bibinfo {pages} {230} (\bibinfo {year} {2007})},\ \Eprint
  {http://arxiv.org/abs/0707.0294} {arXiv:0707.0294 [hep-ph]} \BibitemShut
  {NoStop}%
\bibitem [{\citenamefont {Davies}\ \emph {et~al.}(2020)\citenamefont {Davies},
  \citenamefont {Harrison}, \citenamefont {Lepage}, \citenamefont {Monahan},
  \citenamefont {Shigemitsu},\ and\ \citenamefont {Wingate}}]{Davies:2019gnp}%
  \BibitemOpen
  \bibfield  {author} {\bibinfo {author} {\bibfnamefont {C.~T.~H.}\
  \bibnamefont {Davies}}, \bibinfo {author} {\bibfnamefont {J.}~\bibnamefont
  {Harrison}}, \bibinfo {author} {\bibfnamefont {G.~P.}\ \bibnamefont
  {Lepage}}, \bibinfo {author} {\bibfnamefont {C.~J.}\ \bibnamefont {Monahan}},
  \bibinfo {author} {\bibfnamefont {J.}~\bibnamefont {Shigemitsu}}, \ and\
  \bibinfo {author} {\bibfnamefont {M.}~\bibnamefont {Wingate}} (\bibinfo
  {collaboration} {HPQCD}),\ }\href {\doibase 10.1103/PhysRevLett.124.082001}
  {\bibfield  {journal} {\bibinfo  {journal} {Phys. Rev. Lett.}\ }\textbf
  {\bibinfo {volume} {124}},\ \bibinfo {pages} {082001} (\bibinfo {year}
  {2020})},\ \Eprint {http://arxiv.org/abs/1910.00970} {arXiv:1910.00970
  [hep-lat]} \BibitemShut {NoStop}%
\bibitem [{\citenamefont {Bobeth}\ \emph {et~al.}(2014)\citenamefont {Bobeth},
  \citenamefont {Haisch}, \citenamefont {Lenz}, \citenamefont {Pecjak},\ and\
  \citenamefont {Tetlalmatzi-Xolocotzi}}]{Bobeth:2014rda}%
  \BibitemOpen
  \bibfield  {author} {\bibinfo {author} {\bibfnamefont {C.}~\bibnamefont
  {Bobeth}}, \bibinfo {author} {\bibfnamefont {U.}~\bibnamefont {Haisch}},
  \bibinfo {author} {\bibfnamefont {A.}~\bibnamefont {Lenz}}, \bibinfo {author}
  {\bibfnamefont {B.}~\bibnamefont {Pecjak}}, \ and\ \bibinfo {author}
  {\bibfnamefont {G.}~\bibnamefont {Tetlalmatzi-Xolocotzi}},\ }\href {\doibase
  10.1007/JHEP06(2014)040} {\bibfield  {journal} {\bibinfo  {journal} {JHEP}\
  }\textbf {\bibinfo {volume} {06}},\ \bibinfo {pages} {040} (\bibinfo {year}
  {2014})},\ \Eprint {http://arxiv.org/abs/1404.2531} {arXiv:1404.2531
  [hep-ph]} \BibitemShut {NoStop}%
\bibitem [{\citenamefont {Glashow}\ \emph {et~al.}(1970)\citenamefont
  {Glashow}, \citenamefont {Iliopoulos},\ and\ \citenamefont
  {Maiani}}]{Glashow:1970gm}%
  \BibitemOpen
  \bibfield  {author} {\bibinfo {author} {\bibfnamefont {S.~L.}\ \bibnamefont
  {Glashow}}, \bibinfo {author} {\bibfnamefont {J.}~\bibnamefont {Iliopoulos}},
  \ and\ \bibinfo {author} {\bibfnamefont {L.}~\bibnamefont {Maiani}},\ }\href
  {\doibase 10.1103/PhysRevD.2.1285} {\bibfield  {journal} {\bibinfo  {journal}
  {Phys. Rev. D}\ }\textbf {\bibinfo {volume} {2}},\ \bibinfo {pages} {1285}
  (\bibinfo {year} {1970})}\BibitemShut {NoStop}%
\bibitem [{\citenamefont {Chau}\ and\ \citenamefont
  {Keung}(1984)}]{Chau:1984fp}%
  \BibitemOpen
  \bibfield  {author} {\bibinfo {author} {\bibfnamefont {L.-L.}\ \bibnamefont
  {Chau}}\ and\ \bibinfo {author} {\bibfnamefont {W.-Y.}\ \bibnamefont
  {Keung}},\ }\href {\doibase 10.1103/PhysRevLett.53.1802} {\bibfield
  {journal} {\bibinfo  {journal} {Phys. Rev. Lett.}\ }\textbf {\bibinfo
  {volume} {53}},\ \bibinfo {pages} {1802} (\bibinfo {year}
  {1984})}\BibitemShut {NoStop}%
\bibitem [{Web()}]{WebHFLAV:2024}%
  \BibitemOpen
  \href@noop {} {\ }\bibinfo {note}
  {\url{https://hflav-eos.web.cern.ch/hflav-eos/osc/PDG_2024/\#CPV}}\BibitemShut
  {NoStop}%
\bibitem [{LHC()}]{LHCbseminar:2023}%
  \BibitemOpen
  \href@noop {} {\ }\bibinfo {note}
  {\url{https://indico.cern.ch/event/1281612/}}\BibitemShut {NoStop}%
\bibitem [{\citenamefont {Workman}\ \emph {et~al.}(2022)\citenamefont {Workman}
  \emph {et~al.}}]{ParticleDataGroup:2022pth}%
  \BibitemOpen
  \bibfield  {author} {\bibinfo {author} {\bibfnamefont {R.~L.}\ \bibnamefont
  {Workman}} \emph {et~al.} (\bibinfo {collaboration} {Particle Data Group}),\
  }\href {\doibase 10.1093/ptep/ptac097} {\bibfield  {journal} {\bibinfo
  {journal} {PTEP}\ }\textbf {\bibinfo {volume} {2022}},\ \bibinfo {pages}
  {083C01} (\bibinfo {year} {2022})}\BibitemShut {NoStop}%
\bibitem [{\citenamefont {Huang}\ and\ \citenamefont
  {Zhou}(2021)}]{Huang:2020hdv}%
  \BibitemOpen
  \bibfield  {author} {\bibinfo {author} {\bibfnamefont {G.-y.}\ \bibnamefont
  {Huang}}\ and\ \bibinfo {author} {\bibfnamefont {S.}~\bibnamefont {Zhou}},\
  }\href {\doibase 10.1103/PhysRevD.103.016010} {\bibfield  {journal} {\bibinfo
   {journal} {Phys. Rev. D}\ }\textbf {\bibinfo {volume} {103}},\ \bibinfo
  {pages} {016010} (\bibinfo {year} {2021})},\ \Eprint
  {http://arxiv.org/abs/2009.04851} {arXiv:2009.04851 [hep-ph]} \BibitemShut
  {NoStop}%
\bibitem [{\citenamefont {del Aguila}\ and\ \citenamefont
  {Bowick}(1983)}]{delAguila:1982fs}%
  \BibitemOpen
  \bibfield  {author} {\bibinfo {author} {\bibfnamefont {F.}~\bibnamefont {del
  Aguila}}\ and\ \bibinfo {author} {\bibfnamefont {M.~J.}\ \bibnamefont
  {Bowick}},\ }\href {\doibase 10.1016/0550-3213(83)90316-4} {\bibfield
  {journal} {\bibinfo  {journal} {Nucl. Phys. B}\ }\textbf {\bibinfo {volume}
  {224}},\ \bibinfo {pages} {107} (\bibinfo {year} {1983})}\BibitemShut
  {NoStop}%
\bibitem [{\citenamefont {Barger}\ \emph {et~al.}(1995)\citenamefont {Barger},
  \citenamefont {Berger},\ and\ \citenamefont {Phillips}}]{Barger:1995dd}%
  \BibitemOpen
  \bibfield  {author} {\bibinfo {author} {\bibfnamefont {V.~D.}\ \bibnamefont
  {Barger}}, \bibinfo {author} {\bibfnamefont {M.~S.}\ \bibnamefont {Berger}},
  \ and\ \bibinfo {author} {\bibfnamefont {R.~J.~N.}\ \bibnamefont
  {Phillips}},\ }\href {\doibase 10.1103/PhysRevD.52.1663} {\bibfield
  {journal} {\bibinfo  {journal} {Phys. Rev. D}\ }\textbf {\bibinfo {volume}
  {52}},\ \bibinfo {pages} {1663} (\bibinfo {year} {1995})},\ \Eprint
  {http://arxiv.org/abs/hep-ph/9503204} {arXiv:hep-ph/9503204} \BibitemShut
  {NoStop}%
\bibitem [{\citenamefont {Carena}\ \emph {et~al.}(2008)\citenamefont {Carena},
  \citenamefont {Medina}, \citenamefont {Panes}, \citenamefont {Shah},\ and\
  \citenamefont {Wagner}}]{Carena:2007tn}%
  \BibitemOpen
  \bibfield  {author} {\bibinfo {author} {\bibfnamefont {M.}~\bibnamefont
  {Carena}}, \bibinfo {author} {\bibfnamefont {A.~D.}\ \bibnamefont {Medina}},
  \bibinfo {author} {\bibfnamefont {B.}~\bibnamefont {Panes}}, \bibinfo
  {author} {\bibfnamefont {N.~R.}\ \bibnamefont {Shah}}, \ and\ \bibinfo
  {author} {\bibfnamefont {C.~E.~M.}\ \bibnamefont {Wagner}},\ }\href {\doibase
  10.1103/PhysRevD.77.076003} {\bibfield  {journal} {\bibinfo  {journal} {Phys.
  Rev. D}\ }\textbf {\bibinfo {volume} {77}},\ \bibinfo {pages} {076003}
  (\bibinfo {year} {2008})},\ \Eprint {http://arxiv.org/abs/0712.0095}
  {arXiv:0712.0095 [hep-ph]} \BibitemShut {NoStop}%
\bibitem [{\citenamefont {Matsedonskyi}\ \emph {et~al.}(2013)\citenamefont
  {Matsedonskyi}, \citenamefont {Panico},\ and\ \citenamefont
  {Wulzer}}]{Matsedonskyi:2012ym}%
  \BibitemOpen
  \bibfield  {author} {\bibinfo {author} {\bibfnamefont {O.}~\bibnamefont
  {Matsedonskyi}}, \bibinfo {author} {\bibfnamefont {G.}~\bibnamefont
  {Panico}}, \ and\ \bibinfo {author} {\bibfnamefont {A.}~\bibnamefont
  {Wulzer}},\ }\href {\doibase 10.1007/JHEP01(2013)164} {\bibfield  {journal}
  {\bibinfo  {journal} {JHEP}\ }\textbf {\bibinfo {volume} {01}},\ \bibinfo
  {pages} {164} (\bibinfo {year} {2013})},\ \Eprint
  {http://arxiv.org/abs/1204.6333} {arXiv:1204.6333 [hep-ph]} \BibitemShut
  {NoStop}%
\bibitem [{\citenamefont {Branco}\ and\ \citenamefont
  {Lavoura}(1986)}]{Branco:1986my}%
  \BibitemOpen
  \bibfield  {author} {\bibinfo {author} {\bibfnamefont {G.~C.}\ \bibnamefont
  {Branco}}\ and\ \bibinfo {author} {\bibfnamefont {L.}~\bibnamefont
  {Lavoura}},\ }\href {\doibase 10.1016/0550-3213(86)90060-X} {\bibfield
  {journal} {\bibinfo  {journal} {Nucl. Phys. B}\ }\textbf {\bibinfo {volume}
  {278}},\ \bibinfo {pages} {738} (\bibinfo {year} {1986})}\BibitemShut
  {NoStop}%
\bibitem [{\citenamefont {Langacker}\ and\ \citenamefont
  {London}(1988)}]{Langacker:1988ur}%
  \BibitemOpen
  \bibfield  {author} {\bibinfo {author} {\bibfnamefont {P.}~\bibnamefont
  {Langacker}}\ and\ \bibinfo {author} {\bibfnamefont {D.}~\bibnamefont
  {London}},\ }\href {\doibase 10.1103/PhysRevD.38.886} {\bibfield  {journal}
  {\bibinfo  {journal} {Phys. Rev. D}\ }\textbf {\bibinfo {volume} {38}},\
  \bibinfo {pages} {886} (\bibinfo {year} {1988})}\BibitemShut {NoStop}%
\bibitem [{\citenamefont {Nir}\ and\ \citenamefont
  {Silverman}(1990)}]{Nir:1990yq}%
  \BibitemOpen
  \bibfield  {author} {\bibinfo {author} {\bibfnamefont {Y.}~\bibnamefont
  {Nir}}\ and\ \bibinfo {author} {\bibfnamefont {D.~J.}\ \bibnamefont
  {Silverman}},\ }\href {\doibase 10.1103/PhysRevD.42.1477} {\bibfield
  {journal} {\bibinfo  {journal} {Phys. Rev. D}\ }\textbf {\bibinfo {volume}
  {42}},\ \bibinfo {pages} {1477} (\bibinfo {year} {1990})}\BibitemShut
  {NoStop}%
\bibitem [{\citenamefont {Barenboim}\ and\ \citenamefont
  {Botella}(1998)}]{Barenboim:1997pf}%
  \BibitemOpen
  \bibfield  {author} {\bibinfo {author} {\bibfnamefont {G.}~\bibnamefont
  {Barenboim}}\ and\ \bibinfo {author} {\bibfnamefont {F.~J.}\ \bibnamefont
  {Botella}},\ }\href {\doibase 10.1016/S0370-2693(98)00695-9} {\bibfield
  {journal} {\bibinfo  {journal} {Phys. Lett. B}\ }\textbf {\bibinfo {volume}
  {433}},\ \bibinfo {pages} {385} (\bibinfo {year} {1998})},\ \Eprint
  {http://arxiv.org/abs/hep-ph/9708209} {arXiv:hep-ph/9708209} \BibitemShut
  {NoStop}%
\bibitem [{\citenamefont {Barenboim}\ \emph {et~al.}(1998)\citenamefont
  {Barenboim}, \citenamefont {Botella}, \citenamefont {Branco},\ and\
  \citenamefont {Vives}}]{Barenboim:1997qx}%
  \BibitemOpen
  \bibfield  {author} {\bibinfo {author} {\bibfnamefont {G.}~\bibnamefont
  {Barenboim}}, \bibinfo {author} {\bibfnamefont {F.~J.}\ \bibnamefont
  {Botella}}, \bibinfo {author} {\bibfnamefont {G.~C.}\ \bibnamefont {Branco}},
  \ and\ \bibinfo {author} {\bibfnamefont {O.}~\bibnamefont {Vives}},\ }\href
  {\doibase 10.1016/S0370-2693(97)01515-3} {\bibfield  {journal} {\bibinfo
  {journal} {Phys. Lett. B}\ }\textbf {\bibinfo {volume} {422}},\ \bibinfo
  {pages} {277} (\bibinfo {year} {1998})},\ \Eprint
  {http://arxiv.org/abs/hep-ph/9709369} {arXiv:hep-ph/9709369} \BibitemShut
  {NoStop}%
\bibitem [{\citenamefont {del Aguila}\ \emph {et~al.}(2000)\citenamefont {del
  Aguila}, \citenamefont {Perez-Victoria},\ and\ \citenamefont
  {Santiago}}]{delAguila:2000rc}%
  \BibitemOpen
  \bibfield  {author} {\bibinfo {author} {\bibfnamefont {F.}~\bibnamefont {del
  Aguila}}, \bibinfo {author} {\bibfnamefont {M.}~\bibnamefont
  {Perez-Victoria}}, \ and\ \bibinfo {author} {\bibfnamefont {J.}~\bibnamefont
  {Santiago}},\ }\href {\doibase 10.1088/1126-6708/2000/09/011} {\bibfield
  {journal} {\bibinfo  {journal} {JHEP}\ }\textbf {\bibinfo {volume} {09}},\
  \bibinfo {pages} {011} (\bibinfo {year} {2000})},\ \Eprint
  {http://arxiv.org/abs/hep-ph/0007316} {arXiv:hep-ph/0007316} \BibitemShut
  {NoStop}%
\bibitem [{\citenamefont {Aguilar-Saavedra}\ \emph {et~al.}(2013)\citenamefont
  {Aguilar-Saavedra}, \citenamefont {Benbrik}, \citenamefont {Heinemeyer},\
  and\ \citenamefont {P\'erez-Victoria}}]{Aguilar-Saavedra:2013qpa}%
  \BibitemOpen
  \bibfield  {author} {\bibinfo {author} {\bibfnamefont {J.~A.}\ \bibnamefont
  {Aguilar-Saavedra}}, \bibinfo {author} {\bibfnamefont {R.}~\bibnamefont
  {Benbrik}}, \bibinfo {author} {\bibfnamefont {S.}~\bibnamefont {Heinemeyer}},
  \ and\ \bibinfo {author} {\bibfnamefont {M.}~\bibnamefont
  {P\'erez-Victoria}},\ }\href {\doibase 10.1103/PhysRevD.88.094010} {\bibfield
   {journal} {\bibinfo  {journal} {Phys. Rev. D}\ }\textbf {\bibinfo {volume}
  {88}},\ \bibinfo {pages} {094010} (\bibinfo {year} {2013})},\ \Eprint
  {http://arxiv.org/abs/1306.0572} {arXiv:1306.0572 [hep-ph]} \BibitemShut
  {NoStop}%
\bibitem [{\citenamefont {Ishiwata}\ \emph {et~al.}(2015)\citenamefont
  {Ishiwata}, \citenamefont {Ligeti},\ and\ \citenamefont
  {Wise}}]{Ishiwata:2015cga}%
  \BibitemOpen
  \bibfield  {author} {\bibinfo {author} {\bibfnamefont {K.}~\bibnamefont
  {Ishiwata}}, \bibinfo {author} {\bibfnamefont {Z.}~\bibnamefont {Ligeti}}, \
  and\ \bibinfo {author} {\bibfnamefont {M.~B.}\ \bibnamefont {Wise}},\ }\href
  {\doibase 10.1007/JHEP10(2015)027} {\bibfield  {journal} {\bibinfo  {journal}
  {JHEP}\ }\textbf {\bibinfo {volume} {10}},\ \bibinfo {pages} {027} (\bibinfo
  {year} {2015})},\ \Eprint {http://arxiv.org/abs/1506.03484} {arXiv:1506.03484
  [hep-ph]} \BibitemShut {NoStop}%
\bibitem [{\citenamefont {Alves}\ \emph {et~al.}(2024)\citenamefont {Alves},
  \citenamefont {Branco}, \citenamefont {Cherchiglia}, \citenamefont {Nishi},
  \citenamefont {Penedo}, \citenamefont {Pereira}, \citenamefont {Rebelo},\
  and\ \citenamefont {Silva-Marcos}}]{Alves:2023ufm}%
  \BibitemOpen
  \bibfield  {author} {\bibinfo {author} {\bibfnamefont {J.~a.~M.}\
  \bibnamefont {Alves}}, \bibinfo {author} {\bibfnamefont {G.~C.}\ \bibnamefont
  {Branco}}, \bibinfo {author} {\bibfnamefont {A.~L.}\ \bibnamefont
  {Cherchiglia}}, \bibinfo {author} {\bibfnamefont {C.~C.}\ \bibnamefont
  {Nishi}}, \bibinfo {author} {\bibfnamefont {J.~T.}\ \bibnamefont {Penedo}},
  \bibinfo {author} {\bibfnamefont {P.~M.~F.}\ \bibnamefont {Pereira}},
  \bibinfo {author} {\bibfnamefont {M.~N.}\ \bibnamefont {Rebelo}}, \ and\
  \bibinfo {author} {\bibfnamefont {J.~I.}\ \bibnamefont {Silva-Marcos}},\
  }\href {\doibase 10.1016/j.physrep.2023.12.004} {\bibfield  {journal}
  {\bibinfo  {journal} {Phys. Rept.}\ }\textbf {\bibinfo {volume} {1057}},\
  \bibinfo {pages} {1} (\bibinfo {year} {2024})},\ \Eprint
  {http://arxiv.org/abs/2304.10561} {arXiv:2304.10561 [hep-ph]} \BibitemShut
  {NoStop}%
\bibitem [{\citenamefont {Tumasyan}\ \emph {et~al.}(2023)\citenamefont
  {Tumasyan} \emph {et~al.}}]{CMS:2022fck}%
  \BibitemOpen
  \bibfield  {author} {\bibinfo {author} {\bibfnamefont {A.}~\bibnamefont
  {Tumasyan}} \emph {et~al.} (\bibinfo {collaboration} {CMS}),\ }\href
  {\doibase 10.1007/JHEP07(2023)020} {\bibfield  {journal} {\bibinfo  {journal}
  {JHEP}\ }\textbf {\bibinfo {volume} {07}},\ \bibinfo {pages} {020} (\bibinfo
  {year} {2023})},\ \Eprint {http://arxiv.org/abs/2209.07327} {arXiv:2209.07327
  [hep-ex]} \BibitemShut {NoStop}%
\bibitem [{\citenamefont {Banerjee}\ \emph {et~al.}(2024)\citenamefont
  {Banerjee}, \citenamefont {Bergeaas~Kuutmann}, \citenamefont {Ellajosyula},
  \citenamefont {Enberg}, \citenamefont {Ferretti},\ and\ \citenamefont
  {Panizzi}}]{Banerjee:2024zvg}%
  \BibitemOpen
  \bibfield  {author} {\bibinfo {author} {\bibfnamefont {A.}~\bibnamefont
  {Banerjee}}, \bibinfo {author} {\bibfnamefont {E.}~\bibnamefont
  {Bergeaas~Kuutmann}}, \bibinfo {author} {\bibfnamefont {V.}~\bibnamefont
  {Ellajosyula}}, \bibinfo {author} {\bibfnamefont {R.}~\bibnamefont {Enberg}},
  \bibinfo {author} {\bibfnamefont {G.}~\bibnamefont {Ferretti}}, \ and\
  \bibinfo {author} {\bibfnamefont {L.}~\bibnamefont {Panizzi}},\ }\href@noop
  {} {\  (\bibinfo {year} {2024})},\ \Eprint {http://arxiv.org/abs/2406.09193}
  {arXiv:2406.09193 [hep-ph]} \BibitemShut {NoStop}%
\bibitem [{\citenamefont {J\"ager}\ \emph {et~al.}(2018)\citenamefont
  {J\"ager}, \citenamefont {Kirk}, \citenamefont {Lenz},\ and\ \citenamefont
  {Leslie}}]{Jager:2017gal}%
  \BibitemOpen
  \bibfield  {author} {\bibinfo {author} {\bibfnamefont {S.}~\bibnamefont
  {J\"ager}}, \bibinfo {author} {\bibfnamefont {M.}~\bibnamefont {Kirk}},
  \bibinfo {author} {\bibfnamefont {A.}~\bibnamefont {Lenz}}, \ and\ \bibinfo
  {author} {\bibfnamefont {K.}~\bibnamefont {Leslie}},\ }\href {\doibase
  10.1103/PhysRevD.97.015021} {\bibfield  {journal} {\bibinfo  {journal} {Phys.
  Rev. D}\ }\textbf {\bibinfo {volume} {97}},\ \bibinfo {pages} {015021}
  (\bibinfo {year} {2018})},\ \Eprint {http://arxiv.org/abs/1701.09183}
  {arXiv:1701.09183 [hep-ph]} \BibitemShut {NoStop}%
\bibitem [{\citenamefont {Lenz}\ and\ \citenamefont
  {Tetlalmatzi-Xolocotzi}(2020)}]{Lenz:2019lvd}%
  \BibitemOpen
  \bibfield  {author} {\bibinfo {author} {\bibfnamefont {A.}~\bibnamefont
  {Lenz}}\ and\ \bibinfo {author} {\bibfnamefont {G.}~\bibnamefont
  {Tetlalmatzi-Xolocotzi}},\ }\href {\doibase 10.1007/JHEP07(2020)177}
  {\bibfield  {journal} {\bibinfo  {journal} {JHEP}\ }\textbf {\bibinfo
  {volume} {07}},\ \bibinfo {pages} {177} (\bibinfo {year} {2020})},\ \Eprint
  {http://arxiv.org/abs/1912.07621} {arXiv:1912.07621 [hep-ph]} \BibitemShut
  {NoStop}%
\bibitem [{\citenamefont {J\"ager}\ \emph {et~al.}(2020)\citenamefont
  {J\"ager}, \citenamefont {Kirk}, \citenamefont {Lenz},\ and\ \citenamefont
  {Leslie}}]{Jager:2019bgk}%
  \BibitemOpen
  \bibfield  {author} {\bibinfo {author} {\bibfnamefont {S.}~\bibnamefont
  {J\"ager}}, \bibinfo {author} {\bibfnamefont {M.}~\bibnamefont {Kirk}},
  \bibinfo {author} {\bibfnamefont {A.}~\bibnamefont {Lenz}}, \ and\ \bibinfo
  {author} {\bibfnamefont {K.}~\bibnamefont {Leslie}},\ }\href {\doibase
  10.1007/JHEP03(2020)122} {\bibfield  {journal} {\bibinfo  {journal} {JHEP}\
  }\textbf {\bibinfo {volume} {03}},\ \bibinfo {pages} {122} (\bibinfo {year}
  {2020})},\ \bibinfo {note} {[Erratum: JHEP 04, 094 (2023)]},\ \Eprint
  {http://arxiv.org/abs/1910.12924} {arXiv:1910.12924 [hep-ph]} \BibitemShut
  {NoStop}%
\bibitem [{\citenamefont {Lenz}\ \emph {et~al.}(2023)\citenamefont {Lenz},
  \citenamefont {M\"uller}, \citenamefont {Piscopo},\ and\ \citenamefont
  {Rusov}}]{Lenz:2022pgw}%
  \BibitemOpen
  \bibfield  {author} {\bibinfo {author} {\bibfnamefont {A.}~\bibnamefont
  {Lenz}}, \bibinfo {author} {\bibfnamefont {J.}~\bibnamefont {M\"uller}},
  \bibinfo {author} {\bibfnamefont {M.~L.}\ \bibnamefont {Piscopo}}, \ and\
  \bibinfo {author} {\bibfnamefont {A.~V.}\ \bibnamefont {Rusov}},\ }\href
  {\doibase 10.1007/JHEP09(2023)028} {\bibfield  {journal} {\bibinfo  {journal}
  {JHEP}\ }\textbf {\bibinfo {volume} {09}},\ \bibinfo {pages} {028} (\bibinfo
  {year} {2023})},\ \Eprint {http://arxiv.org/abs/2211.02724} {arXiv:2211.02724
  [hep-ph]} \BibitemShut {NoStop}%
\bibitem [{\citenamefont {Altmannshofer}\ and\ \citenamefont
  {Stangl}(2021)}]{Altmannshofer:2021qrr}%
  \BibitemOpen
  \bibfield  {author} {\bibinfo {author} {\bibfnamefont {W.}~\bibnamefont
  {Altmannshofer}}\ and\ \bibinfo {author} {\bibfnamefont {P.}~\bibnamefont
  {Stangl}},\ }\href {\doibase 10.1140/epjc/s10052-021-09725-1} {\bibfield
  {journal} {\bibinfo  {journal} {Eur. Phys. J. C}\ }\textbf {\bibinfo {volume}
  {81}},\ \bibinfo {pages} {952} (\bibinfo {year} {2021})},\ \Eprint
  {http://arxiv.org/abs/2103.13370} {arXiv:2103.13370 [hep-ph]} \BibitemShut
  {NoStop}%
\bibitem [{\citenamefont {Greljo}\ \emph {et~al.}(2023)\citenamefont {Greljo},
  \citenamefont {Salko}, \citenamefont {Smolkovi\v{c}},\ and\ \citenamefont
  {Stangl}}]{Greljo:2022jac}%
  \BibitemOpen
  \bibfield  {author} {\bibinfo {author} {\bibfnamefont {A.}~\bibnamefont
  {Greljo}}, \bibinfo {author} {\bibfnamefont {J.}~\bibnamefont {Salko}},
  \bibinfo {author} {\bibfnamefont {A.}~\bibnamefont {Smolkovi\v{c}}}, \ and\
  \bibinfo {author} {\bibfnamefont {P.}~\bibnamefont {Stangl}},\ }\href
  {\doibase 10.1007/JHEP05(2023)087} {\bibfield  {journal} {\bibinfo  {journal}
  {JHEP}\ }\textbf {\bibinfo {volume} {05}},\ \bibinfo {pages} {087} (\bibinfo
  {year} {2023})},\ \Eprint {http://arxiv.org/abs/2212.10497} {arXiv:2212.10497
  [hep-ph]} \BibitemShut {NoStop}%
\bibitem [{\citenamefont {Bause}\ \emph {et~al.}(2023)\citenamefont {Bause},
  \citenamefont {Gisbert}, \citenamefont {Golz},\ and\ \citenamefont
  {Hiller}}]{Bause:2022rrs}%
  \BibitemOpen
  \bibfield  {author} {\bibinfo {author} {\bibfnamefont {R.}~\bibnamefont
  {Bause}}, \bibinfo {author} {\bibfnamefont {H.}~\bibnamefont {Gisbert}},
  \bibinfo {author} {\bibfnamefont {M.}~\bibnamefont {Golz}}, \ and\ \bibinfo
  {author} {\bibfnamefont {G.}~\bibnamefont {Hiller}},\ }\href {\doibase
  10.1140/epjc/s10052-023-11586-9} {\bibfield  {journal} {\bibinfo  {journal}
  {Eur. Phys. J. C}\ }\textbf {\bibinfo {volume} {83}},\ \bibinfo {pages} {419}
  (\bibinfo {year} {2023})},\ \Eprint {http://arxiv.org/abs/2209.04457}
  {arXiv:2209.04457 [hep-ph]} \BibitemShut {NoStop}%
\bibitem [{\citenamefont {Bordone}\ and\ \citenamefont
  {Fern\'andez~Navarro}(2023)}]{Bordone:2023ybl}%
  \BibitemOpen
  \bibfield  {author} {\bibinfo {author} {\bibfnamefont {M.}~\bibnamefont
  {Bordone}}\ and\ \bibinfo {author} {\bibfnamefont {M.}~\bibnamefont
  {Fern\'andez~Navarro}},\ }\href {\doibase 10.1140/epjc/s10052-023-12013-9}
  {\bibfield  {journal} {\bibinfo  {journal} {Eur. Phys. J. C}\ }\textbf
  {\bibinfo {volume} {83}},\ \bibinfo {pages} {842} (\bibinfo {year} {2023})},\
  \Eprint {http://arxiv.org/abs/2307.07013} {arXiv:2307.07013 [hep-ph]}
  \BibitemShut {NoStop}%
\bibitem [{\citenamefont {Aaij}\ \emph {et~al.}(2017)\citenamefont {Aaij} \emph
  {et~al.}}]{LHCb:2017myy}%
  \BibitemOpen
  \bibfield  {author} {\bibinfo {author} {\bibfnamefont {R.}~\bibnamefont
  {Aaij}} \emph {et~al.} (\bibinfo {collaboration} {LHCb}),\ }\href {\doibase
  10.1103/PhysRevLett.118.251802} {\bibfield  {journal} {\bibinfo  {journal}
  {Phys. Rev. Lett.}\ }\textbf {\bibinfo {volume} {118}},\ \bibinfo {pages}
  {251802} (\bibinfo {year} {2017})},\ \Eprint
  {http://arxiv.org/abs/1703.02508} {arXiv:1703.02508 [hep-ex]} \BibitemShut
  {NoStop}%
\bibitem [{\citenamefont {Crivellin}\ and\ \citenamefont
  {Kirk}(2023)}]{Crivellin:2023saq}%
  \BibitemOpen
  \bibfield  {author} {\bibinfo {author} {\bibfnamefont {A.}~\bibnamefont
  {Crivellin}}\ and\ \bibinfo {author} {\bibfnamefont {M.}~\bibnamefont
  {Kirk}},\ }\href {\doibase 10.1103/PhysRevD.108.L111701} {\bibfield
  {journal} {\bibinfo  {journal} {Phys. Rev. D}\ }\textbf {\bibinfo {volume}
  {108}},\ \bibinfo {pages} {L111701} (\bibinfo {year} {2023})},\ \Eprint
  {http://arxiv.org/abs/2309.07205} {arXiv:2309.07205 [hep-ph]} \BibitemShut
  {NoStop}%
\bibitem [{\citenamefont {Cornella}\ \emph {et~al.}(2021)\citenamefont
  {Cornella}, \citenamefont {Faroughy}, \citenamefont {Fuentes-Martin},
  \citenamefont {Isidori},\ and\ \citenamefont {Neubert}}]{Cornella:2021sby}%
  \BibitemOpen
  \bibfield  {author} {\bibinfo {author} {\bibfnamefont {C.}~\bibnamefont
  {Cornella}}, \bibinfo {author} {\bibfnamefont {D.~A.}\ \bibnamefont
  {Faroughy}}, \bibinfo {author} {\bibfnamefont {J.}~\bibnamefont
  {Fuentes-Martin}}, \bibinfo {author} {\bibfnamefont {G.}~\bibnamefont
  {Isidori}}, \ and\ \bibinfo {author} {\bibfnamefont {M.}~\bibnamefont
  {Neubert}},\ }\href {\doibase 10.1007/JHEP08(2021)050} {\bibfield  {journal}
  {\bibinfo  {journal} {JHEP}\ }\textbf {\bibinfo {volume} {08}},\ \bibinfo
  {pages} {050} (\bibinfo {year} {2021})},\ \Eprint
  {http://arxiv.org/abs/2103.16558} {arXiv:2103.16558 [hep-ph]} \BibitemShut
  {NoStop}%
\bibitem [{\citenamefont {Aaboud}\ \emph {et~al.}(2018)\citenamefont {Aaboud}
  \emph {et~al.}}]{ATLAS:2017jnp}%
  \BibitemOpen
  \bibfield  {author} {\bibinfo {author} {\bibfnamefont {M.}~\bibnamefont
  {Aaboud}} \emph {et~al.} (\bibinfo {collaboration} {ATLAS}),\ }\href
  {\doibase 10.1140/epjc/s10052-018-5693-4} {\bibfield  {journal} {\bibinfo
  {journal} {Eur. Phys. J. C}\ }\textbf {\bibinfo {volume} {78}},\ \bibinfo
  {pages} {250} (\bibinfo {year} {2018})},\ \Eprint
  {http://arxiv.org/abs/1710.07171} {arXiv:1710.07171 [hep-ex]} \BibitemShut
  {NoStop}%
\bibitem [{\citenamefont {Sirunyan}\ \emph {et~al.}(2018)\citenamefont
  {Sirunyan} \emph {et~al.}}]{CMS:2018mts}%
  \BibitemOpen
  \bibfield  {author} {\bibinfo {author} {\bibfnamefont {A.~M.}\ \bibnamefont
  {Sirunyan}} \emph {et~al.} (\bibinfo {collaboration} {CMS}),\ }\href
  {\doibase 10.1103/PhysRevD.98.112014} {\bibfield  {journal} {\bibinfo
  {journal} {Phys. Rev. D}\ }\textbf {\bibinfo {volume} {98}},\ \bibinfo
  {pages} {112014} (\bibinfo {year} {2018})},\ \Eprint
  {http://arxiv.org/abs/1808.03124} {arXiv:1808.03124 [hep-ex]} \BibitemShut
  {NoStop}%
\bibitem [{\citenamefont {Collaboration}\ \emph {et~al.}(2019)\citenamefont
  {Collaboration} \emph {et~al.}}]{CMS:2019zmd}%
  \BibitemOpen
  \bibfield  {author} {\bibinfo {author} {\bibfnamefont {T.~C.}\ \bibnamefont
  {Collaboration}} \emph {et~al.} (\bibinfo {collaboration} {CMS}),\ }\href
  {\doibase 10.1007/JHEP10(2019)244} {\bibfield  {journal} {\bibinfo  {journal}
  {JHEP}\ }\textbf {\bibinfo {volume} {10}},\ \bibinfo {pages} {244} (\bibinfo
  {year} {2019})},\ \Eprint {http://arxiv.org/abs/1908.04722} {arXiv:1908.04722
  [hep-ex]} \BibitemShut {NoStop}%
\bibitem [{\citenamefont {Aad}\ \emph {et~al.}(2021)\citenamefont {Aad} \emph
  {et~al.}}]{ATLAS:2020syg}%
  \BibitemOpen
  \bibfield  {author} {\bibinfo {author} {\bibfnamefont {G.}~\bibnamefont
  {Aad}} \emph {et~al.} (\bibinfo {collaboration} {ATLAS}),\ }\href {\doibase
  10.1007/JHEP02(2021)143} {\bibfield  {journal} {\bibinfo  {journal} {JHEP}\
  }\textbf {\bibinfo {volume} {02}},\ \bibinfo {pages} {143} (\bibinfo {year}
  {2021})},\ \Eprint {http://arxiv.org/abs/2010.14293} {arXiv:2010.14293
  [hep-ex]} \BibitemShut {NoStop}%
\bibitem [{\citenamefont {Agrawal}\ \emph {et~al.}(2014)\citenamefont
  {Agrawal}, \citenamefont {Blanke},\ and\ \citenamefont
  {Gemmler}}]{Agrawal:2014aoa}%
  \BibitemOpen
  \bibfield  {author} {\bibinfo {author} {\bibfnamefont {P.}~\bibnamefont
  {Agrawal}}, \bibinfo {author} {\bibfnamefont {M.}~\bibnamefont {Blanke}}, \
  and\ \bibinfo {author} {\bibfnamefont {K.}~\bibnamefont {Gemmler}},\ }\href
  {\doibase 10.1007/JHEP10(2014)072} {\bibfield  {journal} {\bibinfo  {journal}
  {JHEP}\ }\textbf {\bibinfo {volume} {10}},\ \bibinfo {pages} {072} (\bibinfo
  {year} {2014})},\ \Eprint {http://arxiv.org/abs/1405.6709} {arXiv:1405.6709
  [hep-ph]} \BibitemShut {NoStop}%
\bibitem [{\citenamefont {Oh}\ and\ \citenamefont {Tandean}(2011)}]{Oh:2010vc}%
  \BibitemOpen
  \bibfield  {author} {\bibinfo {author} {\bibfnamefont {S.}~\bibnamefont
  {Oh}}\ and\ \bibinfo {author} {\bibfnamefont {J.}~\bibnamefont {Tandean}},\
  }\href {\doibase 10.1016/j.physletb.2011.01.030} {\bibfield  {journal}
  {\bibinfo  {journal} {Phys. Lett. B}\ }\textbf {\bibinfo {volume} {697}},\
  \bibinfo {pages} {41} (\bibinfo {year} {2011})},\ \Eprint
  {http://arxiv.org/abs/1008.2153} {arXiv:1008.2153 [hep-ph]} \BibitemShut
  {NoStop}%
\bibitem [{\citenamefont {Crivellin}\ \emph {et~al.}(2022)\citenamefont
  {Crivellin}, \citenamefont {Manzari}, \citenamefont {Altmannshofer},
  \citenamefont {Inguglia}, \citenamefont {Feichtinger},\ and\ \citenamefont
  {Martin~Camalich}}]{Crivellin:2022obd}%
  \BibitemOpen
  \bibfield  {author} {\bibinfo {author} {\bibfnamefont {A.}~\bibnamefont
  {Crivellin}}, \bibinfo {author} {\bibfnamefont {C.~A.}\ \bibnamefont
  {Manzari}}, \bibinfo {author} {\bibfnamefont {W.}~\bibnamefont
  {Altmannshofer}}, \bibinfo {author} {\bibfnamefont {G.}~\bibnamefont
  {Inguglia}}, \bibinfo {author} {\bibfnamefont {P.}~\bibnamefont
  {Feichtinger}}, \ and\ \bibinfo {author} {\bibfnamefont {J.}~\bibnamefont
  {Martin~Camalich}},\ }\href {\doibase 10.1103/PhysRevD.106.L031703}
  {\bibfield  {journal} {\bibinfo  {journal} {Phys. Rev. D}\ }\textbf {\bibinfo
  {volume} {106}},\ \bibinfo {pages} {L031703} (\bibinfo {year} {2022})},\
  \Eprint {http://arxiv.org/abs/2202.12900} {arXiv:2202.12900 [hep-ph]}
  \BibitemShut {NoStop}%
\bibitem [{\citenamefont {Hadjivasiliou}\ \emph {et~al.}(2022)\citenamefont
  {Hadjivasiliou} \emph {et~al.}}]{Belle:2021gmc}%
  \BibitemOpen
  \bibfield  {author} {\bibinfo {author} {\bibfnamefont {C.}~\bibnamefont
  {Hadjivasiliou}} \emph {et~al.} (\bibinfo {collaboration} {Belle}),\ }\href
  {\doibase 10.1103/PhysRevD.105.L051101} {\bibfield  {journal} {\bibinfo
  {journal} {Phys. Rev. D}\ }\textbf {\bibinfo {volume} {105}},\ \bibinfo
  {pages} {L051101} (\bibinfo {year} {2022})},\ \Eprint
  {http://arxiv.org/abs/2110.14086} {arXiv:2110.14086 [hep-ex]} \BibitemShut
  {NoStop}%
\bibitem [{\citenamefont {Lees}\ \emph
  {et~al.}(2023{\natexlab{a}})\citenamefont {Lees} \emph
  {et~al.}}]{BaBar:2023rer}%
  \BibitemOpen
  \bibfield  {author} {\bibinfo {author} {\bibfnamefont {J.~P.}\ \bibnamefont
  {Lees}} \emph {et~al.} (\bibinfo {collaboration} {BaBar}),\ }\href {\doibase
  10.1103/PhysRevD.107.092001} {\bibfield  {journal} {\bibinfo  {journal}
  {Phys. Rev. D}\ }\textbf {\bibinfo {volume} {107}},\ \bibinfo {pages}
  {092001} (\bibinfo {year} {2023}{\natexlab{a}})},\ \Eprint
  {http://arxiv.org/abs/2302.00208} {arXiv:2302.00208 [hep-ex]} \BibitemShut
  {NoStop}%
\bibitem [{\citenamefont {Lees}\ \emph
  {et~al.}(2023{\natexlab{b}})\citenamefont {Lees} \emph
  {et~al.}}]{BaBar:2023dtq}%
  \BibitemOpen
  \bibfield  {author} {\bibinfo {author} {\bibfnamefont {J.~P.}\ \bibnamefont
  {Lees}} \emph {et~al.} (\bibinfo {collaboration} {BaBar}),\ }\href {\doibase
  10.1103/PhysRevLett.131.201801} {\bibfield  {journal} {\bibinfo  {journal}
  {Phys. Rev. Lett.}\ }\textbf {\bibinfo {volume} {131}},\ \bibinfo {pages}
  {201801} (\bibinfo {year} {2023}{\natexlab{b}})},\ \Eprint
  {http://arxiv.org/abs/2306.08490} {arXiv:2306.08490 [hep-ex]} \BibitemShut
  {NoStop}%
\bibitem [{\citenamefont {Khodjamirian}\ and\ \citenamefont
  {Wald}(2022)}]{Khodjamirian:2022vta}%
  \BibitemOpen
  \bibfield  {author} {\bibinfo {author} {\bibfnamefont {A.}~\bibnamefont
  {Khodjamirian}}\ and\ \bibinfo {author} {\bibfnamefont {M.}~\bibnamefont
  {Wald}},\ }\href {\doibase 10.1016/j.physletb.2022.137434} {\bibfield
  {journal} {\bibinfo  {journal} {Phys. Lett. B}\ }\textbf {\bibinfo {volume}
  {834}},\ \bibinfo {pages} {137434} (\bibinfo {year} {2022})},\ \Eprint
  {http://arxiv.org/abs/2206.11601} {arXiv:2206.11601 [hep-ph]} \BibitemShut
  {NoStop}%
\bibitem [{\citenamefont {Elor}\ and\ \citenamefont
  {Guerrera}(2023)}]{Elor:2022jxy}%
  \BibitemOpen
  \bibfield  {author} {\bibinfo {author} {\bibfnamefont {G.}~\bibnamefont
  {Elor}}\ and\ \bibinfo {author} {\bibfnamefont {A.~W.~M.}\ \bibnamefont
  {Guerrera}},\ }\href {\doibase 10.1007/JHEP02(2023)100} {\bibfield  {journal}
  {\bibinfo  {journal} {JHEP}\ }\textbf {\bibinfo {volume} {02}},\ \bibinfo
  {pages} {100} (\bibinfo {year} {2023})},\ \Eprint
  {http://arxiv.org/abs/2211.10553} {arXiv:2211.10553 [hep-ph]} \BibitemShut
  {NoStop}%
\bibitem [{\citenamefont {Boushmelev}\ and\ \citenamefont
  {Wald}(2024)}]{Boushmelev:2023huu}%
  \BibitemOpen
  \bibfield  {author} {\bibinfo {author} {\bibfnamefont {A.}~\bibnamefont
  {Boushmelev}}\ and\ \bibinfo {author} {\bibfnamefont {M.}~\bibnamefont
  {Wald}},\ }\href {\doibase 10.1103/PhysRevD.109.055049} {\bibfield  {journal}
  {\bibinfo  {journal} {Phys. Rev. D}\ }\textbf {\bibinfo {volume} {109}},\
  \bibinfo {pages} {055049} (\bibinfo {year} {2024})},\ \Eprint
  {http://arxiv.org/abs/2311.13482} {arXiv:2311.13482 [hep-ph]} \BibitemShut
  {NoStop}%
\bibitem [{\citenamefont {Shi}\ \emph {et~al.}(2023)\citenamefont {Shi},
  \citenamefont {Xing},\ and\ \citenamefont {Xing}}]{Shi:2023riy}%
  \BibitemOpen
  \bibfield  {author} {\bibinfo {author} {\bibfnamefont {Y.-J.}\ \bibnamefont
  {Shi}}, \bibinfo {author} {\bibfnamefont {Y.}~\bibnamefont {Xing}}, \ and\
  \bibinfo {author} {\bibfnamefont {Z.-P.}\ \bibnamefont {Xing}},\ }\href
  {\doibase 10.1140/epjc/s10052-023-11930-z} {\bibfield  {journal} {\bibinfo
  {journal} {Eur. Phys. J. C}\ }\textbf {\bibinfo {volume} {83}},\ \bibinfo
  {pages} {744} (\bibinfo {year} {2023})},\ \Eprint
  {http://arxiv.org/abs/2305.17622} {arXiv:2305.17622 [hep-ph]} \BibitemShut
  {NoStop}%
\bibitem [{\citenamefont {Bigi}\ and\ \citenamefont
  {Sanda}(2009)}]{Bigi:2000yz}%
  \BibitemOpen
  \bibfield  {author} {\bibinfo {author} {\bibfnamefont {I.~I.}\ \bibnamefont
  {Bigi}}\ and\ \bibinfo {author} {\bibfnamefont {A.~I.}\ \bibnamefont
  {Sanda}},\ }\href {\doibase 10.1017/CBO9780511581014} {\emph {\bibinfo
  {title} {{CP violation}}}},\ Vol.~\bibinfo {volume} {9}\ (\bibinfo
  {publisher} {Cambridge University Press},\ \bibinfo {year}
  {2009})\BibitemShut {NoStop}%
\bibitem [{\citenamefont {Romao}\ and\ \citenamefont
  {Silva}(2012)}]{Romao:2012pq}%
  \BibitemOpen
  \bibfield  {author} {\bibinfo {author} {\bibfnamefont {J.~C.}\ \bibnamefont
  {Romao}}\ and\ \bibinfo {author} {\bibfnamefont {J.~P.}\ \bibnamefont
  {Silva}},\ }\href {\doibase 10.1142/S0217751X12300256} {\bibfield  {journal}
  {\bibinfo  {journal} {Int. J. Mod. Phys. A}\ }\textbf {\bibinfo {volume}
  {27}},\ \bibinfo {pages} {1230025} (\bibinfo {year} {2012})},\ \Eprint
  {http://arxiv.org/abs/1209.6213} {arXiv:1209.6213 [hep-ph]} \BibitemShut
  {NoStop}%
\bibitem [{\citenamefont {Buchalla}\ \emph {et~al.}(1996)\citenamefont
  {Buchalla}, \citenamefont {Buras},\ and\ \citenamefont
  {Lautenbacher}}]{Buchalla:1995vs}%
  \BibitemOpen
  \bibfield  {author} {\bibinfo {author} {\bibfnamefont {G.}~\bibnamefont
  {Buchalla}}, \bibinfo {author} {\bibfnamefont {A.~J.}\ \bibnamefont {Buras}},
  \ and\ \bibinfo {author} {\bibfnamefont {M.~E.}\ \bibnamefont
  {Lautenbacher}},\ }\href {\doibase 10.1103/RevModPhys.68.1125} {\bibfield
  {journal} {\bibinfo  {journal} {Rev. Mod. Phys.}\ }\textbf {\bibinfo {volume}
  {68}},\ \bibinfo {pages} {1125} (\bibinfo {year} {1996})},\ \Eprint
  {http://arxiv.org/abs/hep-ph/9512380} {arXiv:hep-ph/9512380} \BibitemShut
  {NoStop}%
\bibitem [{\citenamefont {Cheng}(1982)}]{Cheng:1982hq}%
  \BibitemOpen
  \bibfield  {author} {\bibinfo {author} {\bibfnamefont {H.-Y.}\ \bibnamefont
  {Cheng}},\ }\href {\doibase 10.1103/PhysRevD.26.143} {\bibfield  {journal}
  {\bibinfo  {journal} {Phys. Rev. D}\ }\textbf {\bibinfo {volume} {26}},\
  \bibinfo {pages} {143} (\bibinfo {year} {1982})}\BibitemShut {NoStop}%
\bibitem [{\citenamefont {Buras}\ \emph {et~al.}(1984)\citenamefont {Buras},
  \citenamefont {Slominski},\ and\ \citenamefont {Steger}}]{Buras:1984pq}%
  \BibitemOpen
  \bibfield  {author} {\bibinfo {author} {\bibfnamefont {A.~J.}\ \bibnamefont
  {Buras}}, \bibinfo {author} {\bibfnamefont {W.}~\bibnamefont {Slominski}}, \
  and\ \bibinfo {author} {\bibfnamefont {H.}~\bibnamefont {Steger}},\ }\href
  {\doibase 10.1016/0550-3213(84)90437-1} {\bibfield  {journal} {\bibinfo
  {journal} {Nucl. Phys. B}\ }\textbf {\bibinfo {volume} {245}},\ \bibinfo
  {pages} {369} (\bibinfo {year} {1984})}\BibitemShut {NoStop}%
\bibitem [{\citenamefont {Ellis}\ \emph {et~al.}(1977)\citenamefont {Ellis},
  \citenamefont {Gaillard}, \citenamefont {Nanopoulos},\ and\ \citenamefont
  {Rudaz}}]{Ellis:1977uk}%
  \BibitemOpen
  \bibfield  {author} {\bibinfo {author} {\bibfnamefont {J.~R.}\ \bibnamefont
  {Ellis}}, \bibinfo {author} {\bibfnamefont {M.~K.}\ \bibnamefont {Gaillard}},
  \bibinfo {author} {\bibfnamefont {D.~V.}\ \bibnamefont {Nanopoulos}}, \ and\
  \bibinfo {author} {\bibfnamefont {S.}~\bibnamefont {Rudaz}},\ }\href
  {\doibase 10.1016/0550-3213(77)90374-1} {\bibfield  {journal} {\bibinfo
  {journal} {Nucl. Phys. B}\ }\textbf {\bibinfo {volume} {131}},\ \bibinfo
  {pages} {285} (\bibinfo {year} {1977})},\ \bibinfo {note} {[Erratum:
  Nucl.Phys.B 132, 541 (1978)]}\BibitemShut {NoStop}%
\bibitem [{\citenamefont {Hagelin}(1981)}]{Hagelin:1981zk}%
  \BibitemOpen
  \bibfield  {author} {\bibinfo {author} {\bibfnamefont {J.~S.}\ \bibnamefont
  {Hagelin}},\ }\href {\doibase 10.1016/0550-3213(81)90521-6} {\bibfield
  {journal} {\bibinfo  {journal} {Nucl. Phys. B}\ }\textbf {\bibinfo {volume}
  {193}},\ \bibinfo {pages} {123} (\bibinfo {year} {1981})}\BibitemShut
  {NoStop}%
\bibitem [{\citenamefont {Franco}\ \emph {et~al.}(1982)\citenamefont {Franco},
  \citenamefont {Lusignoli},\ and\ \citenamefont {Pugliese}}]{Franco:1981ea}%
  \BibitemOpen
  \bibfield  {author} {\bibinfo {author} {\bibfnamefont {E.}~\bibnamefont
  {Franco}}, \bibinfo {author} {\bibfnamefont {M.}~\bibnamefont {Lusignoli}}, \
  and\ \bibinfo {author} {\bibfnamefont {A.}~\bibnamefont {Pugliese}},\ }\href
  {\doibase 10.1016/0550-3213(82)90018-9} {\bibfield  {journal} {\bibinfo
  {journal} {Nucl. Phys. B}\ }\textbf {\bibinfo {volume} {194}},\ \bibinfo
  {pages} {403} (\bibinfo {year} {1982})}\BibitemShut {NoStop}%
\bibitem [{\citenamefont {Chau}(1983)}]{Chau:1982da}%
  \BibitemOpen
  \bibfield  {author} {\bibinfo {author} {\bibfnamefont {L.-L.}\ \bibnamefont
  {Chau}},\ }\href {\doibase 10.1016/0370-1573(83)90043-1} {\bibfield
  {journal} {\bibinfo  {journal} {Phys. Rept.}\ }\textbf {\bibinfo {volume}
  {95}},\ \bibinfo {pages} {1} (\bibinfo {year} {1983})}\BibitemShut {NoStop}%
\bibitem [{\citenamefont {Khoze}\ \emph {et~al.}(1987)\citenamefont {Khoze},
  \citenamefont {Shifman}, \citenamefont {Uraltsev},\ and\ \citenamefont
  {Voloshin}}]{Khoze:1986fa}%
  \BibitemOpen
  \bibfield  {author} {\bibinfo {author} {\bibfnamefont {V.~A.}\ \bibnamefont
  {Khoze}}, \bibinfo {author} {\bibfnamefont {M.~A.}\ \bibnamefont {Shifman}},
  \bibinfo {author} {\bibfnamefont {N.~G.}\ \bibnamefont {Uraltsev}}, \ and\
  \bibinfo {author} {\bibfnamefont {M.~B.}\ \bibnamefont {Voloshin}},\
  }\href@noop {} {\bibfield  {journal} {\bibinfo  {journal} {Sov. J. Nucl.
  Phys.}\ }\textbf {\bibinfo {volume} {46}},\ \bibinfo {pages} {112} (\bibinfo
  {year} {1987})}\BibitemShut {NoStop}%
\bibitem [{\citenamefont {Dziurda}\ \emph {et~al.}(2024)\citenamefont {Dziurda}
  \emph {et~al.}}]{Dziurda:2024hrg}%
  \BibitemOpen
  \bibfield  {author} {\bibinfo {author} {\bibfnamefont {A.}~\bibnamefont
  {Dziurda}} \emph {et~al.},\ }in\ \href@noop {} {\emph {\bibinfo {booktitle}
  {{12th International Workshop on the CKM Unitarity Triangle}}}}\ (\bibinfo
  {year} {2024})\ \Eprint {http://arxiv.org/abs/2404.03945} {arXiv:2404.03945
  [hep-ph]} \BibitemShut {NoStop}%
\bibitem [{\citenamefont {Aad}\ \emph {et~al.}(2012)\citenamefont {Aad} \emph
  {et~al.}}]{ATLAS:2012yve}%
  \BibitemOpen
  \bibfield  {author} {\bibinfo {author} {\bibfnamefont {G.}~\bibnamefont
  {Aad}} \emph {et~al.} (\bibinfo {collaboration} {ATLAS}),\ }\href {\doibase
  10.1016/j.physletb.2012.08.020} {\bibfield  {journal} {\bibinfo  {journal}
  {Phys. Lett. B}\ }\textbf {\bibinfo {volume} {716}},\ \bibinfo {pages} {1}
  (\bibinfo {year} {2012})},\ \Eprint {http://arxiv.org/abs/1207.7214}
  {arXiv:1207.7214 [hep-ex]} \BibitemShut {NoStop}%
\bibitem [{\citenamefont {Chatrchyan}\ \emph {et~al.}(2012)\citenamefont
  {Chatrchyan} \emph {et~al.}}]{CMS:2012qbp}%
  \BibitemOpen
  \bibfield  {author} {\bibinfo {author} {\bibfnamefont {S.}~\bibnamefont
  {Chatrchyan}} \emph {et~al.} (\bibinfo {collaboration} {CMS}),\ }\href
  {\doibase 10.1016/j.physletb.2012.08.021} {\bibfield  {journal} {\bibinfo
  {journal} {Phys. Lett. B}\ }\textbf {\bibinfo {volume} {716}},\ \bibinfo
  {pages} {30} (\bibinfo {year} {2012})},\ \Eprint
  {http://arxiv.org/abs/1207.7235} {arXiv:1207.7235 [hep-ex]} \BibitemShut
  {NoStop}%
\end{thebibliography}%

\appendix

\section{Details on the computation of $M_{12}^{q,\mathrm{SM}}$ and $\Gamma_{12}^{q,\mathrm{SM}}$}\label{App:DetailsSM}
The theoretical determination of $M_{12}^{q,\mathrm{SM}}$ and $\Gamma_{12}^{q,\mathrm{SM}}$ is broadly addressed in the literature. However, different choices regarding the definition of the relevant matrix elements and normalization factors are usually encountered. In the light of this, we consider it might be useful for the reader to settle down which convention we have adopted and compare with other references in order to avoid potential confusion on that respect. Furthermore, we take the opportunity to provide additional details on the technical aspects and the physical meaning of the different approximations used in the computation of the mixing matrix elements. 

As previously introduced, $M_{12}^{q}$ and $\Gamma_{12}^{q}$ are the off-diagonal elements of the $2\times 2$ Hamiltonian in Eq.~\eqref{eq:Hamiltonian2x2}, which acts on the Hilbert space of the $\{|B_q\rangle, |\bar{B}_q\rangle\}$ mesons and controls their time evolution. The effects of the underlying fundamental physics can be encoded into the matrix elements of this Hamiltonian $\mathcal{H}^q$ using the framework of perturbation theory, in particular:
\begin{equation}\label{eq:M12-perturbation-theory}
    M_{12}^q = \langle B_q | \mathcal{H}_W | \bar{B}_q \rangle + \sum_n \mathrm{P} \frac{\langle B_q | \mathcal{H}_W | n \rangle \langle n | \mathcal{H}_W | \bar{B}_q \rangle}{M_{B_q} - E_n} + \dots,
\end{equation}
\begin{equation}\label{eq:Gamma12-perturbation-theory}
    \Gamma_{12}^q = 2\pi \sum_n \delta(M_{B_q} - E_n) \langle B_q | \mathcal{H}_W | n \rangle \langle n | \mathcal{H}_W | \bar{B}_q \rangle + \dots,
\end{equation}
where the summation over the $n$ states does not include the $B_q$ and $\bar{B}_q$ mesons themselves, and P stands for ``principal part''. In the SM, it is clear that $\mathcal{H}_W$ denotes the standard weak interactions, which play the role of the perturbation in the previous expansion and are responsible for the flavor-changing transitions we are interested in. Likewise, it is important to remark that Eqs.~\eqref{eq:M12-perturbation-theory} and \eqref{eq:Gamma12-perturbation-theory} are obtained using a non-covariant normalization of the states, that is, $\langle \vec{p} | \vec{p}^{\,\prime} \rangle = (2\pi)^3\delta^{(3)}(\vec{p}-\vec{p}^{\,\prime})$, and they are usually written in this way in the literature \cite{Branco:1999fs,Bigi:2000yz}. In this work, we are adopting a covariant (relativistic) normalization, $\langle \vec{p} | \vec{p}^{\,\prime} \rangle = 2E_{\vec{p}}\,(2\pi)^3\delta^{(3)}(\vec{p}-\vec{p}^{\,\prime})$; therefore, at some point one has to correct the previous equations with a factor $(2M_{B_q})^{-1}$, since we are considering the mesons to be in their rest frame.

Let us first analyze the contributions to $M_{12}^{q,\mathrm{SM}}$. It is clear that the first order term in the perturbative expansion represents a $\Delta B = 2$ transition $\bar{B}_q \rightarrow B_q$. In the SM, the leading contribution mediating this process arises at fourth order in the weak coupling, $\mathcal{O}(g^4)$, through box diagrams with virtual (off-shell) up-type quarks and $W$ bosons exchange, as depicted in Fig.~\ref{fig:Box-SM}. This means that $M_{12}^{q,\mathrm{SM}}$ is sensitive to heavy virtual particles running in the loop, and thus to potential heavy NP effects as well. 

Once the one-loop integral is performed, one can insert the resulting local $\mathcal{H}_{\mathrm{dis,eff}}^{\Delta B = 2}$ into the perturbative expansion and compute $M_{12}^{q,\mathrm{SM}}$. The details of the strong interactions that bind the quarks together into the meson states will be encoded appropriately in the relevant matrix elements of the dimension-6 local operator involving the quark fields. Note that the subscript ``dis'' in the effective Hamiltonian states for \emph{dispersive}, as $M_{12}^{q,\mathrm{SM}}$ arises from the real part of the box diagrams.

This first order term in the perturbative expansion of $M_{12}^{q,\mathrm{SM}}$ is usually referred to as \textit{short distance} contribution. The reason is because $\Delta t \sim (\Delta E)^{-1} = (M - m_b)^{-1} \ll \Lambda_{\mathrm{QCD}}^{-1}$ (being $M = M_W, m_t$), that is, the time scale of the interaction is much smaller than the characteristic scale of strong interactions. 

On the other hand, the second order term in Eq.~\eqref{eq:M12-perturbation-theory} represents two $\Delta B = 1$ transitions involving low-mass (on-shell) intermediate states that are common to both $B_q$ and $\bar{B}_q$, and thus they represent \textit{long distance} contributions. These are further suppressed by $M_{B_q} \sim m_b$ and will be neglected in the computation of $M_{12}^{q,\mathrm{SM}}$.

A simplifying assumption can be carried out when computing the dispersive part of the box diagrams, namely, taking the limit of massless external particles and zero external momentum. The latter approximation is justified because the relevant contribution to the one-loop integral lies around the interval centered at $q \sim M$ (being $M$ the largest scale in the box, i.e., $m_t$ or $M_W$ for the SM case), which is a much larger energy scale than the one associated to the external states. Taking this into account and following \cite{Romao:2012pq} with the choice $\eta = 1$, the SM amplitude in terms of spinors can be written as
\begin{equation}\label{eq:amplitudeM12SM}
    \begin{split}
          i\mathcal{M}_M^{\mathrm{SM}} =& -i \frac{G_F^2 M_W^2}{2\pi^2}\mathcal{F}_0\\
          &\times\bigg\{\left[\bar{u}_q(0)\gamma^\mu L v_b(0)\right]\left[\bar{v}_q(0)\gamma_\mu L u_b(0)\right]\delta_{\alpha_1\alpha_2}\delta_{\beta_1\beta_2} \\
          &-\left[\bar{v}_q(0)\gamma^\mu L v_b(0)\right]\left[\bar{u}_q(0)\gamma_\mu L u_b(0)\right]\delta_{\alpha_1\beta_1}\delta_{\alpha_2\beta_2}\bigg\}\\
          =&\ i\frac{G_F^2 M_W^2}{2\pi^2}\mathcal{F}_0 \left[\bar{v}_q(0)\gamma^\mu L v_b(0)\right]\left[\bar{u}_q(0)\gamma_\mu L u_b(0)\right]\\
          &\times\left[\delta_{\alpha_1\beta_1}\delta_{\alpha_2\beta_2}+\delta_{\alpha_1\alpha_2}\delta_{\beta_1\beta_2}\right],
    \end{split}
\end{equation}
being $L = (1 - \gamma_5)/2$, and where
\begin{equation}
    \mathcal{F}_0 = (\lambda_{bq}^t)^2 S_0(x_t) + (\lambda_{bq}^c)^2 S_0(x_c) + 2\lambda_{bq}^c \lambda_{bq}^t S_0(x_c,x_t),
\end{equation}
with $\lambda^\alpha_{bq} \equiv V_{\alpha b}V_{\alpha q}^{*}$, $x_\alpha = m_\alpha^2/M_W^2$, and 
\begin{equation}
    \begin{split}
        S_0(x,y) = xy\bigg[ &-\frac{3}{4(1-x)(1-y)} \\
        &+\left(1-2x+\frac{x^2}{4}\right)\frac{\ln{x}}{(1-x)^2(x-y)}\\
        &+\left(1-2y+\frac{y^2}{4}\right)\frac{\ln{y}}{(1-y)^2(y-x)}\bigg],
    \end{split}
\end{equation}
\begin{equation}
    S_0(x) \equiv \lim_{y \rightarrow x} S_0(x,y).
\end{equation}
The limit of $m_u \rightarrow 0$ has been considered. Color indices $\alpha_1$, $\alpha_2$, $\beta_1$, and $\beta_2$ in Eq.~\eqref{eq:amplitudeM12SM} correspond to the assignment, in Fig.~\ref{fig:Box-SM}: $\alpha_1$ to the up-left incoming $b$, and then $\beta_1$, $\beta_2$, $\alpha_2$ clockwise to the external $q$, $\bar b$, $\bar q$. Numerically, the dominant contribution comes from virtual top quarks running in the loop because $m_t \gg m_c$, and thus we only keep this term in the following. It is easy to see that the first product of spinor bilinears accounts for the diagram in Fig.~\ref{sfig:Box1-SM}, while the second one arises from the topology in Fig.~\ref{sfig:Box2-SM}. They are Fierz-related, yielding the final compact expression in Eq.~\eqref{eq:amplitudeM12SM}. In terms of an effective $\Delta B = 2$ Hamiltonian, $\mathcal{M}_M^{\mathrm{SM}}$ arises from the matrix element of a single dimension-6 local operator $\left[\bar{q}_{\alpha}(x)\gamma^\mu L b_{\alpha}(x)\right]\left[\bar{q}_{\beta}(x)\gamma_\mu L b_{\beta}(x)\right]$. With two pairs of identical operators, contraction of the fields with external particle states requires some care, yielding
\begin{equation}
\begin{split}
    &\langle B_q|\left[\bar{q}_{\alpha}(x)\gamma^\mu L b_{\alpha}(x)\right]\left[\bar{q}_{\beta}(x)\gamma_\mu L b_{\beta}(x)\right]|\bar B_q\rangle\\ &\ =
    2\left[\bar{v}_q\gamma^\mu L v_b\right]\left[\bar{u}_q\gamma_\mu L u_b\right]\left[\delta_{\alpha_1\beta_1}\delta_{\alpha_2\beta_2}+\delta_{\alpha_1\alpha_2}\delta_{\beta_1\beta_2}\right].
\end{split}
\end{equation}
One can finally write
\begin{equation}
    \mathcal{H}_{\mathrm{dis,eff}}^{\Delta B = 2}(x) = \frac{G_F^2 M_W^2}{16\pi^2}\mathcal{F}_0 O_{V-A}^{\Delta B = 2} + \mathrm{h.c.},
\end{equation}
with the dimension-6 local operator $O_{V-A}^{\Delta B = 2} \equiv \bar{q}_{\alpha}(x)\gamma^\mu (1-\gamma_5) b_{\alpha}(x) \bar{q}_{\beta}(x)\gamma_\mu (1-\gamma_5) b_{\beta}(x)$. Therefore, one can obtain $M_{12}^{q,\mathrm{SM}}$, at lowest order in perturbation theory, as
\begin{equation}\label{eq:M12-matrix-element}
    M_{12}^{q,\mathrm{SM}} = \frac{1}{2M_{B_q}} \langle B_q | \mathcal{H}_{W,\mathrm{eff}}^{\Delta B = 2} | \bar{B}_q \rangle.
\end{equation}
Note that we have inserted at this point a $(2M_{B_q})^{-1}$ factor, so that the meson states that appear in Eq.~\eqref{eq:M12-matrix-element} are covariantly normalized.

In principle, in order to compute the matrix element of the operator $O_{V-A}^{\Delta B = 2}$ between these meson states one has to rely on the vacuum insertion approximation. This means to separate the matrix element of our four-quark operator into the product of two matrix elements of quark bilinears by inserting solely the vacuum state, instead of a complete set of states. The correction to this approximation is expressed in terms of a bag parameter that accounts for all other states neglected in the sum. Therefore, in the relativistic normalization approach, one can write (see App.~C of~\cite{Branco:1999fs}):
\begin{equation}\label{eq:VIA}
    \langle B_q | O_{V-A}^{\Delta B = 2} | \bar{B}_q \rangle = -\frac{8}{3} e^{i(\xi_b - \xi_q - \xi_{B_q})} M_{B_q}^2 f_{B_q}^2 B_{B_q},
\end{equation}
where $\xi_{B_q}$, $\xi_b$ and $\xi_q$ are the CP transformation phases of the $B_q$ meson state and the quark fields, $M_{B_q}$ is the common mass of the $B$ mesons, $f_{B_q}$ is the $B$ meson decay constant encoding all the information about strong interactions, and $B_{B_q}$ is the bag factor. The meson decay constant and the bag parameter are determined non-perturbatively in the framework of QCD sum rules or lattice QCD. One usually assumes a CP transformation in which $\xi_b = \xi_q = 0$. Furthermore, in this work we are assuming $\xi_{B_q} = \pi$, i.e., $\mathcal{CP} | B_q \rangle = - | \bar{B}_q \rangle$. With this choices, we can finally write the result for $M_{12}^{q,\mathrm{SM}}$:
\begin{equation}
    M_{12}^{q,\mathrm{SM}} = \frac{G_F^2 M_W^2}{12\pi^2} (\lambda_{bq}^t)^2 S_0 (x_t) M_{B_q} f_{B_q}^2 B_{B_q} \hat{\eta}_B.
\end{equation}
For completeness, we have included in the final result the factor $\hat{\eta}_B$ which takes into account short distance (perturbative) QCD corrections at two loops. In our convention, all renormalization scale and scheme dependence is translated into the bag parameter.

At this point, we find useful to point out that the majority of references addressing the study of neutral meson mixing make use of the covariant normalization of the states, such as \cite{Buchalla:1995vs,Ciuchini:2003ww,Lenz:2006hd} and more recent references. However, for the sake of clarity, we should highlight that other works like \cite{Cheng:1982hq,Buras:1984pq,Branco:1999fs,Bigi:2000yz} opt to write Eqs.~\eqref{eq:M12-matrix-element} and \eqref{eq:VIA} maintaining the non-covariant normalization of the states, such that
\begin{equation}
    M_{12}^{q,\mathrm{SM}} = \langle B_q | \mathcal{H}_{W,\mathrm{eff}}^{\Delta B = 2} | \bar{B}_q \rangle,
\end{equation}
and
\begin{equation}
    \langle B_q | O_{V-A}^{\Delta B = 2} | \bar{B}_q \rangle = -\frac{8}{6} e^{i(\xi_b - \xi_q - \xi_{B_q})} M_{B_q} f_{B_q}^2 B_{B_q}.
\end{equation}
Likewise, there might be different choices for the CP transformation phase $\xi_{B_q}$; e.g., \cite{Bigi:2000yz} sets $\xi_{B_q} = 0$.

Let us now move to the analysis of $\Gamma_{12}^{q,\mathrm{SM}}$. It accounts for the $\bar{B}_q \rightarrow B_q$ transition via real (on-shell) intermediate states, i.e., through decay modes that are common to $B_q$ and $\bar{B}_q$. Therefore, $\Gamma_{12}^{q,\mathrm{SM}}$ is sensitive to light particles with masses below $M_{B_q} \sim m_b$, such as the up and charm quarks. 

The theoretical determination of $\Gamma_{12}^{q,\mathrm{SM}}$ is cumbersome since it is not the matrix element of a local operator, as one can readily check in Eq.~\eqref{eq:Gamma12-perturbation-theory}. Alternatively, one relies on the validity of the so-called heavy quark expansion, in which $\Gamma_{12}^{q,\mathrm{SM}}$ is expressed as a power series in $\Lambda/m_b$, being $\Lambda$ a mass scale denoting non-perturbative effects, and the strong coupling $\alpha_s$ (see, e.g., \cite{Lenz:2014jha} for a detailed description). Although this is a well-known approach where higher order corrections have already been calculated, it might be instructive for the reader to briefly review the details on the computation of the lowest order contribution to $\Gamma_{12}^{q,\mathrm{SM}}$ in the previous expansion \cite{Ellis:1977uk,Hagelin:1981zk,Franco:1981ea,Chau:1982da,Buras:1984pq,Khoze:1986fa}, that is, at $\mathcal{O}(\Lambda^3/m_b^3)$ and $\mathcal{O}(\alpha_s^0)$ (LO) in QCD.

To this aim, one has to extract the \textit{absorptive} (imaginary) part of the box diagrams in Fig.~\ref{fig:Box-SM}. In that sense, a simplifying assumption can be carried out, namely, neglecting the momentum dependence in the bosonic propagators, since they are bounded to be $q \lesssim m_b \ll M_W$. Furthermore, we take the momentum of external light quarks $k,k^\prime \rightarrow 0$.\footnote{This approximation is justified because $p^2 = p^{\prime 2} = m_b^2 \gg k^2 = k^{\prime 2} = m_q^2$.} With these approximations, one is lead to find that the absorptive part of the amplitude is:
\begin{equation}\label{eq:amplitudeGamma12SM}
    \begin{split}
        i\mathcal{M}_\Gamma^{\mathrm{SM}}=&\ i\frac{G_F^2}{2\pi^2}\sum_{i,j=u,c} \lambda_{bq}^{i}\lambda_{bq}^{j}\\
        \times &\bigg\{A_{ij}\bigg(\left[\bar{v}_q(0) R v_b(p^\prime)\right]\left[\bar{u}_q(0) R u_b(p)\right] \\
        &+\left[\bar{v}_q(0) R u_b(p)\right]\left[\bar{u}_q(0) R v_b(p^\prime)\right]\bigg)\\
        &+B_{ij}\bigg(\left[\bar{v}_q(0)\gamma^\mu L v_b(p^\prime)\right]\left[\bar{u}_q(0)\gamma_\mu L u_b(p)\right] \\
        &+\left[\bar{v}_q(0)\gamma^\mu L u_b(p)\right]\left[\bar{u}_q(0)\gamma_\mu L v_b(p^\prime)\right]\bigg)\bigg\},
    \end{split}
\end{equation}
where $R = (1 + \gamma_5)/2$, and with
\begin{equation}
    \begin{split}
        A_{ij} =&\ \pi \frac{2(m_b^4-2m_i^4-2m_j^4+m_b^2m_i^2+m_b^2m_j^2+4m_i^2m_j^2)}{3m_b^4}\\
        &\times \sqrt{\lambda(m_b^2,m_i^2,m_j^2)},
    \end{split}
\end{equation}
\begin{equation}
        B_{ij} = -\pi \frac{\lambda(m_b^2,m_i^2,m_j^2)\sqrt{\lambda(m_b^2,m_i^2,m_j^2)}}{3m_b^4},
\end{equation}
being $\lambda(x,y,z)=x^2+y^2+z^2-2xy-2xz-2yz$ the Källen function. It is important to note that the sum only includes the contributions from up and charm quarks. In this case, besides $V-A$ spinor bilinears, we get additional $S+P$ currents that would be obtained from the dimension-6 operator $\left[\bar{q}(x) R b(x)\right]\left[\bar{q}(x) R b(x)\right]$, with the corresponding contraction of the fields to the external particles states. The construction of the effective Hamiltonian is analogous to the previous one:
\begin{equation}
    \begin{split}
        &\mathcal{H}_{\mathrm{abs,eff}}^{\Delta B = 2}(x) =\\
        &=-\frac{G_F^2}{16\pi^2} \sum_{i,j=u,c} \lambda_{bq}^{i}\lambda_{bq}^{j} \left(A_{ij}O_{S+P}^{\Delta B = 2} + B_{ij}O_{V-A}^{\Delta B = 2}\right) + \mathrm{h.c.},
    \end{split} 
\end{equation}
where $O_{S+P}^{\Delta B = 2} \equiv \left[\bar{q}(x) (1+\gamma_5) b(x)\right]\left[\bar{q}(x) (1+\gamma_5) b(x)\right]$. Then, one can obtain $\Gamma_{12}^{q,\mathrm{SM}}$ from
\begin{equation}\label{eq:Gamma12-matrix-element}
    \frac{\Gamma_{12}^{q,\mathrm{SM}}}{2} = -\frac{1}{2M_{B_q}} \langle B_q | \mathcal{H}_{\mathrm{abs,eff}}^{\Delta B = 2} | \bar{B}_q \rangle.
\end{equation}
The previous formula constitutes an assumption to reduce a complicated hadron phase space problem into a manageable one-loop integral. Again, we are using the covariant normalization of the meson states. 

The matrix element of the operator $O_{S+P}^{\Delta B = 2}$ is given by
\begin{equation}\label{eq:VIA2}
    \begin{split}
        \langle B_q | O_{S+P}^{\Delta B = 2} | \bar{B}_q \rangle =&\ \frac{5}{3} e^{i(\xi_b - \xi_q - \xi_{B_q})} \frac{M_{B_q}^4 f_{B_q}^2}{(m_b + m_q)^2} (B_{B_q})_S^\prime \\
        \simeq&\ \frac{5}{3} e^{i(\xi_b - \xi_q - \xi_{B_q})} M_{B_q}^2 f_{B_q}^2 (B_{B_q})_S^\prime,
    \end{split}
\end{equation}
where the $(B_{B_q})_S^\prime$ is the bag factor accounting for the deviation from the vacuum insertion approximation. The notation used here is intended to be easily matched with the one used in \cite{Artuso:2015swg}, namely $B_S^\prime$. Setting $\xi_{B_q} = \pi$ and $\xi_{b} = \xi_{q} = 0$, the result for $\Gamma_{12}^{q,\mathrm{SM}}$ turns out to be:
\begin{equation}
    \begin{split}
        \Gamma_{12}^{q,\mathrm{SM}} =& -\sum_{i,j=u,c} \lambda_{bq}^{i}\lambda_{bq}^{j} \frac{G_F^2 M_{B_q} f_{B_q}^2 m_b^2}{24\pi} \\
        &\times \bigg\{ \frac{8}{3} B_{B_q} G_{ij} +\frac{5}{3} (B_{B_q})_S^\prime G_{ij}^S \bigg\},
    \end{split}
\end{equation}
with
\begin{equation}
        G_{ij} = \frac{\lambda(m_b^2,m_i^2,m_j^2)\sqrt{\lambda(m_b^2,m_i^2,m_j^2)}}{2m_b^6},
\end{equation}
\begin{equation}
    \begin{split}
        G_{ij}^S =&\ \frac{2(m_b^4-2m_i^4-2m_j^4+m_b^2m_i^2+m_b^2m_j^2+4m_i^2m_j^2)}{m_b^6}\\
        &\times \sqrt{\lambda(m_b^2,m_i^2,m_j^2)}.
    \end{split}
\end{equation}
Given in this form, one can easily compare the result with Eqs. (58) and (71) from \cite{Artuso:2015swg} in the context of the heavy quark expansion. At this order, $\Lambda^3 \sim 16\pi^2 M_{B_q} f_{B_q}^2 \times$ ``numerical suppression factor'' ~\cite{Lenz:2014jha}. Currently, higher order QCD corrections are only known for the leading term in the $\Lambda/m_b$ expansion, while the contribution to the subleading powers is only known at LO in QCD. For a recent update on the state of the art, see \cite{Dziurda:2024hrg}, Table 6 from \cite{Albrecht:2024oyn}, and references therein.  

For completeness, we include here the alternative form for Eqs.~\eqref{eq:Gamma12-matrix-element} and \eqref{eq:VIA2} that makes use of the non-covariant normalization of the states, i.e.,
\begin{equation}
    \frac{\Gamma_{12}^{q,\mathrm{SM}}}{2} = - \langle B_q | \mathcal{H}_{\mathrm{abs,eff}}^{\Delta B = 2} | \bar{B}_q \rangle,
\end{equation}
and
\begin{equation}
    \langle B_q | O_{S+P}^{\Delta B = 2} | \bar{B}_q \rangle \simeq \frac{5}{6} e^{i(\xi_b - \xi_q - \xi_{B_q})} M_{B_q} f_{B_q}^2 (B_{B_q})_S^\prime.
\end{equation}

\section{Enhancements of $A_{\rm SL}^q$ in detail}\label{app:G12M12analysis}
The enhancement of $A^d_{\mathrm{SL}}$ and $A^s_{\mathrm{SL}}$ with respect to the SM has been discussed in detail in the different sections of this work. It is nevertheless interesting to analyze further how that enhancement is actually achieved, considering that different enhancing mechanisms have been invoked. Although the physically meaningful, rephasing invariant, quantities of interest are the semileptonic asymmetries, one might wish to uncover if their enhancement arises from the modifications introduced in $\Gamma_{12}^q$, $M_{12}^q$, or both. It may seem that such an exercise is to some extent futile and meaningless, since their moduli are rephasing invariant, but not their arguments. In this sense the following discussion is to be understood as addressing their room for variation within the parametric freedom available in each scenario. Although this is, in principle, an exercise within a given phase convention, it might shed some additional light on the inner workings of the main results. We proceed as follows.

\begin{figure*}[htbp]
\begin{center}
\subfloat[\label{sfig:NP33-argM12-vs-argG12-normalized-Bd}]{\includegraphics[width=0.45\textwidth]{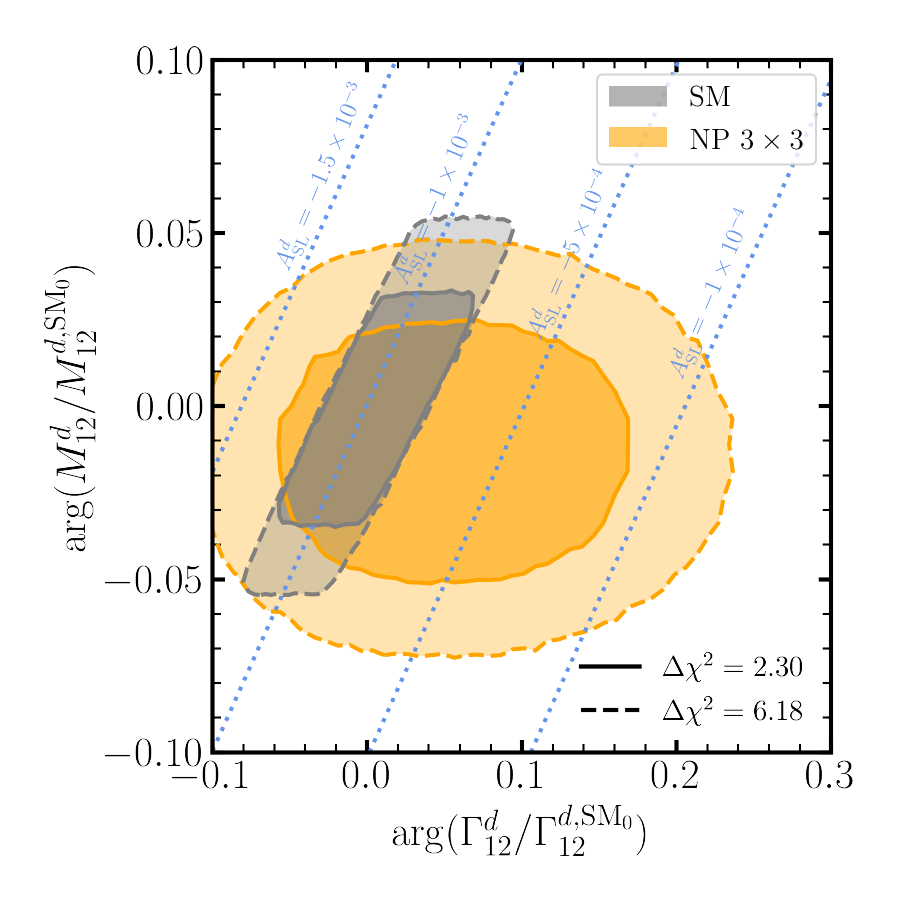}}
\subfloat[\label{sfig:DVLQ-argM12-vs-argG12-normalized-Bd}]{\includegraphics[width=0.45\textwidth]{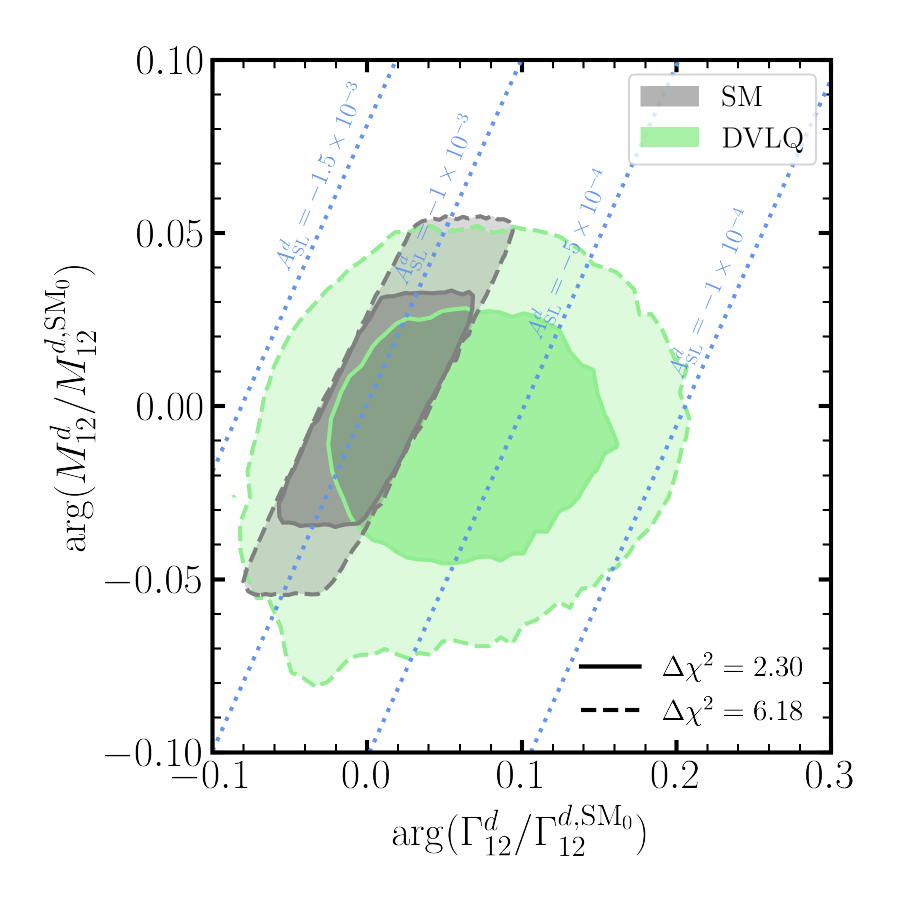}}\\
\subfloat[\label{sfig:NP33-argM12-vs-argG12-normalized-Bs}]{\includegraphics[width=0.45\textwidth]{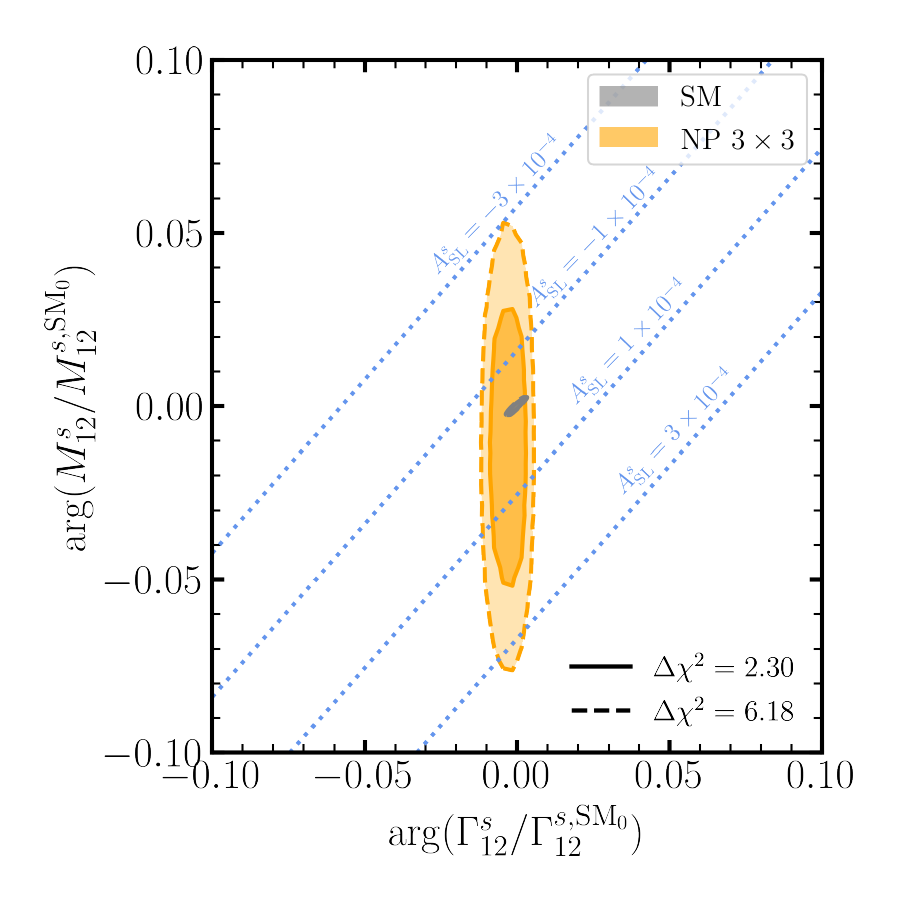}}
\subfloat[\label{sfig:DVLQ-argM12-vs-argG12-normalized-Bs}]{\includegraphics[width=0.45\textwidth]{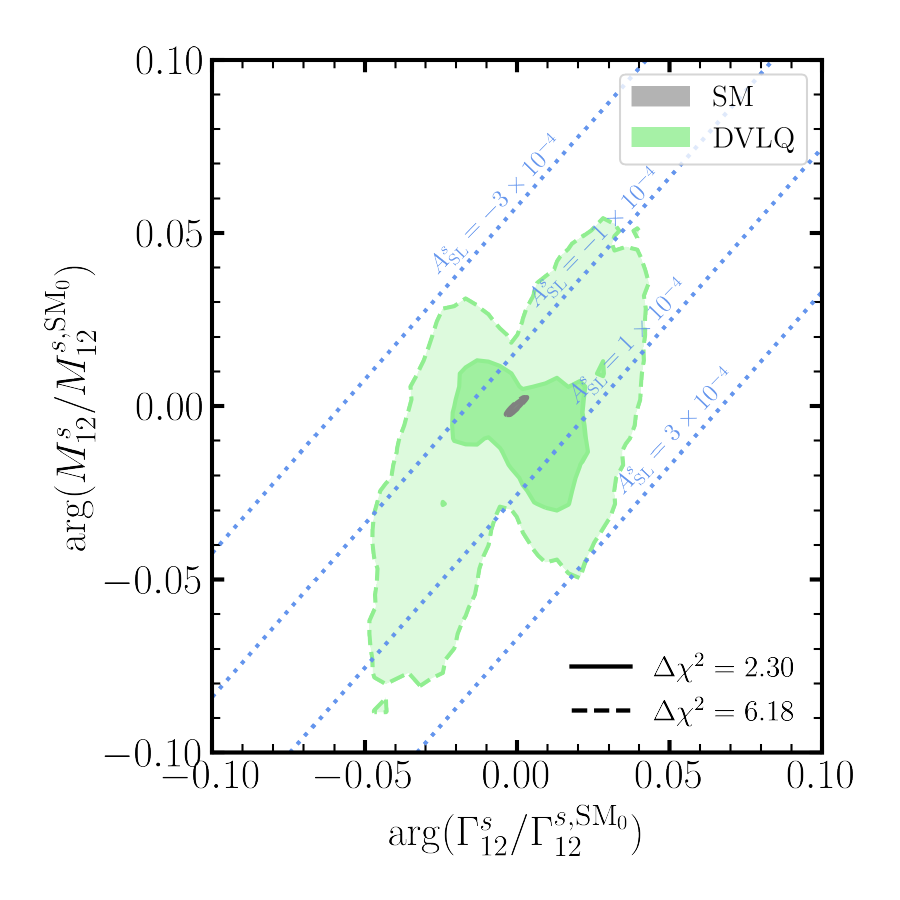}}\\
\caption{\justifying Allowed regions in the $\arg(M_{12}^q/M_{12}^{q,\mathrm{SM}_0})$ vs.~$\arg(\Gamma_{12}^q/\Gamma_{12}^{q,\mathrm{SM}_0})$ plane for $B_d$ (upper panels) and $B_s$ (lower panels) regarding the analyses of Sec.~\ref{SEC:NPM12q} and \ref{SEC:VLQ}, together with the gray SM region as discussed in the text. Dashed contours of the semileptonic asymmetries are included for reference: they are represented assuming $|M_{12}^q/M_{12}^{q,\mathrm{SM}_0}|=|\Gamma_{12}^q/\Gamma_{12}^{q,\mathrm{SM}_0}|=1$. Notice the change of scale in the horizontal axis from the upper panels to the lower ones.}\label{fig:argM12-vs-argG12-normalized}
\end{center}
\end{figure*}

From a strictly SM analysis with the same set of constraints used in the analyses BSM, we take the best fit point values of the CKM parameters, of the lattice parameters $f_{B_d}^2 B_{B_d}$ and $f_{B_s}^2 B_{B_s}$, and of the parameters\footnote{On this respect, since the SM fit is insensitive to these parameters (in the considered ranges, their profile likelihoods obtained in such a fit are simply flat), it would be rather meaningless to consider best fit values of $a_q$, $b_q$, $c_q$, and we choose to take the central values in Eqs.~\eqref{eq:cdadbd}--\eqref{eq:csasbs}. The following results would be essentially unchanged had we used best fit values.} $a_q$, $b_q$, $c_q$ entering $\Gamma_{12}^q$. With these input values, we compute reference SM values of $\Gamma_{12}^q$ and $M_{12}^q$, hereafter $\Gamma_{12}^{q,\mathrm{SM}_0}$ and $M_{12}^{q,\mathrm{SM}_0}$. We can now consider in our analyses $\Gamma_{12}^q/\Gamma_{12}^{q,\mathrm{SM}_0}$ and $M_{12}^q/M_{12}^{q,\mathrm{SM}_0}$, with $\Gamma_{12}^q$ and $M_{12}^q$ in the corresponding scenario. This will provide information on the room that each scenario allows for deviations with respect to a SM reference value. For completeness, the same is also done for a ``SM itself'' analysis. The most relevant results are shown in Fig.~\ref{fig:argM12-vs-argG12-normalized}. For the vector-like quark scenarios, we only display the DVLQ case since the corresponding results for the UVLQ case are not significantly different. 

For clarity in the discussion to follow, let us recall first some basic implications of $3\times 3$ unitarity of the CKM matrix. In the $bd$ unitarity triangle relation $\lambda^t_{bd}+\lambda^c_{bd}+\lambda^u_{bd}=0$ all three sides have size $\mathcal O(\lambda^3)$, while in the corresponding $bs$ relation $\lambda^t_{bs}+\lambda^c_{bs}+\lambda^u_{bs}=0$, $\lambda^t_{bs}\sim \mathcal O(\lambda^2)$, $\lambda^c_{bs}\sim \mathcal O(\lambda^2)$ and $\lambda^u_{bs}\sim \mathcal O(\lambda^4)$ (the triangle is ``squashed''). In terms of the room for variation that the tree-level dominated experimental data (moduli in the first two rows of CKM and the phase $\gamma$) allows, the $bd$ triangle can allow for larger absolute deformations (in particular the $\mathcal O(1)$ relative phase between $\lambda^t_{bd}$ and $\lambda^c_{bd}$, that is, $\beta$) than the $bs$ triangle can allow for (in particular the $\mathcal O(\lambda^2)$ relative phase between $\lambda^t_{bs}$ and $\lambda^c_{bs}$, that is, $\beta_s$).

We have
\begin{multline}
\frac{\Gamma_{12}^q}{\Gamma_{12}^{q,\mathrm{SM}_0}}=-[\Gamma_{12}^{q,uc}(\lambda^c_{bq}+\lambda^u_{bq})^2+(\Gamma_{12}^{q,cc}-\Gamma_{12}^{q,uc})(\lambda^c_{bq})^2\\ +(\Gamma_{12}^{q,uu}-\Gamma_{12}^{q,uc})(\lambda^u_{bq})^2]/\Gamma_{12}^{q,\mathrm{SM}_0},
\end{multline}
with $\Gamma_{12}^{q,cc}\simeq \Gamma_{12}^{q,uc}\simeq \Gamma_{12}^{q,uu}$ as commented in Sec.~\ref{SEC:BMesongeneral}; the first term gives the leading behaviour for both $q=d,s$. Through $3\times 3$ unitarity of CKM one can obviously identify the first term as $\Gamma_{12}^{q,uc}(\lambda^t_{bq})^2$; the point is that its phase, within the room for variation of CKM, has $\mathcal O(1)$ room for variation in the $B_d$ system while only $\mathcal O(\lambda^2)$ room for variation in the $B_s$ system. In $A^d_{\mathrm{SL}}$ and $A^s_{\mathrm{SL}}$ this variation is of course matched by $M_{12}^q\propto (\lambda^t_{bq})^2$ as the SM regions in Fig.~\ref{fig:argM12-vs-argG12-normalized} illustrate according to expectations.
\begin{center}
    \underline{\textit{$B_d$ system}}
\end{center}
Moving into the results shown in Fig.~\ref{fig:argM12-vs-argG12-normalized}, let us start with the $B_d$ system --Figs.~\ref{sfig:NP33-argM12-vs-argG12-normalized-Bd} and \ref{sfig:DVLQ-argM12-vs-argG12-normalized-Bd}--. The span of the SM region (displayed in both subfigures for comparison), essentially due to the uncertainties in the experimental constraints, is below the $\pm 0.1$ level in both $\arg(\Gamma_{12}^d/\Gamma_{12}^{d,\mathrm{SM}_0})$ and $\arg(M_{12}^d/M_{12}^{d,\mathrm{SM}_0})$. Most importantly, that region is aligned with the diagonal $\arg(\Gamma_{12}^d/\Gamma_{12}^{d,\mathrm{SM}_0})=\arg(M_{12}^d/M_{12}^{d,\mathrm{SM}_0})$, further illustrating the reduced range of variation of $A^d_{\mathrm{SL}}$ within the SM. Beyond the SM, it is interesting to notice that the allowed regions in the NP $3\times 3$ case in Fig.~\ref{sfig:NP33-argM12-vs-argG12-normalized-Bd} and in the DVLQ case in Fig.~\ref{sfig:DVLQ-argM12-vs-argG12-normalized-Bd} are rather similar: $\arg(M_{12}^d/M_{12}^{d,\mathrm{SM}_0})$ basically spans the same range as in the SM, while $\arg(\Gamma_{12}^d/\Gamma_{12}^{d,\mathrm{SM}_0})$ covers a much larger range. This is somehow unexpected. What Fig.~\ref{sfig:NP33-argM12-vs-argG12-normalized-Bd} shows is that, in the NP $3\times 3$ case, the freedom introduced by the new parameter $\phi^\Delta_d$ in $M_{12}^d$ does not enlarge the room for variation of $\arg(M_{12}^d)$ itself, but allows instead for sizable room for variation of $\arg(\Gamma_{12}^d)$, leading to the enhancement of $A^d_{\mathrm{SL}}$. This is so because the allowed ``deformations'' of the $bd$ unitarity triangle aforementioned allow for increased room in $\arg(\Gamma_{12}^d/\Gamma_{12}^{d,\mathrm{SM}_0})$ while the freedom in $\phi^\Delta_d$ can put ``in place'' $\arg(M_{12}^d)$ to satisfy the $J/\Psi K_S$ CP asymmetry constraint, which is quite restrictive. It does appear, observing Fig.~\ref{sfig:DVLQ-argM12-vs-argG12-normalized-Bd}, that a similar rationale accounts for the enhancement with respect to the SM of $A^d_{\mathrm{SL}}$ in the UVLQ and DVLQ cases, with the role of $\phi^\Delta_d$ played by the new contributions to $M_{12}^d$.
\begin{center}
    \underline{\textit{$B_s$ system}}
\end{center}
We now turn to the $B_s$ system. The SM allowed region displayed in Figs.~\ref{sfig:NP33-argM12-vs-argG12-normalized-Bs} and \ref{sfig:DVLQ-argM12-vs-argG12-normalized-Bs} is tiny, in line with the expectations discussed before. In the cases beyond the SM, the situation differs clearly from the $B_d$ system, and is not similar in the NP $3\times 3$ case and the VLQ scenarios. In the NP $3\times 3$ case, although there is some room for variation of $\arg(\Gamma_{12}^s/\Gamma_{12}^{s,\mathrm{SM}_0})$ at the $10^{-2}$ level, the relevant feature is the room for variation of $\arg(M_{12}^s/M_{12}^{s,\mathrm{SM}_0})$. Contrary to the $B_d$ system, the constraint imposed by the CP asymmetry in $J/\Psi \Phi$ is much less restrictive, and this is exploited by the freedom introduced through $\phi^\Delta_s$ to generate variations in $\arg(M_{12}^s/M_{12}^{s,\mathrm{SM}_0})$ at the $\pm 0.05$ level. In the DVLQ case, Fig.~\ref{sfig:DVLQ-argM12-vs-argG12-normalized-Bs}, in addition to the effect in $\arg(M_{12}^s/M_{12}^{s,\mathrm{SM}_0})$, sizable deviations in $\arg(\Gamma_{12}^s/\Gamma_{12}^{s,\mathrm{SM}_0})$ at almost the same level are also present. The allowed region is not a simple ellipse as in the NP $3\times 3$ case of Fig.~\ref{sfig:NP33-argM12-vs-argG12-normalized-Bs}, and this might point towards the more involved interplay of deviations of $3\times 3$ unitarity in CKM contributing new terms to $M_{12}^s$, and misaligning $\Gamma_{12}^s$ with respect to the SM: what is apparent is that deviations in $M_{12}^s$, in $\Gamma_{12}^s$, or in both $M_{12}^s$ and $\Gamma_{12}^s$, are present and can enhance $A^s_{\mathrm{SL}}$.

As a final comment, concerning $|M_{12}^q/M_{12}^{q,\mathrm{SM}_0}|$ and $|\Gamma_{12}^q/\Gamma_{12}^{q,\mathrm{SM}_0}|$, it is worth mentioning that we obtain
\begin{equation}
\begin{aligned}
& |M_{12}^d/M_{12}^{d,\mathrm{SM}_0}| = 1.00 \pm 0.01\,,\\
& |\Gamma_{12}^d/\Gamma_{12}^{d,\mathrm{SM}_0}| = 1.00\pm 0.15\,,\\
& |M_{12}^s/M_{12}^{s,\mathrm{SM}_0}| = 1.000 \pm 0.001\,,\\
& |\Gamma_{12}^s/\Gamma_{12}^{s,\mathrm{SM}_0}| = 1.00\pm 0.08\,.
\end{aligned}
\end{equation}
For $|M_{12}^q/M_{12}^{q,\mathrm{SM}_0}|$ these results are as expected from the values of $\Delta M_q$, while $|\Gamma_{12}^q/\Gamma_{12}^{q,\mathrm{SM}_0}|$ have larger --$\mathcal O(10\%)$-- room for variation: such a variation, on its own, can only account for an $\mathcal O(10\%)$ variation of $A^d_{\mathrm{SL}}$ and $A^s_{\mathrm{SL}}$ with respect to the SM values, making it clear that it is rather the arguments of $M_{12}^q$ and $\Gamma_{12}^q$ which are of interest to explore enhancements of the semileptonic asymmetries.

\section{Models with vector-like quark singlets}\label{App:ModelsVLQ}
We provide additional details concerning the models that include vector-like quark singlets in our analyses. Starting from the SM electroweak gauge group, $G \equiv SU(2)_L \otimes U(1)_Y$, the quark quantum numbers read
\begin{equation}
    \begin{split}
        &Q_L = \begin{pmatrix} P_L \\ N_L \end{pmatrix} \sim (2,1/6),\\
        &p_{L/R} \sim (1,2/3),\\
        &n_{L/R} \sim (1,-1/3),
    \end{split}
\end{equation}
where uppercase and lowercase letters are used to denote $SU(2)_L$ doublets and singlets, respectively. In addition to the SM matter content, we consider one up-type and one down-type vector-like singlets. In this way, we end up with a 4-dimensional flavor space that consists of 3 chiral quark generations and 1 vector-like generation. 

The Yukawa Lagrangian in the quark sector is then
\begin{equation}
    \begin{split}
         -\mathcal{L}_Y^{\mathrm{quark}} = &\ \bar{Q}_L Y_d n_R \Phi + \bar{Q}_L Y_u p_R \tilde{\Phi} \\
         &+ \bar{n}_L \mu_d n_R + \bar{p}_L \mu_u p_R + \mathrm{h.c.},
    \end{split}
\end{equation}
being $Y_d$ and $Y_u$ $3\times4$ complex matrices in flavor space, and $\mu_d$ and $\mu_u$ the new $1\times4$ complex Yukawa couplings that are allowed by the gauge symmetry. The scalar doublet $\Phi$ and its C-conjugate $\tilde{\Phi} \equiv i\sigma_2 \Phi^{*}$ remain the same as in the SM. The diagonalization of the mass matrices is carried out through $4\times4$ unitary transformations $U_{fX}$ (with $f = d, u$, and $X = L, R$) of the quark flavor fields into their mass eigenstates, namely
\begin{align}
    \begin{pmatrix} N_L \\ n_L \end{pmatrix} = U_{dL} d_L, \quad n_R = U_{dR} d_R, \\
    \begin{pmatrix} P_L \\ p_L \end{pmatrix} = U_{uL} u_L, \quad p_R = U_{uR} u_R,
\end{align}
with $U_{fX}^\dagger U_{fX} = U_{fX} U_{fX}^\dagger = 1_{4\times4}$. It is convenient to organize the left transformations $U_{fL}$ as
\begin{equation}
    U_{dL} = \begin{pmatrix} A_{dL} \\ B_{dL} \end{pmatrix}, \quad U_{uL} = \begin{pmatrix} A_{uL} \\ B_{uL} \end{pmatrix},
\end{equation}
so that $A_{fL}$ and $B_{fL}$ are $3\times4$ and $1\times4$ blocks related to the mixing of the components of the doublets and the singlets, respectively. 

Regarding the charged current sector, since it only involves $SU(2)_L$ doublet fields, one can readily check that the $4\times4$ CKM mixing matrix is given by
\begin{equation}
    V_L = A_{uL}^\dagger A_{dL},
\end{equation}
which is not unitary. In fact, the Hermitian combinations
\begin{align}
    &U_L \equiv V_L V_L^\dagger,\\
    &D_L \equiv V_L^\dagger V_L,\label{eq:DL-matrix}
\end{align}
will control the appearance of FCNC at tree-level.

The scalar and gauge interactions of quarks in these models are:
\begin{align}
        \mathcal{L}_{G^\pm} =& -\frac{\sqrt{2}}{v} \bar{u}_L V_L \mathcal{D}_d d_R G^{+} + \mathrm{h.c.} \notag \\
        &+ \frac{\sqrt{2}}{v} \bar{d}_L V_L^\dagger \mathcal{D}_u u_R G^{-} + \mathrm{h.c.},
\end{align}
\begin{align}
    \mathcal{L}_{G^0} =& -\frac{i}{v} \bar{d}_L\mathcal{D}_d d_R G^{0} + \frac{i}{v} \bar{u}_L \mathcal{D}_u u_R G^{0} +\mathrm{h.c.} \notag \\
    &-\frac{i}{v} \bar{d}_L (D_L - 1) \mathcal{D}_d d_R G^{0} +\mathrm{h.c.} \notag \\
    &+ \frac{i}{v} \bar{u}_L (U_L - 1) \mathcal{D}_u u_R G^{0} +\mathrm{h.c.},
\end{align}
\begin{align}
    \mathcal{L}_{h} =& -\frac{1}{v} \bar{d}_L\mathcal{D}_d d_R h - \frac{1}{v} \bar{u}_L \mathcal{D}_u u_R h +\mathrm{h.c.} \notag \\
    &-\frac{1}{v} \bar{d}_L (D_L - 1) \mathcal{D}_d d_R h +\mathrm{h.c.} \notag\\
    &-\frac{1}{v} \bar{u}_L (U_L - 1) \mathcal{D}_u u_R h +\mathrm{h.c.},
\end{align}
\begin{equation}
    \mathcal{L}_{W^\pm} = -\frac{g}{\sqrt{2}} W_\mu^\dagger \bar{u}_L V_L \gamma^\mu d_L + \mathrm{h.c.},
\end{equation}
\begin{align}
    \mathcal{L}_{Z} =& -\frac{g}{c_W} \bigg\{ -s_W^2 J_{\mathrm{EM}}^\mu + \frac{1}{2}\left(\bar{u}_L \gamma^\mu u_L - \bar{d}_L \gamma^\mu d_L \right) \notag \\
    &+ \frac{1}{2}\left[\bar{u}_L \gamma^\mu (U_L - 1) u_L - \bar{d}_L \gamma^\mu (D_L - 1)d_L\right] \bigg\} Z_\mu,
\end{align}
\begin{equation}
    \mathcal{L}_{A} = -e J_{\mathrm{EM}}^\mu A_\mu,
\end{equation}
where $v$ is the vacuum expectation value of the scalar doublet, $g$ is the $SU(2)_L$ coupling constant, $c_W \equiv \cos{\theta_W}$ is the weak mixing angle, $J_{\mathrm{EM}}^\mu$ stands for the electromagnetic current, and $\mathcal{D}_d$ and $\mathcal{D}_u$ are the diagonal mass matrices of down-type and up-type quarks, respectively. $G^0$ and $G^{\pm}$ denote the neutral and charged would-be Goldstone bosons, while $h$ is identified as the 125 GeV Higgs boson \cite{ATLAS:2012yve,CMS:2012qbp}. Notice the appearance of FCNC at tree-level mediated by the $Z$ boson and the neutral scalars $h$ and $G^0$, whose intensity is directly proportional to the deviation of unitarity of the CKM matrix, that is, $(D_L - 1)$ and $(U_L - 1)$.

Let us now particularize for the case of just one up-type or just one down-type vector-like singlet. 
\begin{center}
    \underline{\textit{UVLQ model: one extra up-type vector-like singlet}}
\end{center}
The CKM matrix has dimensions $4\times3$ and satisfies:
\begin{equation}
    V_L = A_{uL}^\dagger U_{dL}, \quad U_L = 1_{4\times4} -  B_{uL}^\dagger B_{uL}, \quad D_L = 1_{3\times3}.
\end{equation}
Although $V_L$ is not unitary, it is embedded in a larger $4\times4$ unitary matrix such that the unitarity relations between its columns still hold.

\begin{center}
    \underline{\textit{DVLQ model: one extra down-type vector-like singlet}}
\end{center}
The CKM matrix has dimensions $3\times4$ and satisfies:
\begin{equation}\label{eq:DVLQ-unit}
    V_L = U_{uL}^\dagger A_{dL}, \quad U_L = 1_{3\times3}, \quad D_L = 1_{4\times4} - B_{dL}^\dagger B_{dL}.
\end{equation}
In this case, $V_L$ is part of a larger $4\times4$ unitary matrix such that one can still apply the unitarity relations between rows.

\section{New contributions to $M_{12}^q$ in vector-like quark models}\label{app:M12VLQ}
In this appendix, we summarize how $M_{12}^q$ gets modified due to the inclusion of vector-like quarks to the SM matter content. In particular, we consider separately the UVLQ and DVLQ models with one additional up-type or down-type vector-like quark singlet, respectively.

On the one hand, in the UVLQ scenario the new contributions to $B_q$--$\bar{B}_q$ meson mixing arise from SM-like box diagrams, depicted in Fig.~\ref{fig:Box-SM}, with an additional heavy $T$ quark propagating in the loop. It is important to notice that, despite the CKM mixing matrix is no longer unitary, the orthogonality relation $\lambda_{bq}^{u} + \lambda_{bq}^{c} + \lambda_{bq}^{t} + \lambda_{bq}^{T} = 0$ holds, which ensures the gauge independence of the box diagrams as in the SM. All in all, $M_{12}^{q,\mathrm{UVLQ}}$ is given by
\begin{equation}\label{eq:M12up}
    \frac{M_{12}^{q,\mathrm{UVLQ}}}{M_{12}^{q,\mathrm{SM-like}}} = 1 + \frac{\lambda_{bq}^T}{\lambda_{bq}^t} \frac{C_1^\mathrm{up}(x_t,x_T)}{S_0(x_t)} + \left(\frac{\lambda_{bq}^T}{\lambda_{bq}^t}\right)^2 \frac{C_2^\mathrm{up}(x_T)}{S_0(x_t)},
\end{equation}
where
\begin{equation}
    C_1^\mathrm{up}(x_t,x_T) = 2 S_0(x_t,x_T), \quad  C_2^\mathrm{up}(x_T) = S_0(x_T),
\end{equation}
and $M_{12}^{q,\mathrm{SM-like}}$ the SM-like result computed in App.~\ref{App:DetailsSM} in the limit $m_u \rightarrow 0$, although in this case one could have different values for the CKM couplings since $3\times 3$ unitarity is not imposed. In Eq.~\eqref{eq:M12up}, we have only kept the terms that are numerically relevant, that is, those involving $t$ and $T$ quarks (with $m_T > 1.6$ TeV). As we can check, the new contributions in this scenario $C_1^{\mathrm{up}}$ and $C_2^{\mathrm{up}}$ appear linearly and quadratically in the deviation of unitarity $\lambda_{bq}^T$, respectively. Note that, in this calculation the two-loop perturbative QCD corrections, encoded in $\hat{\eta}_B$, are assumed to be similar to the SM case.

On the other hand, models that include one additional down-type vector-like singlet have been addressed in \cite{Barenboim:1997pf}, where a detailed discussion on the relevant contributions to $\Delta B = 2$ processes is presented. The deviation of unitarity of the CKM matrix implies that $\lambda_{bq}^{u} + \lambda_{bq}^{c} + \lambda_{bq}^{t} = (D_L)_{qb}$. In this scenario, we have new tree-level diagrams, as shown in Figs.~\ref{sfig:DVLQ-tree-channel-s} and \ref{sfig:DVLQ-tree-channel-t}, involving the flavor-changing coupling of the $Z$ boson to down-type quarks.\footnote{Diagrams involving neutral scalar bosons are suppressed by small external quark masses.} Therefore, they generate a contribution proportional to $G_F (D_L)_{qb}^2$. Additionally, box diagrams are still present, and their contribution can be split in three pieces: (i) SM-like, (ii) $\propto G_F \alpha (D_L)_{qb}$, and (iii) $\propto G_F \alpha (D_L)_{qb}^2$. The latter consists of a radiative correction to tree-level diagrams and thus it is neglected. On the other hand, the second piece is linear in $(D_L)_{qb}$, which means that, in the decoupling limit of the heavy down quark, it goes to zero more slowly than tree-level diagrams quadratic in $(D_L)_{qb}^2$. So, a priori, this second piece should not be neglected, at least in the limit of small $(D_L)_{qb}$. Finally, one should worry about cancelling the gauge dependence of the box contributions that are linear in $(D_L)_{qb}$. To this aim, different penguin topologies represented in Figs.~\ref{sfig:DVLQ-blob-left} and \ref{sfig:DVLQ-blob-up} must be added. They consist of a $Z$ tree-level flavor-changing vertex and another one where the change of flavor is obtained through the exchange of $W$ bosons at one loop. Then, summing up all the previous contributions, both tree-level and one-loop diagrams, $M_{12}^{q,\mathrm{DVLQ}}$ is given by
\begin{equation}\label{eq:M12down}
    \frac{M_{12}^{q,\mathrm{DVLQ}}}{M_{12}^{q,\mathrm{SM-like}}} = 1 + \frac{(D_L)_{qb}}{\lambda_{bq}^t} \frac{C_1^\mathrm{down}(x_t)}{S_0(x_t)} + \left(\frac{(D_L)_{qb}}{\lambda_{bq}^t}\right)^2 \frac{C_2^\mathrm{down}}{S_0(x_t)},
\end{equation}
where 
\begin{equation}
    C_1^\mathrm{down}(x_t) = -4 Y(x_t), \quad  C_2^\mathrm{down} = \frac{2\sqrt{2}\pi^2}{G_F M_W^2},
\end{equation}
and
\begin{equation}
    Y(x) = \frac{x}{4(x-1)}\left[x - 4 + \frac{3x\ln{x}}{x-1}\right],
\end{equation}
a well-known Inami-Lim function. Again, only those terms that are numerically relevant appear in Eq.~\eqref{eq:M12down}, where two new terms linear and quadratic in the deviation of unitarity $(D_L)_{qb}$ arise. 

\section{New contributions to $\Gamma_{12}^q$ and $M_{12}^q$ in the minimal $B$-Mesogenesis realization}\label{app:formulaeBMesogenesis}
For the interested reader, we collect in the following appendix the results of the computation of the diagrams in Fig.~\ref{fig:Box-BSM_psi} and Fig.~\ref{fig:Box-BSM_diquark}, together with the complete expressions of the relevant functions used in Sec.~\ref{SEC:G12q_BMesogenesis}.

On the one hand, the purely new physics contribution to $\Gamma_{12}^q$ and $M_{12}^q$ arising from the interactions that involve both the dark sector antibaryon and the color-triplet scalar (see Fig.~\ref{fig:Box-BSM_psi}) are given by:
\begin{align}
    \Gamma_{12}^{q,\mathrm{NP}}(\psi) &= -\frac{f_{B_q}^2 M_{B_q}}{256\pi} \frac{(y_{\psi q}y_{\psi b}^{*})^2 m_b^2}{M_Y^4} \\
    &\times \left(1-\frac{2}{3}\frac{m_\psi^2}{m_b^2}\right)\sqrt{1-4\frac{m_\psi^2}{m_b^2}}, \nonumber
\end{align}
and
\begin{equation}
    M_{12}^{q,\mathrm{NP}}(\psi) = \frac{f_{B_q}^2 M_{B_q}}{384\pi^2} \frac{(y_{\psi q}y_{\psi b}^{*})^2}{M_Y^2} G(x_{\psi Y}),
\end{equation}
with $x_{\psi Y} = m_\psi^2/M_Y^2$, and
\begin{equation}
    G(x) = \frac{1+x}{(1-x)^2}+\frac{2x\ln{x}}{(1-x)^3}.
\end{equation}
For the range of interest where $m_\psi \lesssim m_b/2$ and $M_Y > 500\,{\rm GeV}$, one has $G(x) \sim 1$. 

On the other hand, the couplings of the color-triplet scalar to up-type and down-type quarks lead to (see Fig.~\ref{fig:Box-BSM_diquark}):
\begin{align}
    \Gamma_{12}^{q,\mathrm{NP}}(\cancel{\psi}) &= \frac{f_{B_q}^2 M_{B_q}}{384\pi^2} \sum_{i,j=u,c} \frac{\pi \sqrt{\lambda(m_b^2,m_i^2,m_j^2)}}{m_b^2} \\ 
    &\times \left[ (V_{ib}V_{jq}^{*} y_{iq}y_{jb}^{*}) \frac{m_i m_j}{M_W^2 M_Y^2} 8 g^2 \right. \nonumber \\ 
    &\left.\,\,\,\,\,\,+\,(y_{iq}y_{ib}^{*} y_{jq}y_{jb}^{*}) \frac{m_b^2}{12M_Y^4} (8g_{2}^{ij}-5 g_{3}^{ij})\right]\,,\nonumber
\end{align}
and
\begin{align}
    M_{12}^{q,\mathrm{NP}}(\cancel{\psi}) = &-\frac{f_{B_q}^2 M_{B_q}}{384\pi^2} \\ &\times\sum_{i,j = u,c,t} \left[(V_{ib}V_{jq}^{*} y_{iq}y_{jb}^{*}) \frac{m_i m_j}{M_W^2 M_Y^2} g^2 f_{1}^{ij} \right. \nonumber\\ 
    &\left.\,\,\,\,\,\, \,\,\,\,\,\,\,\,\,\,\,\,\,\,\,\,\,\,\,\, -\,(y_{iq}y_{ib}^{*} y_{jq}y_{jb}^{*}) \frac{1}{M_Y^2} f_{2}^{ij}\right]\,\nonumber,
\end{align}
where the functions $g_2^{ij}$, $g_3^{ij}$, $f_1^{ij}$, and $f_2^{ij}$ only depend on the masses. On the one hand,
\begin{equation}
    g_2^{ij} = - \frac{\lambda(m_b^2,m_i^2,m_j^2)}{m_b^4},
\end{equation}

\begin{equation}
    g_3^{ij} = \frac{2(m_b^4 - 2m_i^4 - 2m_j^4 + m_b^2 m_i^2 + m_b^2 m_j^2 + 4 m_i^2 m_j^2)}{m_b^4}.
\end{equation}
Numerically, we find $g_{2}^{uu} = -1$, $g_{2}^{uc}=g_{2}^{cu} = -0.8$, $g_{2}^{cc} = -0.6$, while $g_{3}^{uu} = 2$, $g_{3}^{uc}=g_{3}^{cu} = 2.16$, $g_{3}^{cc} = 2.41$. On the other hand,

\begin{align}
    f_1^{ij} &= \frac{x_{iW} (x_{iW} - 4)\ln{x_{iY}}}{(x_{iW} - 1)(x_{iY} - 1)(x_{iW} - x_{jW})} \nonumber \\
    &+ \frac{x_{jW} (x_{jW} - 4)\ln{x_{jY}}}{(x_{jW} - 1)(x_{jY} - 1)(x_{jW} - x_{iW})} \nonumber\\
    &- \frac{3 \ln{x_{WY}}}{(x_{iW} - 1)(x_{jW} - 1)(x_{WY} - 1)}, \nonumber
\end{align}
\begin{align}
    f_2^{ij} &= \frac{1}{(x_{iY} - 1)(x_{jY} - 1)} \\
    &+ \frac{x_{iY}^2\ln{x_{iY}}}{(x_{iY} - x_{jY})(x_{iY} - 1)^2} \nonumber\\
    &+ \frac{x_{jY}^2\ln{x_{jY}}}{(x_{jY} - x_{iY})(x_{jY} - 1)^2}, \nonumber
\end{align}
with the usual definition $x_{ij} = m_i^2/m_j^2$, and $\lambda(x,y,z)$ the Källen function again. 
A rough approximation for these functions reads:
\begin{align}
& f_{1}^{ii}\simeq - 6\log m_i/M_Y\,, \\
& f_{1}^{tt} \simeq - 2\log m_t/M_Y\,,\\
& f_{1}^{ij}(i\neq j) \simeq (2/3) \log x_i/M_Y \log x_j/M_Y \,,\\
& f_2^{ij} \simeq -1\,.
\end{align}

\end{document}